\documentclass{aa}  

\usepackage[]{natbib}
\usepackage{lscape}
\usepackage{longtable}
\usepackage{float}
\usepackage{cuted}
\usepackage{caption}
\usepackage{multirow}
\usepackage{threeparttable}
\usepackage{nicefrac}
\usepackage{graphicx}
\usepackage{txfonts}
\usepackage{upgreek}

\usepackage{color}

\begin{document}

   \title{CHANG-ES XIII: Transport processes and the magnetic fields of NGC 4666 - indication of a reversing disk magnetic field}

   \author{Y. Stein \inst{1,2}
          \and
          R.-J. Dettmar \inst{2,3}
          \and
          J. Irwin \inst{4}
          \and
          R. Beck \inst{5}
          \and
          M. We\.zgowiec \inst{6}
          \and
          A. Miskolczi \inst{2}
          \and
          M. Krause \inst{5}
          \and
          V. Heesen \inst{7}
           \and
          T. Wiegert \inst{4}
         \and
         G. Heald \inst{8}
         \and
          R.A.M. Walterbos \inst{9}
         \and
         J.-T. Li \inst{10}
         \and
         M. Soida \inst{6}
          }

   \institute{Observatoire astronomique de Strasbourg, Universit\'e de Strasbourg, CNRS, UMR 7550, 11 rue de l'Universit\'e, 67000 Strasbourg, France; \email{yelena.stein@astro.unistra.fr.}
              \and
       Ruhr-University Bochum, Faculty of Physics and Astronomy, Astronomisches Institut (AIRUB), Germany.
              \and
       Ruhr-University Bochum, Faculty of Physics and Astronomy, Research Department of Plasmas with Complex Interactions, Germany.
              \and
              Department of Physics, Engineering, and  Astronomy, Queen's University, Kingston, Ontario, Canada, K7L 3N6.
              \and
              Max-Planck-Institut f\"ur Radioastronomie, Auf dem H\"ugel 69, 53121 Bonn, Germany.
              \and
              Obserwatorium Astronomiczne Uniwersytetu Jagiello\'nskiego, ul. Orla 171, 30-244 Krak\'ow, Poland.
               \and
               Universit\"at Hamburg, Hamburger Sternwarte, Gojenbergsweg 112, 21029 Hamburg, Germany.             
              \and
              CSIRO Astronomy and Space Science, 26 Dick Perry Avenue, Kensington, WA 6151, Australia.
              \and
              Department of Astronomy, New Mexico State University, PO Box 30001, MSC 4500, Las Cruces, NM, 88003, U.S.A.
              \and
              Department of Astronomy, University of Michigan, 311 West Hall, 1085 S. University Ave, Ann Arbor, MI, 48109-1107, U.S.A.}

   \date{Received October 26, 2018; accepted December 3, 2018}

  \abstract
   {The observation of total and linearly polarized synchrotron radiation of spiral galaxies in the radio continuum reveals the distribution and structure of their magnetic fields. By observing these, information about the proposed dynamo processes that preserve the large-scale magnetic fields in spiral galaxies can be gained. Additionally, by analyzing the synchrotron intensity, the transport processes of cosmic rays into the halo of edge-on spiral galaxies can be investigated.}
   {We analyze the magnetic field geometry and the transport processes of the cosmic rays of the edge-on spiral starburst galaxy NGC~4666 from CHANG-ES radio data in two frequencies; 6 GHz (C-band) and 1.5 GHz (L-band). Supplementary X-ray data are used to investigate the hot gas in NGC~4666.}
   {We determine the radio scale heights of total power emission at both frequencies for this galaxy. We show the magnetic field orientations derived from the polarization data. Using rotation measure (RM) synthesis we further study the behavior of the RM values along the disk in C-band to investigate the large-scale magnetic-field pattern. We use the revised equipartition formula to calculate a map of the magnetic field strength. Furthermore, we model the processes of cosmic-ray transport into the halo with the 1D {\small SPINNAKER} model.}
   {The extended radio halo of NGC 4666 is box-shaped and is probably produced by the previously observed supernova-driven superwind. This is supported by our finding of an advective cosmic-ray transport such as that expected for a galactic wind. The scale-height analysis revealed an asymmetric halo above and below the disk as well as between the two sides of the major axis. A central point source as well as a bubble structure is seen in the radio data for the first time. Our X-ray data show a box-shaped hot halo around NGC 4666 and furthermore confirm the AGN nature of the central source. NGC~4666 has a large-scale X-shaped magnetic field in the halo, as has been observed in other edge-on galaxies. The analysis furthermore revealed that the disk of NGC 4666 shows hints of field reversals along its radius, which is the first detection of this phenomenon in an external galaxy.}
   {}


\titlerunning{CHANG-ES XIII: Transport Processes and the magnetic fields of NGC 4666}
\authorrunning{Stein et al.}

   \maketitle
%

\section{Introduction}
Strong star formation leads to outflow of matter over the whole disk of a galaxy \citep[e.g.,][]{habeetal1981}. In addition to gas, cosmic rays (CRs) and magnetic fields play an important role in the disk-halo interaction of galaxies, which leads to the formation of radio halos in galaxies \citep[e.g.,][]{parker1992}. Observations of radio halos in polarization reveal that star-forming galaxies often show X-shaped magnetic-field structures if observed edge-on \citep[][]{tullmannetal2000, krause2009}. Several mechanisms to generate and maintain large-scale regular magnetic fields in spiral galaxies have been proposed. One mechanism is the mean-field $\upalpha - \upomega$ dynamo \citep[see e.g.,][]{ruzmaikinetal1988, becketal1996, chamandy2016}. Shear motions due to differential rotation combined with the coriolis force acting on the vertical turbulent gas motions amplify and order large-scale regular magnetic fields from turbulent magnetic fields in spiral galaxies \citep[e.g.,][]{arshakianetal2009}. The solutions of the $\alpha - \omega$ dynamo equation in the thin-disk approximation are different dynamo modes, like the axisymmetric spiral magnetic field (ASS) or the bisymmetric spiral magnetic field (BSS). These fields in the galactic disk are generally accompanied by poloidal magnetic fields of even or odd symmetry extending into the galactic halo. However, these magnetic fields are, according to the thin-disk $\alpha - \omega$ dynamo, a factor of about ten weaker than the disk fields and, hence, cannot explain the halo fields that are observed to be almost as strong as the disk fields. One reason for this discrepancy could be that the magneto-ionic disk of galaxies is not thin, due to outflows for example.

The complex behavior of galactic dynamos is not fully understood yet, and the influence of galactic outflows
in particular is still a topic of discussion. Outflows can order a turbulent-disk magnetic field such that it evolves into a regular halo field in dwarf galaxies \citep{mosssokoloff2017}. In Milky Way-type spiral galaxies, galactic outflows lead to two countervailing effects: They are crucial to remove small-scale helicity and hence to avoid quenching of dynamo action, but also lead to field losses into the halo that are responsible for the saturation of the dynamo \citep[e.g.,][]{brendreetal2015,chamandy2015}.

In this paper, the radio halo of the edge-on spiral starburst galaxy NGC~4666 is investigated to obtain further clues about the magnetic fields and processes in spiral galaxies with high star formation rates (SFRs). With the data of NGC~4666 from the Continuum HAlos in Nearby Galaxies - an Evla Survey (CHANG-ES), observed with the Karl G. Jansky Very Large Array (VLA), it is possible to analyze the galaxy in C-band (6\,GHz) and L-band (1.5\,GHz) in terms of linear polarization, the magnetic field structure, and transport processes. 

NGC~4666 is an actively star-forming spiral galaxy at a distance of 27.5\,Mpc \citep{wiegertetal2015}. The basic galaxy parameters are listed in Table~\ref{tab:bparameter}. In the optical spectral range the star-forming disk is slightly visible. Therefore, the inclination of NGC~4666 is less than 90$^\circ$. NGC~4666 is a starburst galaxy \citep{lehnert1996} and is  considered a superwind galaxy with a prominent X-ray halo \citep{dahlemetal1997,tuellmannI06}. An outflow cone is associated with a galactic superwind, which can be described as a global outflow powered by the combined effect of supernova (SN) explosions and stellar winds associated with powerful starbursts \citep{heckmannetal1993}.

In \citet{heesenetal2018}, archival Very Large Array (VLA) data of NGC 4666 and other galaxies were used to model the CR transport with the 1D transport model {\small SPINNAKER}. For NGC~4666, transport by advection was concluded with a very high advection speed between 500\,km~s$^{-1}$ and 700\,km~s$^{-1}$ supporting the starburst phase of this galaxy.

With an SFR of 7.3\,M$_{\odot}$ yr$^{-1}$ \citep{wiegertetal2015}, NGC~4666 is very similar to the CHANG-ES galaxy NGC~5775. Like other starburst galaxies, NGC~4666 is also a member of an interacting system. It is part of a small group of galaxies containing also NGC~4632 and NGC~4668 \citep[][]{garcia1993}. A small dwarf galaxy, which is also part of the group, was discovered by \citet{walteretal2004}. The HI data of \citet[][]{walteretal2004} show strong interaction between NGC~4666 and NGC~4668 as well as with the dwarf galaxy. The strong starburst is triggered by ongoing far-field gravitational interactions \citep{walteretal2004}. The H$\upalpha$ image and the radio continuum maps \citep{dahlemetal1997} show a homogeneous distribution of many star-forming regions across the disk of NGC~4666. This is different from a nuclear starburst galaxy with a 1\,kpc diameter of high star formation in the center. It is not clear why the star formation in NGC~4666 is widely spread along the disk and therefore looks different from many other starburst galaxies \citep{walteretal2004}. Furthermore, the H$\upalpha$ image \citep{dahlemetal1997} shows outflow of filaments, which is represented by the diffuse ionized gas (DIG) and caused by the strong starburst. The X-ray data \citep{tullmannetal2000} indicate a large X-ray halo. The outflow of the hot X-ray emitting gas is comparable to the size of the disk and may even extend radially beyond the H$\upalpha$ filaments. Additionally, a nuclear outflow (bubble structure to the southeast) is visible in the X-ray image. 

NGC~4666 hosts a modestly active galactic nucleus (AGN)  \citep{dahlemetal1997,persicetal2004}, which is probably highly obscured \citep{dudiketal2005}. Based on BeppoSAX and XMM Newton data, \citet{persicetal2004} found that the star burst, which extends over most of the disk, and AGN activities coexist in NGC~4666. They found a prominent emission line from Fe-K$\upalpha$ at $\approx$ 6.4~keV from the nuclear region. Additionally, they observed the presence of a flat continuum that is in agreement with a model in which the continuum originates from the reflection of the primary continuum by the cold inner wall of the circumnuclear torus in the nuclear region. They conclude the existence of a strongly absorbed (i.e., Compton-thick) AGN. The optical emission lines of spectroscopic measurements from the core region of \citet{dahlemetal1997} also suggest a central AGN. In a further study of high-resolution X-ray imaging of nearby low-ionisation nuclear emission-line regions (LINERs) observed by {\it Chandra}, a nuclear point source of NGC~4666 was not detected \citep{dudiketal2005}. They concluded either a lack of an energetically significant AGN or a highly obscured AGN with internal absorptions reaching 1.1~$\times$~$10^{23}$\,cm$^{-2}$ to 8.4~$\times$~$10^{24}$\,cm$^{-2}$, which would imply that the AGN shows luminosities between 2~$\times$~10$^{38}$\,erg s$^{-1}$ and 9.5~$\times$~10$^{42}$\,erg s$^{-1}$. No nuclear (point source) flux density has been determined so far in the radio wavelength regime as the resolution of the observations was insufficient.

\begin{threeparttable}[h]
\centering
\caption[Basic galaxy parameters.]{Basic galaxy parameters}  
\label{tab:bparameter}
\begin{tabular}{lc} 
\hline \hline
Galaxy                                                          &  NGC 4666  \\ \hline
RA [J2000]                                                      & 12h 45m 08.6s \tnote{1} \\
DEC [J2000]                                                     & -00$^\circ$ 27' 43" \tnote{1}\\
Distance [Mpc]                                  & 27.5 \tnote{2}  \\
Inclination [$^\circ$]                          & 85~$\pm$~2 \tnote{3}           \\
PA [$^\circ$]                                           &               40 \tnote{3}                \\
Major Axis [arcmin]                     &               4.6 \tnote{1}            \\
Minor Axis [arcmin]                     &               1.3 \tnote{1}           \\
v$_{\text{sys}}$ [km s$^{-1}$]  & 1517 \tnote{4}                 \\
v$_{\text{rot}}$ [km s$^{-1}$]  & 195 \tnote{5}          \\
SFR [M$_{\odot}$~yr$^{-1}$]         &  7.3 \tnote{6}            \\
SFRD [10$^{-3}$\,M$_{\odot}$~yr$^{-1}$~kpc$^{-2}$]& 8.9 \tnote{6}       \\
Classification                          & SABc\tnote{7} \\
Approaching side                        & northeast    \\
Receding side                           & southwest    \\

\hline
\end{tabular}
\begin{tablenotes}
\footnotesize
\item \textbf{References:}
\item[1] NASA/IPAC Extragalactic Database (NED, https://ned.ipac.caltech.edu)
\item[2] \citet{wiegertetal2015}
\item[3] estimated in this work 
\item[4] \citet{mathewsonetal1992} 
\item[5] \citet{mathewsonford1996}
\item[6] SFR and SFR density (SFRD) from \citet{wiegertetal2015}
\item[7] \citet{irwinetal2012}
\end{tablenotes}
\end{threeparttable}
\normalsize
\vspace{0.5cm}


Here, we investigate the radio properties of NGC~4666 and its central source. We present the first flux measurements of the central source from the high-resolution radio continuum data of CHANG-ES. We further analyze the data on the radio halo, scale heights, and the linear polarization. Additional archival VLA C-band data of NGC~4666 complement the polarization data. With supplementary X-ray data from XMM-Newton we are able to support the AGN classification of the central source and analyze the far-extended X-ray halo, which seems to be correlated to the radio halo. With the radio data we are able to investigate the disk magnetic field, which is axisymmetric with hints of one reversal. After the thermal/nonthermal separation, the magnetic field strength is determined. We further use the 1D {\small SPINNAKER} model for cosmic-ray transport in NGC~4666.

The paper is organized as follows. Section~2 gives an overview of the data and the adapted technique of rotation measure (RM) synthesis. In Section 3 the results of Stokes~I, the polarization using RM synthesis, the disk field of NGC~4666, and the maps of the magnetic field strength using equipartition are presented. The 1D transport model is applied to the data. The magnetic field distribution from that model is then compared to the equipartition field. In Section 4 the results are summarized and conclusions are drawn.

\section{Observations and data analysis}
\subsection{VLA data}
\subsubsection{CHANG-ES data}
The radio continuum data are part of the CHANG-ES survey observed with the Karl G. Jansky VLA \citep[][]{irwinetal2012}. 
Observations were obtained in B-, C-, and D-configurations at L-band (1.5\,GHz, 500\,MHz bandwidth, with a gap of 144\,MHz width where strong radio frequency interference (RFI) is located), and in the C- and D-configurations at C-band (6\,GHz, 2\,GHz bandwidth). We used 2048 spectral channels in 32 spectral windows (spws) at 1.5\,GHz and 1024 channels in 16 spws at 6\,GHz. All polarization products (Stokes I, Q, U, and V) were obtained. The D-configuration data \citep{wiegertetal2015} are public\footnote{CHANG-ES data release I available at www.queensu.ca/changes} and the C-configuration data will become public soon (Walterbos et al. in prep).

\begin{table}
\centering
\caption{Observation parameters of CHANG-ES} 
\label{tab:oparameter}
\begin{tabular}{l c c c} 
\hline\hline   
Dataset &Observing Date         & Time on Source\\
      &                       & (before flagging)\\ 
\hline
L-band B-configuration &  10.06.2012  & 2 hr\\
L-band C-configuration &  30.03.2012    & 30 min\\
L-band D-configuration &  30.12.2012   & 20 min\\
C-band C-configuration&  23.02.2012   & 3 hr\\
C-band D-configuration & 19.12.2012   & 40 min\\
\hline
\end{tabular} 
\end{table}

The data reduction for Stokes I (total power) and Stokes Q and U (linear polarization) was carried out for all five data sets of NGC~4666 (see Table~\ref{tab:oparameter}) separately, using the Common Astronomy Software Applications (CASA) package \citep[][]{mcmullinetal2007} and following the calibration procedures as described in the CHANG-ES paper by \citet{irwinetal2013}. We used J1331+3030 (3C286) as the primary calibrator, J1246-0730 as the secondary calibrator, and J1407+2827 as the zero polarization calibrator. 
The calibrated data from the different configurations were then combined for C-band and L-band and these combined data were used for imaging Stokes I and the polarization.

The Stokes I maps were produced by cleaning with a robust weighting parameter of zero \citep{briggs1995}. The polarization and magnetic-field-orientation maps were created from the Stokes Q and U maps, which were cleaned with a robust parameter of two in order to be more sensitive to faint structures. The achieved rms from the combined C-band data for Stokes I is 4.6\,$\mu$Jy/beam with a beam of 3.0"~$\times$~3.5". The resulting rms of the combined L-band data for Stokes I is 30\,$\mu$Jy/beam with a beam of 11.8"~$\times$~13.6". Subsequently, smoothed images of Stokes I of both bands were produced to match the resolution of the polarization maps of 7", as well as the resolution of the maps from RM synthesis of 18". 

\subsubsection{Archival VLA data}
Archival VLA D-configuration C-band observations exist for the galaxy NGC~4666 (Program AD326). These observations were obtained before the upgrade of the VLA. The central frequency was 4.86\,GHz with a bandwidth of 2x50\,MHz. The calibration was done using the Astronomical Image Processing System (AIPS\footnote{www.aips.nrao.edu}) with J1331+305 (3C286) as the primary calibrator and J1246-075 as the secondary calibrator.
The primary calibrator was also used for polarisation-angle correction  and the secondary calibrator for polarisation leakage term determination. 

The Stokes I maps were produced by cleaning with robust zero weighting. The polarization and magnetic-field-orientation maps were created from the Stokes Q and Stokes U maps, which were also cleaned with robust zero weighting. The application of RM synthesis on these data is not possible as they consist of just two channels.

\begin{table}
\centering
\caption{Observation parameters of the archival VLA data} 
\label{tab:oparameter_oldVLA}
\begin{tabular}{l c c c} 
\hline\hline   
Dataset &Observing date         & Time on source\\
  &                                      & (before flagging)\\ 
\hline
C-band D-configuration & 20.12.1993   & 378 min\\
\hline
\end{tabular} 
\end{table}

\subsection{XMM data}
To study the nature of the possible central source, as well as of the galactic energy budget via analysis of the emission from the hot gas, XMM-Newton archive data for NGC\,4666 were used (see Table~\ref{xdat} for the parameters of the observations). The data were processed using the SAS 15.0.0 package \citep{gabriel04} with standard reduction procedures. The tasks \texttt{epchain} and \texttt{emchain} helped to obtain event lists for two EPIC-MOS cameras \citep{turner01} and the EPIC-pn camera \citep{strueder01}. 
The event lists were then carefully filtered for periods of intense background radiation. From the output data an image in the soft energy range of 0.2-1\,keV was produced, along with the exposure map (without vignetting correction) masked for the acceptable detector area using the {\it images} script\footnote{http://xmm.esac.esa.int/external/xmm\_science/\\gallery/utils/images.shtml}, modified by the authors to allow adaptive smoothing.

Furthermore, a spectral analysis was performed. The spectra of each region were created using all three EPIC cameras. The background spectra were obtained using blank sky-event lists \citep[see][]{carter07}, filtered using the same procedures as for the source-event lists.
For each spectrum, response matrices and effective area files were produced.
Next, including these ancillary files, spectra from all three EPIC cameras and the corresponding background blank sky spectra were merged 
using the SAS task \texttt{epicspeccombine} into a final background subtracted source spectrum. Finally, the spectra were fitted using XSPEC~12 \citep{arnaud96}.

\begin{table}[ht]
\centering
  \caption{Parameters of the XMM-Newton X-ray observations of NGC~4666}
\label{xdat}
   \begin{tabular}{lccccc}
\hline\hline
Obs. ID                            & 0110980201  \\
Date of observations              & 27.06.2002  \\
Column density $N_{\rm H}$ [10$^{20}$ cm$^{-2}$] \tablefootmark{a}& 1.73   \\
MOS filter                        & thin   \\
MOS obs. mode                     & Full Frame   \\
Total/clean MOS time [ks]         & 115.5/115.5 \\
pn filter                         & thin   \\
pn obs. mode                      & Extended Full Frame  \\
Total/clean pn time [ks]          & 54.4/54.4   \\
\hline
\end{tabular}
\tablefoot{
\tablefoottext{a}{Weighted average value after LAB Survey of Galactic \ion{H}{i} \citet{kalberla05}.}
}
\end{table}

\subsection{RM synthesis}
While propagating through a magnetized plasma, electromagnetic waves experience a frequency dependent rotation of the polarization angle. This effect is called Faraday rotation and is caused by the different propagation speeds of the left and right circular polarized waves. The rotation of the polarization angle $\chi$ is proportional to the wavelength ($\lambda$) squared and the RM:
\begin{equation}
\chi = \text{RM} \, \lambda^2 \, .
\end{equation}

If different regions along the line of sight emit polarized intensity and/or rotate the polarization angle, RM has to be replaced by the Faraday depth $\Phi$ \citep{burn1966}:
\begin{equation}
\Phi = 0.81 \int\limits_{r_0}^{0}\frac{n_e}{cm^{-3}} \ \frac{\text{B}_{\|}}{\mu G} \ \frac{\text{d}r}{\text{pc}}\ \text{rad} \ \text{m}^{-2} \, ,
\end{equation}
with the electron density $n_e$ and the magnetic field integrated along the line of sight B$_{\|}$.

With the method of RM synthesis applied to wide-band multi-channel receiver data, multiple sources along the line of sight can be measured \citep{brentjensdebruyn2005}.
The basic idea was introduced by \citet{burn1966} who defined a complex ``Faraday dispersion function'', which connects the complex polarized surface brightness in Faraday depth space via a Fourier transform with the dependence of the complex polarized surface brightness in $\lambda^2$-space. When applying this technique, several parameters are important, which are presented in Table \ref{tab:RMpar}.

In order to find out if each channel has to be imaged for Stokes Q and U or if it is sufficient to image every spw, we calculated the rotation over both frequency bands. There are 32 spws in L-band and 16 spws in C-band. Each spw contains 64 channels before flagging. Since the rotation of the electromagnetic wave is calculated as  $\lambda^2$ multiplied by the RM, the rotation of the polarization angle $\chi$ can be estimated. A median value of $|RM| \approx$~100 rad m$^{-2}$ was found in external galaxies \citep[e.g.,][]{fletcheretal2004}. With the L-band ranging in frequency from 1.2\, to 1.75\,GHz (wavelength range between $\lambda$~=~0.25\, and $\lambda$~=~0.17\,m), and the C-band ranging in frequency from 5 to 7\,GHz (wavelength range between $\lambda$~=~0.06\,m and $\lambda$~=~0.042\,m) the rotation of the polarization angle $\Delta \chi$ can be calculated:
\begin{align}
\Delta \chi &= \text{RM} \, (\lambda_1^2 - \lambda_2^2) = \text{RM} \, \Delta \lambda^2 \\ \nonumber
\Delta \chi_{\text{L-band}} &= 100\,\text{rad m}^{-2} \, (0.0625 - 0.0289)\,\text{m}^2 \\ 
                            &= 3.36\,\text{rad} = 192.5^{\circ}\\ 
\Delta \chi_{\text{C-band}} &= 0.184\,\text{rad} = 10.5^{\circ}.
\label{eq:rmwinkelL}
\end{align}

Standard imaging averages over the entire bandwidth, which in L-band leads to strong depolarization across the frequency band and no information on the RM is gained. By imaging each spw separately, the rotation of the polarization vector (in the above example $192.5^{\circ}$) can be lowered by a factor of 32 in L-band, which is sufficient (in the above example the rotation within the individual spw is then $\sim6^{\circ}$). Due to the wavelength dependence, this effect is not as strong in C-band.

\begin{threeparttable}
\centering
\caption{Rotation measure synthesis parameters}  
\label{tab:RMpar}
\begin{tabular}{l c c} \hline\hline
 \                                                                                                                                                                                                                              & C-band     & L-band   \\ \hline
Bandwidth [GHz]                                                                                                                                                                                                                                                                 & 2         & 0.5     \\
$\upnu_{\text{min}}$ to $\upnu_{\text{max}}$ [GHz]                                                                                                                      & 5 to 7    & 1.2 to 1.75 [gap]\\
$\Delta \lambda^2$ = $\lambda_1^2 - \lambda_2^2$ [m$^2$]                                                                                                        & 0.00184         & 0.0336        \\
$\delta \phi$ = $\frac{2 \sqrt{3}}{\Delta \lambda^2}$ [rad m$^{-2}$]           & 1882      & 103    \\ \hline
$\lambda_{min}$ [m]                                                                                                                                                                                                                                                     & 0.042     & 0.17    \\
max$_{\text{scale}}$  = $\frac{\pi}{\lambda_{min}^2}$ [rad m$^{-2}$]                                            & 1781      & 109    \\ \hline
spw-width $\delta$f [MHz]                                                                                                                                                                                                                       & 125       & 16        \\
$\delta \lambda^2$ [m$^2$]                                                                                                                                                                        & 0.0001735 & 0.001447    \\
RM$_{\text{max}}$ = $\frac{\sqrt{3}}{\delta \lambda^2} $ [rad m$^{-2}$]                                                                                 & 9983      & 1197       \\
\hline
\end{tabular}
\begin{tablenotes}
\footnotesize
\item \textbf{Notes:}
\begin{itemize}
\item Channel width: $\delta \lambda^2$, width of the $\lambda^2$ distribution: $\Delta \lambda^2$, shortest wavelength squared: $\lambda_{min}^2$  
\item FWHM of the resolution (RMSF) in $\Phi$ space: $\delta \phi$ (rad m$^{-2}$) 
\item Largest scale in $\Phi$ space to which the observation is sensitive: max-scale (rad m$^{-2}$)
\item Maximum observable RM (resp. $\Phi$): RM$_{\text{max}}$ (rad m$^{-2}$).
(from \cite{brentjensdebruyn2005}).
\end{itemize}
\end{tablenotes}
\end{threeparttable}
\normalsize
\vspace{0.5cm} 

From the data in \citet{rm2014}, the Galactic foreground RM in the direction to NGC~4666 was determined. This value (RM$_{\text{foreground}}~=~-5.8~\pm~1.3$~rad m$^{-2}$) was subtracted from the final RM cube.

To adopt RM synthesis, Stokes Q and Stokes U images for each spw were produced. These have to be convolved to the largest beam in the frequency range (the beam size being dependent on frequency). Due to RFI in L-band, spws~0,~1,~and~2 were flagged before all images were merged together into an image cube with the frequency as the third axis. This was done for the Stokes Q and U images separately. Then the script "RMsynth" (B. Adebahr, private communication, based on \citet{brentjensdebruyn2005}) was applied. A Fourier transformation was performed on the complex flux densities of the cubes in $\lambda^2$-space. The result was a cube with the two image axes and $\Phi$ as the third axis. Then the RM of a single source along the line of sight was determined by fitting, for example, a parabola to the main peak of F($\Phi$) \citep{brentjensdebruyn2005}. The result was a cube of Stokes Q and U, as well as a polarization cube, which was then further analyzed to obtain, for example, the RM, the polarized intensity (PI), and the polarization angle (PA), as well as their corresponding error maps. The RM map represents the fitted peak position of each pixel, again assuming there is only one component along the line of sight.

\section{Results and Discussion}

\subsection{Radio continuum Stokes I}
\begin{figure}
        \centering
                \includegraphics[width=1.0\hsize]{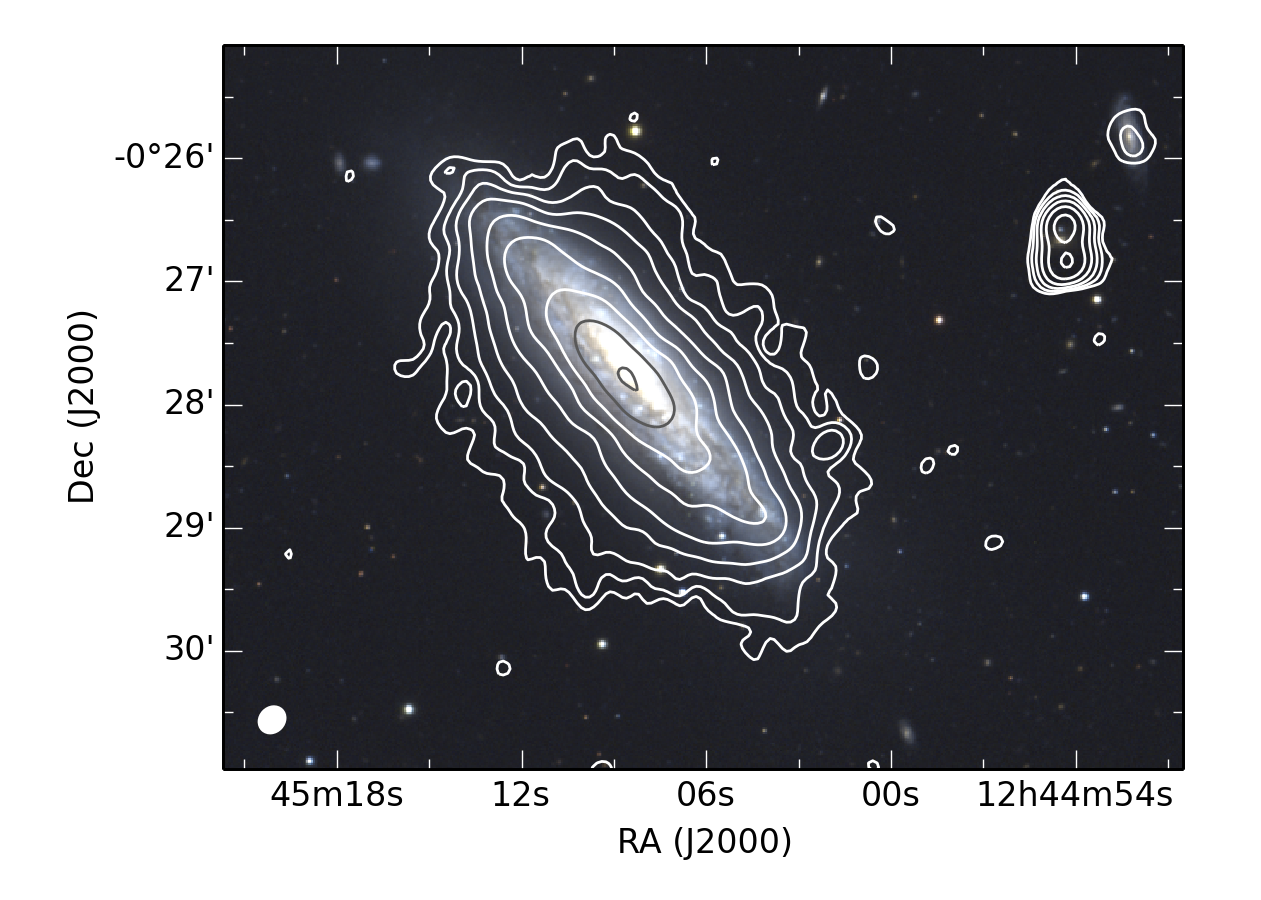}
        \caption{NGC~4666 Stokes I image of L-band combined (B-, C-, and D-configuration) with a robust zero weighting, overlayed onto the optical SDSS image made from the u, g, r filters. Contours start at a 3$\upsigma$ level with an rms noise $\upsigma$ of 30\,$\upmu$Jy/beam and increase in powers of 2 (up to 256). The beam size is 11.8" $\times$ 13.6", shown in the bottom-left corner.}
        \label{fig:N4666_Lcomb_color}
\end{figure}

The total power map (Stokes I) of L-band is shown in Figure~\ref{fig:N4666_Lcomb_color} with total intensity contours overlayed on the optical Sloan Digital Sky Survey (SDSS)\footnote{www.sdss.org} image made from the ugr filters using the formulas of Lupton 2005\footnote{www.sdss.org/dr12/algorithms/sdssUBVRITransform/\#Lupton2005}. The radio halo in the combined configuration of L-band (Figure~\ref{fig:N4666_Lcomb_color}) reaches up to 9 kpc above and below the plane of the galaxy as measured to the 3-$\upsigma$ level corrected for the beam size. The box-like appearance of the radio halo is quite striking. It is evident that the extent of the radio continuum radiation beyond the star-forming disk only occurs in the vertical direction of the galaxy. In Appendix A, we present L-band contours from the B-, C-, and D-configurations of the VLA, which result in different resolutions. The C-band data are presented in the following section.

\citet{dahlemetal2006} investigated possible parameters influencing the existence of radio halos. They found that galaxies that show radio halos also have high energy input rates into their interstellar medium (ISM) and high average dust temperatures. These latter authors report that NGC~4666 shows the highest total nonthermal radio power at 1.49\,GHz in their sample with a value of 34.1~$\times$~10$^{21}$\,W~Hz$^{-1}$. \citet{liwang2013a} calculated the total supernova (SN) mechanical energy injection rate to be 10.4~$\times$~10$^{38}$\,erg~s$^{-1}$, which is one of the highest in their sample.

The radius of star formation in NGC~4666, which was derived in \citet{dahlemetal2006} via the radial extent of H$\upalpha$ emission, is quite high in comparison to the optical extent of the galaxy. The radial extent of star formation was determined in this study to be 14.2\,kpc (scaled to the distance used in this paper) in comparison to the radius of the major axis of 18\,kpc, showing that NGC~4666 has widespread star formation across almost the entire disk. This is very different from a classical starburst galaxy with a central starburst. The SFR of 7.3 M$_{\odot}$~yr$^{-1}$ \citep{wiegertetal2015} is the highest among the 35 edge-on galaxies in the CHANG-ES sample. The star formation-driven winds originating from the wide-spread star forming regions may be an explanation to the box-like appearance of the halo as seen in Fig.~\ref{fig:N4666_Lcomb_color}. 

The CHANG-ES observations of the galaxy NGC~5775 (Heald et al. in prep.) provide further evidence corroborating the explanation for the box-like radio halo from above. This galaxy is the only galaxy in the CHANG-ES sample that is comparable to NGC~4666 with regard to its size and SFR, with the second highest SFR of 5.3~M$_{\odot}$~yr$^{-1}$ \citep{wiegertetal2015}. Furthermore, neither is strongly dominated by its central object and in both the star formation is widely spread over almost the entire disk. NGC~5775 is also defined as a starburst and superwind galaxy and shows a similarly boxy radio halo. The wind could be confined by the magnetic field lines of the disk, which mostly follow the spiral structure and reach into the halo \citep{henriksenirwin2016}, thus resulting in a box-like structure of the radio halo. NGC~4666 is a remarkable example of a clear connection between the star-formation distribution in the disk and the morphology of the radio halo.

\subsection{Integrated spectral distribution}

\begin{figure}
\begin{center}
\includegraphics[width=1.0\hsize]{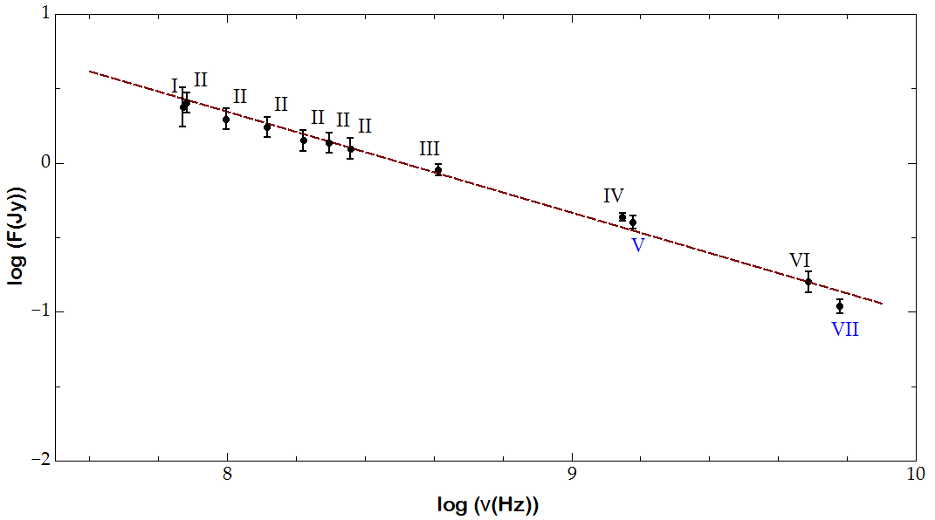}
        \caption{NGC~4666 flux density measurements in the radio regime with a fit that gives a spectral slope of $\upalpha~=~-0.68~\pm~0.04$. I: VLA Low-Frequency Sky Survey \citep[VLSS,][]{cohenetal2007}, II: GaLactic and Extragalactic All-sky MWA survey \citep[GLEAM,][]{Hurley-WalkeretalCat2016} , III: The Molonglo Reference Catalogue of Radio Sources \citep[MRC,][]{largeetal1981}, IV: The NRAO VLA Sky Survey \citep[NVSS,][]{condonetal1998}, V: CHANG-ES (this work), VI: The Parkes-MIT-NRAO surveys \citep[PMN,][]{griffithetal1995}, VII: CHANG-ES (this work).}
        \label{fig:N4666_Fluxmeasure}
\end{center}
\end{figure}

The spectral index behavior of the integrated flux densities $I$ of NGC~4666 is shown in Figure~\ref{fig:N4666_Fluxmeasure}. Other integrated flux-density measurements from the literature were selected from the NASA/IPAC Extragalactic Database (NED)\footnote{https://ned.ipac.caltech.edu} and VizieR\footnote{http://vizier.u-strasbg.fr/viz-bin/VizieR}, complementing the two CHANG-ES flux densities from this work (Table~\ref{tab:fluxmeasuren4666}); only literature values with corresponding errors were used. The fitted spectral slope is $\upalpha$~=~$-$0.68~$\pm$~0.04 (where $I \propto \upnu^\upalpha$). This spectral index is typical for radio spectra with superposition of synchrotron and thermal radiation. The spectral index through the higher-frequency data shows hints of a steeper index. There is no indication of flattening towards the lower frequencies from the different GLEAM data.

\begin{table}
\begin{center}
\begin{threeparttable}[t]
\caption{Flux densities of NGC~4666}
\label{tab:fluxmeasuren4666}
\begin{tabular}{lcccc} \hline \hline
 \#   &$\upnu$ &  Flux density & log($\upnu$)& log(Flux dens.)\\
  & [GHz] &       [Jy] &         &   \                           \\
\hline
I         & 0.074  &  2.380 $\pm$ 0.360 & 7.87 & 0.38 $\pm$ 0.13\\
II   & 0.076   & 2.559 $\pm$ 0,204 & 7.88  & 0.41 $\pm$ 0.07 \\
II   & 0.099   & 1.989 $\pm$ 0.159  & 8.00 & 0.30 $\pm$ 0.07 \\
II   & 0.130  & 1.756 $\pm$ 0.14048  &8.11&     0.24 $\pm$ 0.07 \\
II   & 0.166 &  1.428 $\pm$ 0.11424      &8.22& 0.15 $\pm$ 0.07 \\
II   & 0.197 &  1.377 $\pm$ 0.11016      &8.29& 0.14 $\pm$ 0.07 \\
II   & 0.227 &  1.258 $\pm$ 0.10064      &8.36& 0.10 $\pm$ 0.07 \\
III    & 0.408  &       0.910 $\pm$ 0.040       &8.61   & -0.04 $\pm$ 0.03      \\
IV              & 1.400  &  0.437 $\pm$ 0.014   &9.15   & -0.35 $\pm$ 0.02\\
V               & 1.580  &  0.402 $\pm$ 0.020   &9.20 & -0.40   $\pm$ 0.04 \\
VI              & 4.850  &      0.161 $\pm$ 0.013 &9.68 & -0.79 $\pm$ 0.07\\
VII             & 6.000  &      0.111 $\pm$ 0.006 &     9.78& -0.96 $\pm$ 0.05    \\
\hline
\end{tabular}
\small
\begin{tablenotes}
\item \textbf{References:}\\
I:  The VLA Low-Frequency Sky Survey \citep[VLSS,][]{cohenetal2007}\\
II: GaLactic and Extragalactic All-sky MWA survey \citep[GLEAM,][]{Hurley-WalkeretalCat2016}
III: The Molonglo Reference Catalogue of Radio Sources \citep[MRC,][]{largeetal1981}\\
IV: The NRAO VLA Sky Survey \citep[NVSS,][]{condonetal1998}\\
V: CHANG-ES (this work)\\
VI: The Parkes-MIT-NRAO surveys \citep[PMN,][]{griffithetal1995}\\
VII: CHANG-ES (this work)
\end{tablenotes}
\end{threeparttable}
\end{center}
\end{table}
\normalsize

\subsection{Central point source}
\subsubsection{Radio continuum}
\begin{figure}
 \captionsetup{margin=10pt}
        \centering
                \includegraphics[width=0.98\hsize]{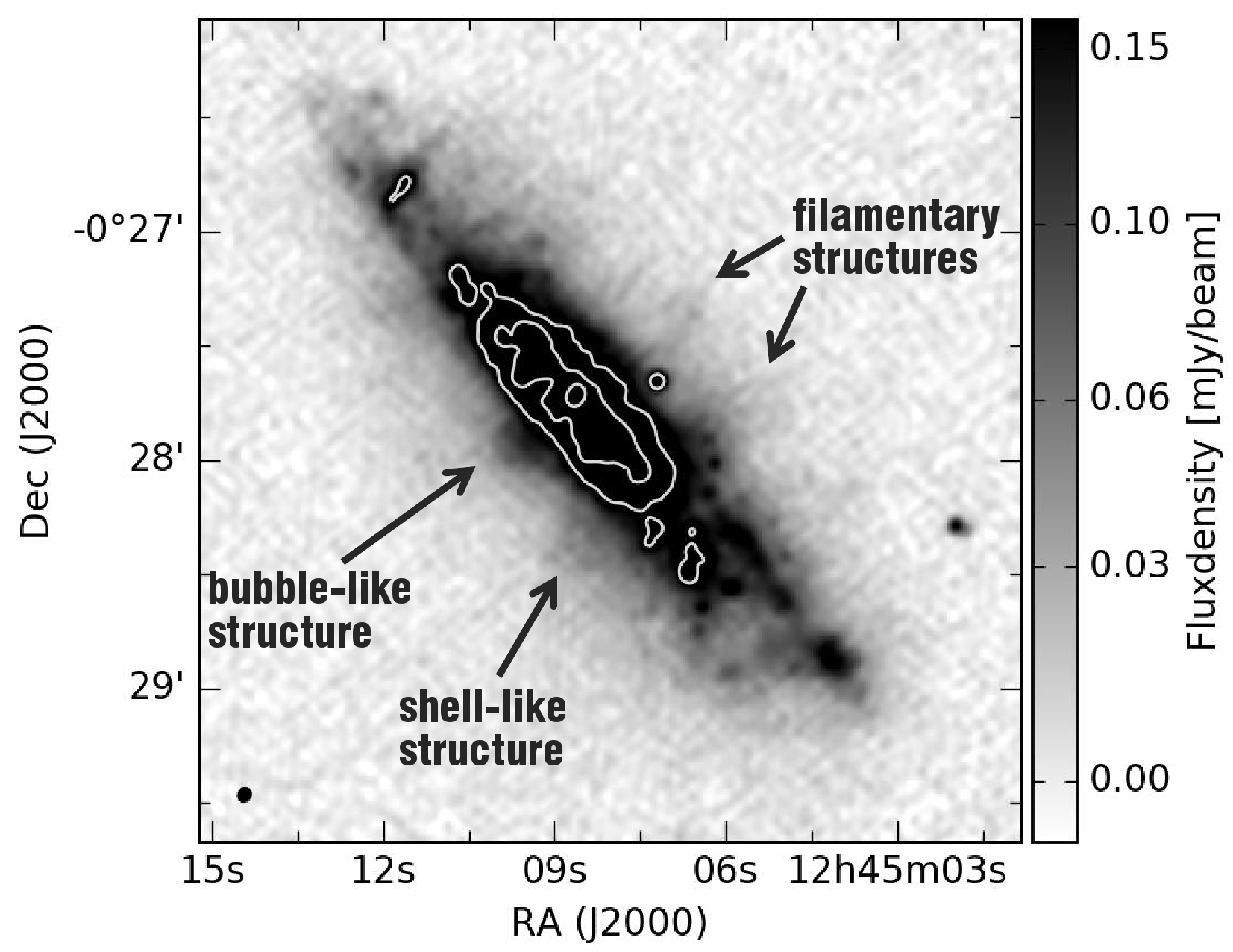}
        \caption[NGC~4666 C-band image (C- and D-configurations)]{NGC~4666 C-band image (C- and D-configuration) in gray scale with an rms noise $\upsigma$ of 4.6\,$\upmu$Jy/beam and a beam size of 3.0" $\times$ 3.5" (see bottom left), using robust zero weighting. The light-gray contours show the inner region with the radio point source from the same data. The corresponding flux densities are 0.22, 0.44, and 0.88\,mJy/beam. We note the bubble-like structure in the eastern direction from the point source and the filamentary structures (threads) in the northern halo.}
        \label{fig:N4666_Ccomb_bw}
\end{figure}

The central point-like source is seen in both radio bands (at the J2000 position RA~12h 45m 08.62s, DEC~-00$^{\circ}$ 27' 43.2"): in the high-resolution B-configuration of L-band (see Appendix, Fig.~\ref{fig:N4666-BCDL_all}, first panel) and in the combined image of C-band (Figure~\ref{fig:N4666_Ccomb_bw}), as well as in the C-configuration alone, which is used for the analysis performed here. These are the first detections in the radio regime. To measure the flux density of the central source, the images of both bands were smoothed to the same beam size of 4" to allow for a better comparison of the measurements and to ensure consistency. The flux density was then measured via a Gaussian fit to the intensity profile after subtraction of the diffuse flux density from the galaxy disk. We chose a circular region of two times the size of the beam (at least twice the beam is recommended). The resulting error was very small\footnote{From the help file of imfit (https://casa.nrao.edu/docs/taskref/imfit-task.html): "Fitting a zero level offset that is not fixed will tend to cause the reported parameter uncertainties to be slightly underestimated".} and was therefore analyzed further. The chosen region was varied from twice the beam size to four times the beam size. The difference in the flux density measurements was then taken to calculate a mean error.

\begin{table}
\centering
\begin{threeparttable}
\caption{Flux densities of the central source of NGC~4666}  
\label{tab:centralsourcemeasure_n4666}
\begin{tabular}{lcccc}
 \hline \hline
Band &Flux density      &  Beam&  Region size \\
   & [mJy]        & ["]           & ["]             \\
 \hline
L-band B-conf.  &  4.4 $\pm$ 0.4                        & 4      & 8     \\
C-band C-conf.  &  1.8 $\pm$ 0.2                        & 4      & 8 \\
\hline
\end{tabular}
\end{threeparttable}
\end{table}

The resulting radio spectral index of the nuclear flux density measurements is $\upalpha~=~-~0.67~\pm~0.12$, which indicates a synchrotron source. As some AGNs show circular polarisation, Stokes~V was imaged but did not show any signal in either band.

In Figure~\ref{fig:N4666_Ccomb_bw} the combined (C- + D-configurations) C-band data are shown in gray scale with contours overlaid showing the location of the central radio source. In the eastern direction from the point source (gray contour) a bubble-like structure is seen that may indicate a jet. This further strengthens the argument that the central source is an AGN. Other interesting features of NGC~4666 seen in Fig.~\ref{fig:N4666_Ccomb_bw} are the filamentary structures to the western side of the galaxy. These are comparable to H$\upalpha$ filaments emerging from the disk. With the new high-resolution CHANG-ES data this structure is now seen in the radio regime. Since the galaxy is classified as a superwind and starburst galaxy, these structures are probably best explained by a galactic wind from the disk and up into the halo. Therefore, the magnetic fields as well as the CRs are transported with the wind. This would result in advection being the main transport process of CRs, which agrees with what is found for this galaxy by \citet{heesenetal2018}; this is further discussed in Section~\ref{subsec:spinteractive}. Similar filamentary radio structures described as threads were found in the center of the Milky Way by \citet{larosaetal2000}.

The western and eastern side of the galaxy halo look very different in terms of this filamentary structure. Beside the bubble-like structure to the east there are also thin features of radio intensity in the halo on this side of the galaxy. They are not filamentary, but shell-like structures. A possible reason for this is the ongoing interaction mainly occurring with NGC~4668 to the eastern side of the galaxy which could bend this side of the halo. Therefore, the two sides are affected differently.

\subsubsection{XMM data}
 
The analysis of the X-ray emission from the central source was performed using additional components in the model fitted to the spectrum from the disk region (Table~\ref{models}). These include an absorbed power law and a simple Gaussian, with the latter accounting for a weak but clearly visible iron Fe-K{$\upalpha$} line at around 6.4\,keV. This line is considered to be a typical fingerprint of an AGN. 
Table~\ref{agn} presents the parameters of the central source obtained from the fit, including the unabsorbed luminosity. 

\begin{table}
\centering
    \caption{Model type and reduced $\chi_{\rm red}^2$}
       \begin{tabular}{lcc}
\hline
Region  & Model type & $\chi_{\rm red}^2$ \\
\hline
Disk & wabs(mekal+mekal+powerlaw & 1.23 \\
    &   +wabs(powerlaw+gauss)) &                        \\
East & wabs(mekal+mekal+powerlaw) & 1.13 \\
West & wabs(mekal+mekal+powerlaw) & 0.96 \\
\hline
\end{tabular}
\label{models}
\end{table}

\begin{table}
\centering
\caption{Characteristics of the central source in NGC~4666}
\label{agn}
\begin{tabular}{llcccrr}
\hline\hline
Internal absorption [10$^{22}$ cm$^{-2}$]& 1.72$^{+1.38}_{-0.71}$        \\
\vspace{5pt}
Photon Index                            & 2.06$^{+0.78}_{-0.31}$        \\
\vspace{5pt}
Iron line [keV]                         & 6.36$\pm$0.05                 \\
\vspace{5pt}
$\sigma$ [keV]                          & 0.10$^{+0.78}_{-0.31}$        \\
\vspace{5pt}                    
Luminosity [10$^{40}$ erg/s]            & 4.68$^{+7.48}_{-1.64}$        \\
\hline
\end{tabular}
\end{table}

\subsection{Further analysis of the XMM data}
\begin{figure}
        \centering
                \includegraphics[width=0.9\hsize]{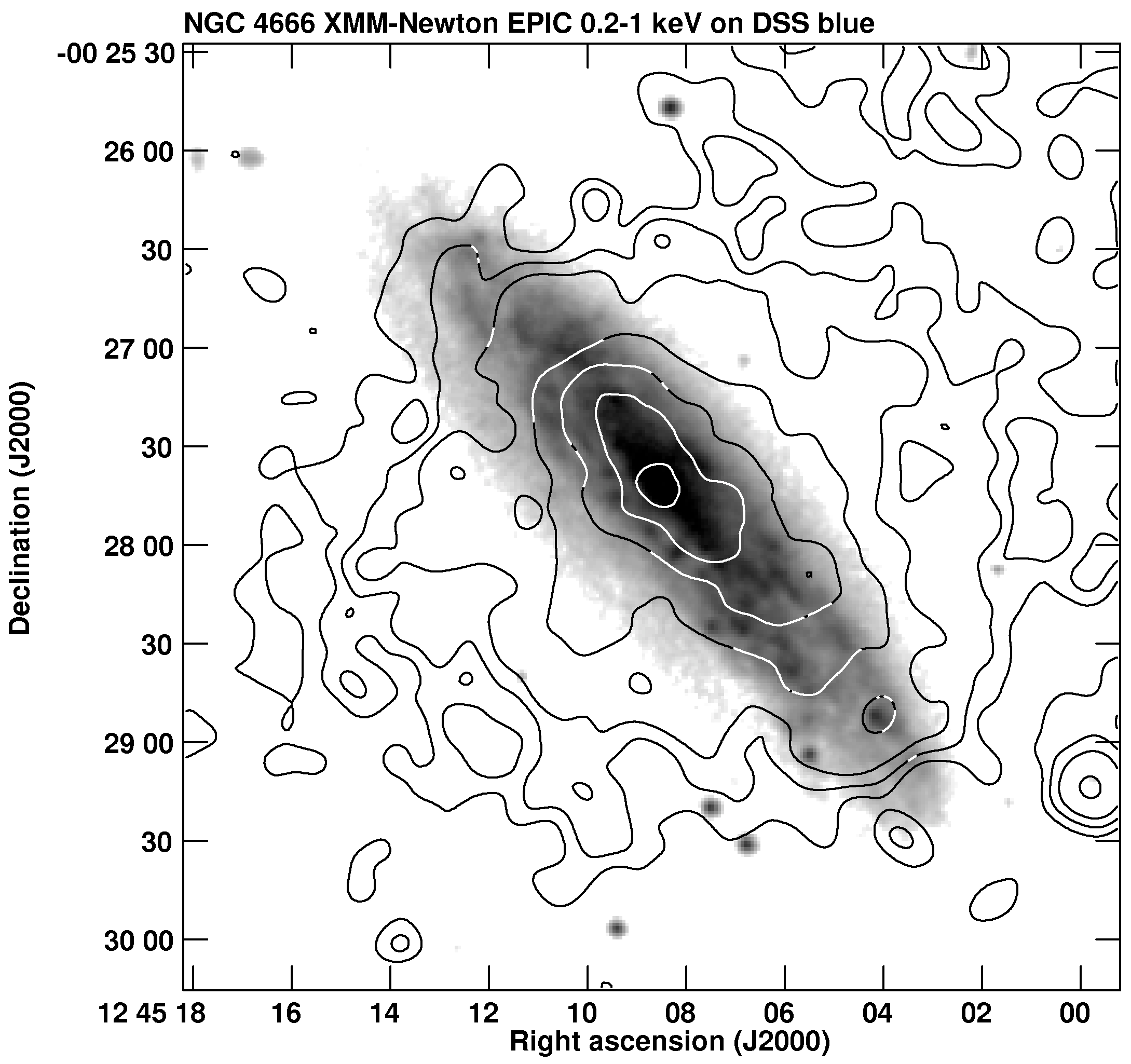}
        \caption{Map of soft X-ray emission from NGC~4666 in the 0.2 - 1 keV band overlaid onto the DSS blue image. The contours are 3, 5, 8, 16, 32, 64, 128 $\times\ \upsigma$. The map is adaptively smoothed with the largest scale of 10$\arcsec$. }
        \label{ximg}
\end{figure}

\begin{figure}
        \centering
                \includegraphics[width=0.75\hsize]{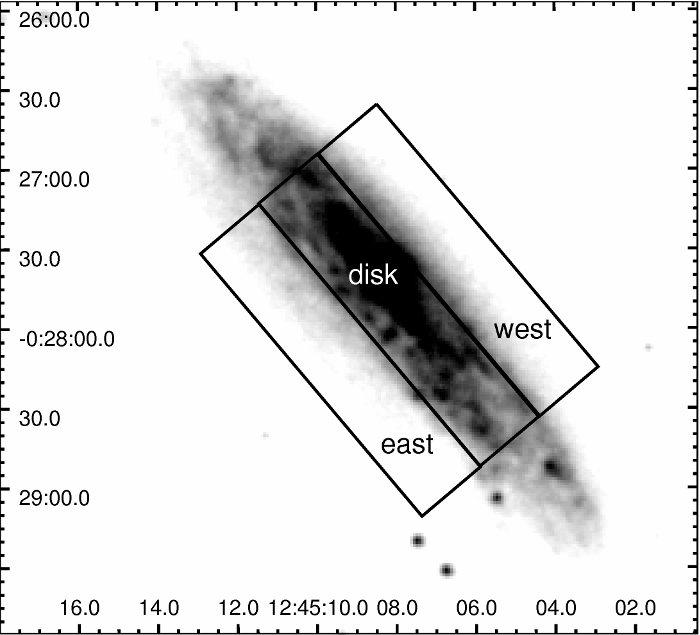}
        \caption {The regions used for the spectral analysis of the disk of NGC~4666 on the DSS blue image. }
        \label{ximg2}
\end{figure}

Figure~\ref{ximg} shows the large-scale emission from the hot gas in NGC~4666. While the brightest emission comes from the central parts of the disk and the nuclear point-like source, the vertical outflows reach up to 9\,kpc into the halo. The hot gas around NGC~4666 is also box-shaped as seen in radio. Interestingly, the bubble-like structure from the radio map coincides with a feature in the X-rays to the southeast. The spectral analysis was performed for three regions of NGC\,4666: The central disk region was chosen to supplement the analysis of the
magnetic fields of NGC 4666 in radio and has a boxsize of 140"~$\times$~28". Two additional regions of the same size on both sides of the disk region (east and west) were selected in order to look for possible changes in the energy budget of the ISM in the disk outskirts, as well as to study the emission for the hot gas in the galactic halo. All regions are presented in Figure~\ref{ximg2}.

 A complex
model was used for the spectral analysis of the emission from the disk region. 
It consisted of two gaseous components represented by a {\it mekal} model, which is an emission spectrum from hot diffuse gas based on the model calculations of Mewe and Kaastra \citep{mewe85,kaastra92}, accounting for the emission from the hot gas in the disk and the halo of NGC~4666, and a power-law model to account for the unresolved point sources in the galactic disk. As mentioned above, an additional absorbed power law and a simple Gaussian component were also used to account for the emission from the central source.

\begin{figure}
\begin{center}
\resizebox{\hsize}{!}{\includegraphics[angle=-90]{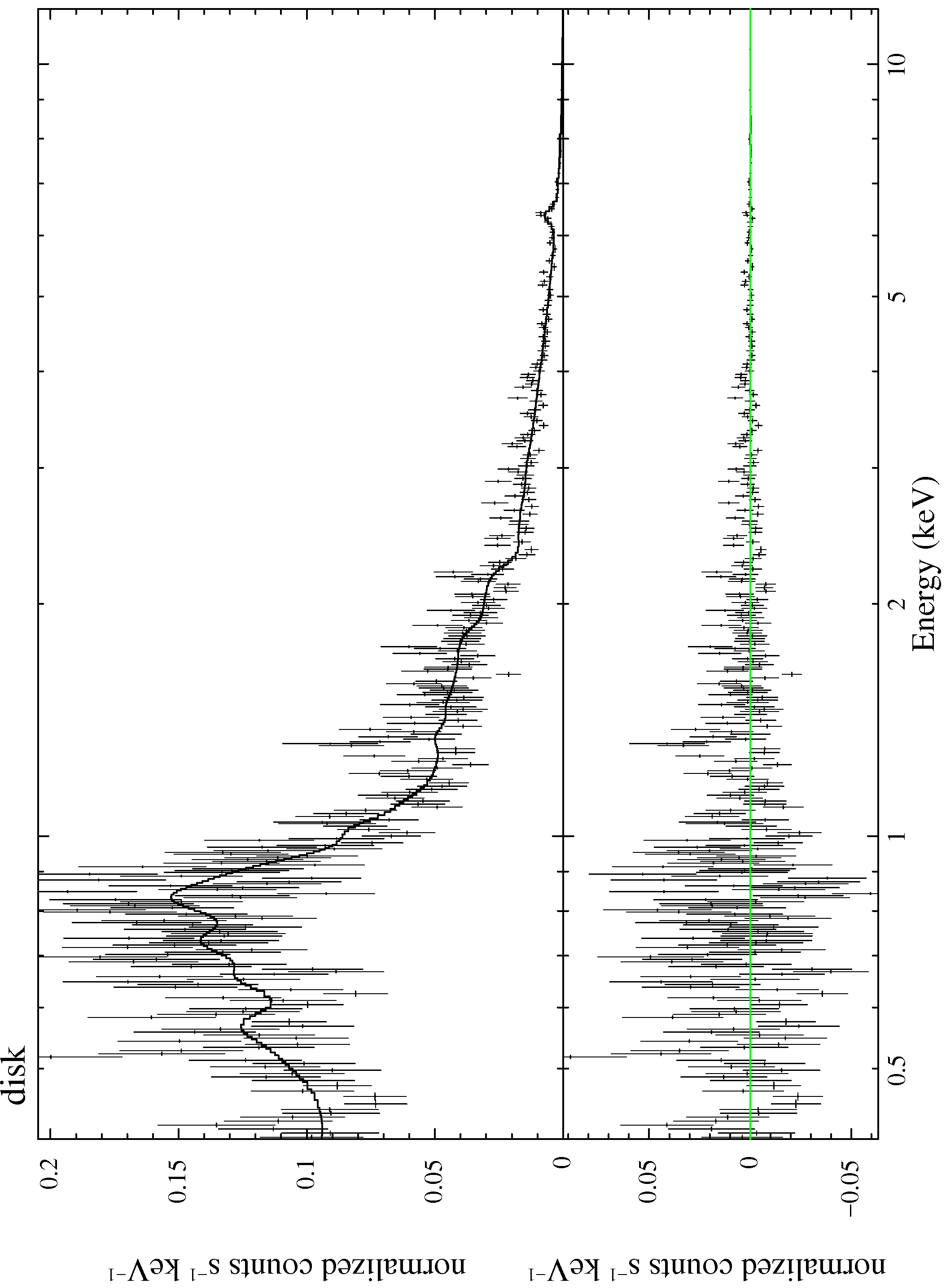}}
\resizebox{\hsize}{!}{\includegraphics[angle=-90]{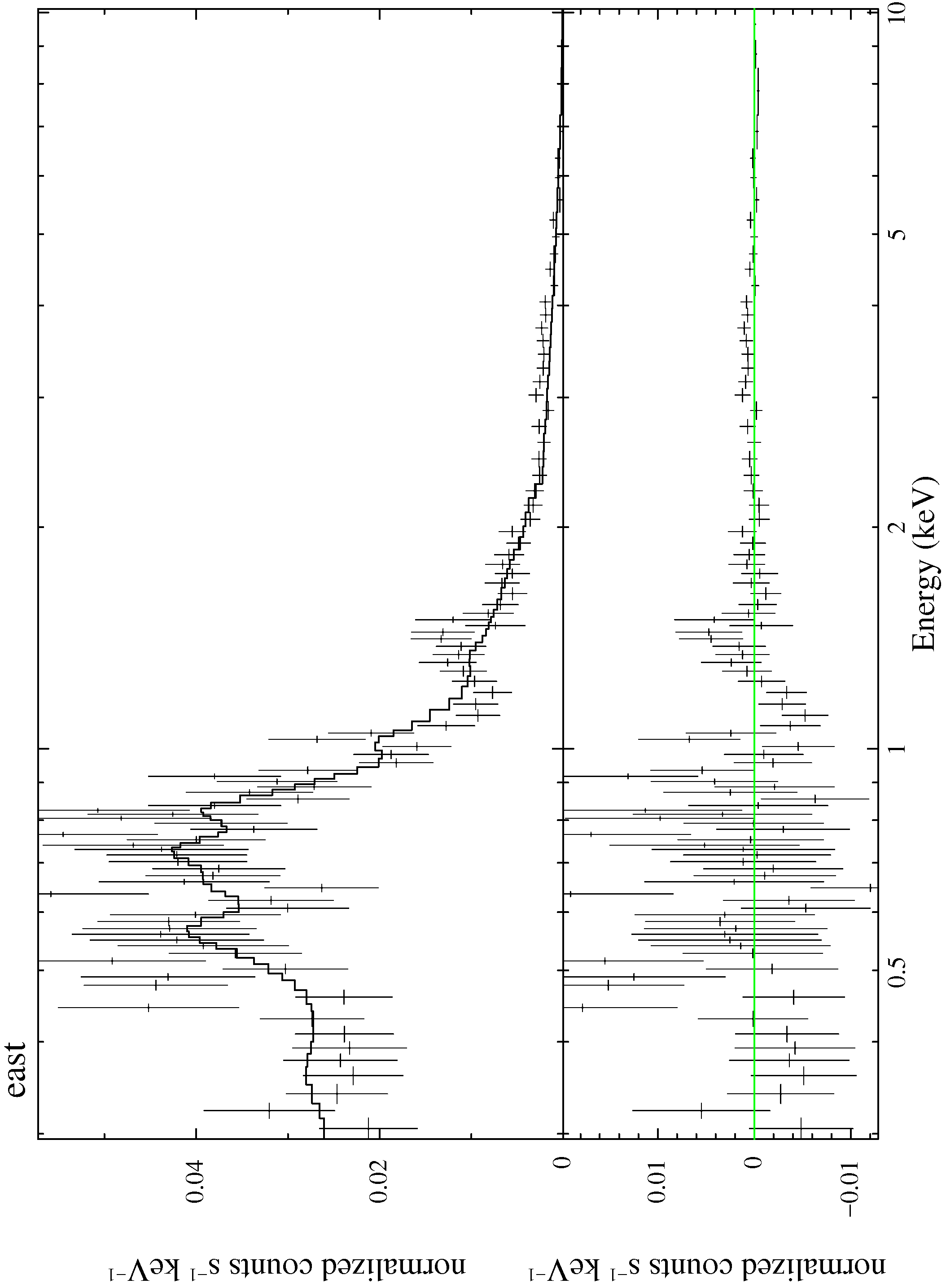}}
\resizebox{\hsize}{!}{\includegraphics[angle=-90]{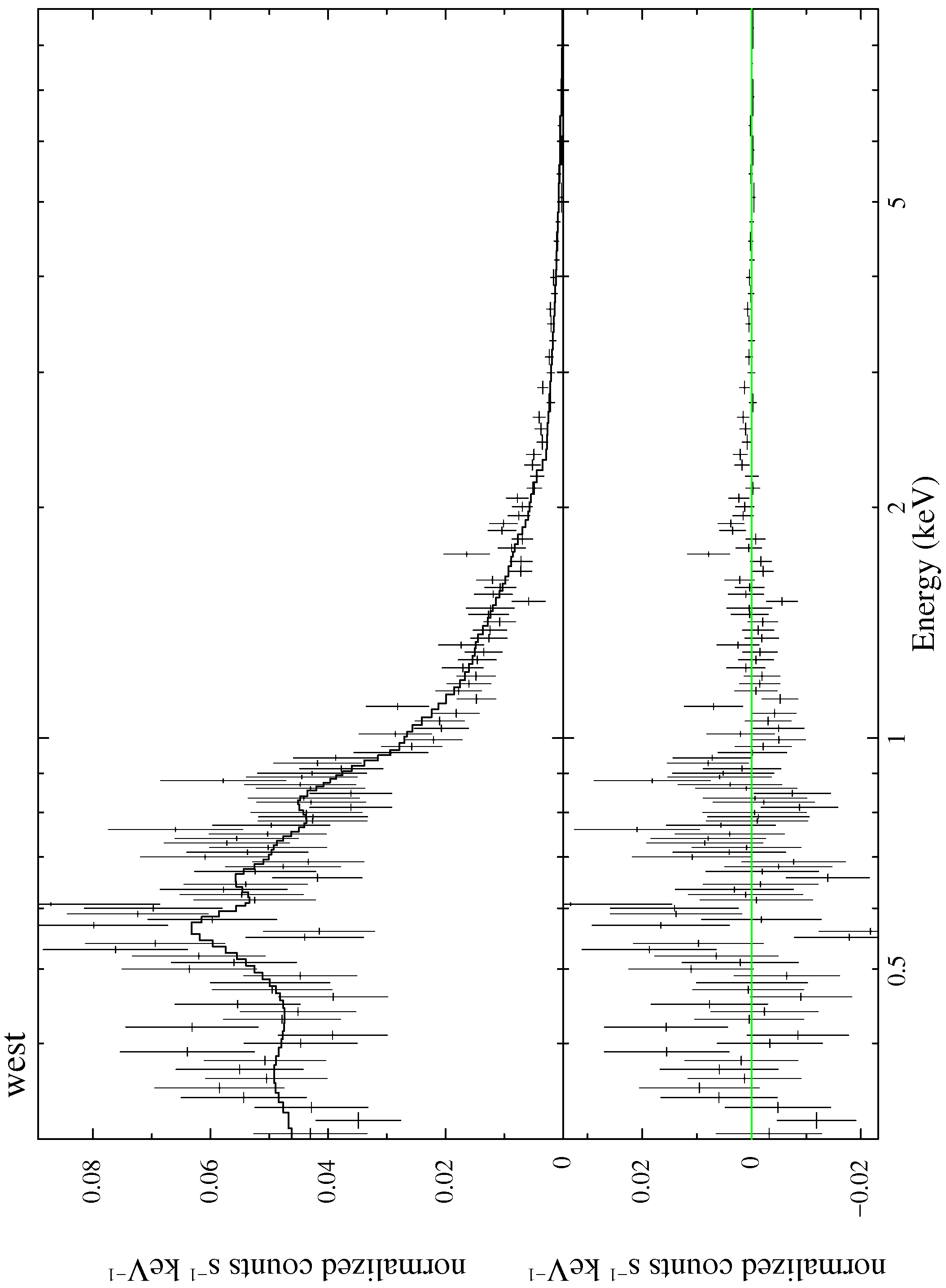}}
\end{center}
\caption{Model fits to the regions of NGC~4666 (see Tables~\ref{models},~\ref{agn}, and~\ref{xparams}).}
\label{spectra}
\end{figure}

The two remaining spectra (east and west regions) were fitted with a model consisting of two {\it mekal}s (hot gas from the disk and the halo), as well as a power law (unresolved disk sources). The spectra and the fitted models are presented in Figure~\ref{spectra} and Table~\ref{models}, and their fitted parameters in Tables~\ref{agn}~and~\ref{xparams}. 

\begin{table}
\caption{Model fit parameters for the regions in NGC~4666}
\label{xparams}
\centering
\begin{tabular}{clcccrr}
\hline\hline
Region          & kT$_1$                     & kT$_2$                   &       Photon                  \\
                & [keV]                      & [keV]                    &       Index                   \\
\hline
\vspace{5pt}
Disk            & 0.19$\pm$0.03              & 0.59$^{+0.03}_{-0.05}$   & 1.68$^{+0.79}_{-0.28}$        \\
\vspace{5pt}
East            & 0.17$\pm$0.04              & 0.43$^{+0.10}_{-0.09}$   & 1.47$\pm$0.14                 \\
\vspace{5pt}
West            & 0.20$^{+0.04}_{-0.03}$     & 0.56$^{+0.14}_{-0.23}$   & 1.84$^{+0.11}_{-0.12}$        \\
\hline
\end{tabular}
\end{table}

\begin{figure*}
\begin{minipage}[c]{0.43\textwidth}
        \centering
        \includegraphics[width=1.0\hsize]{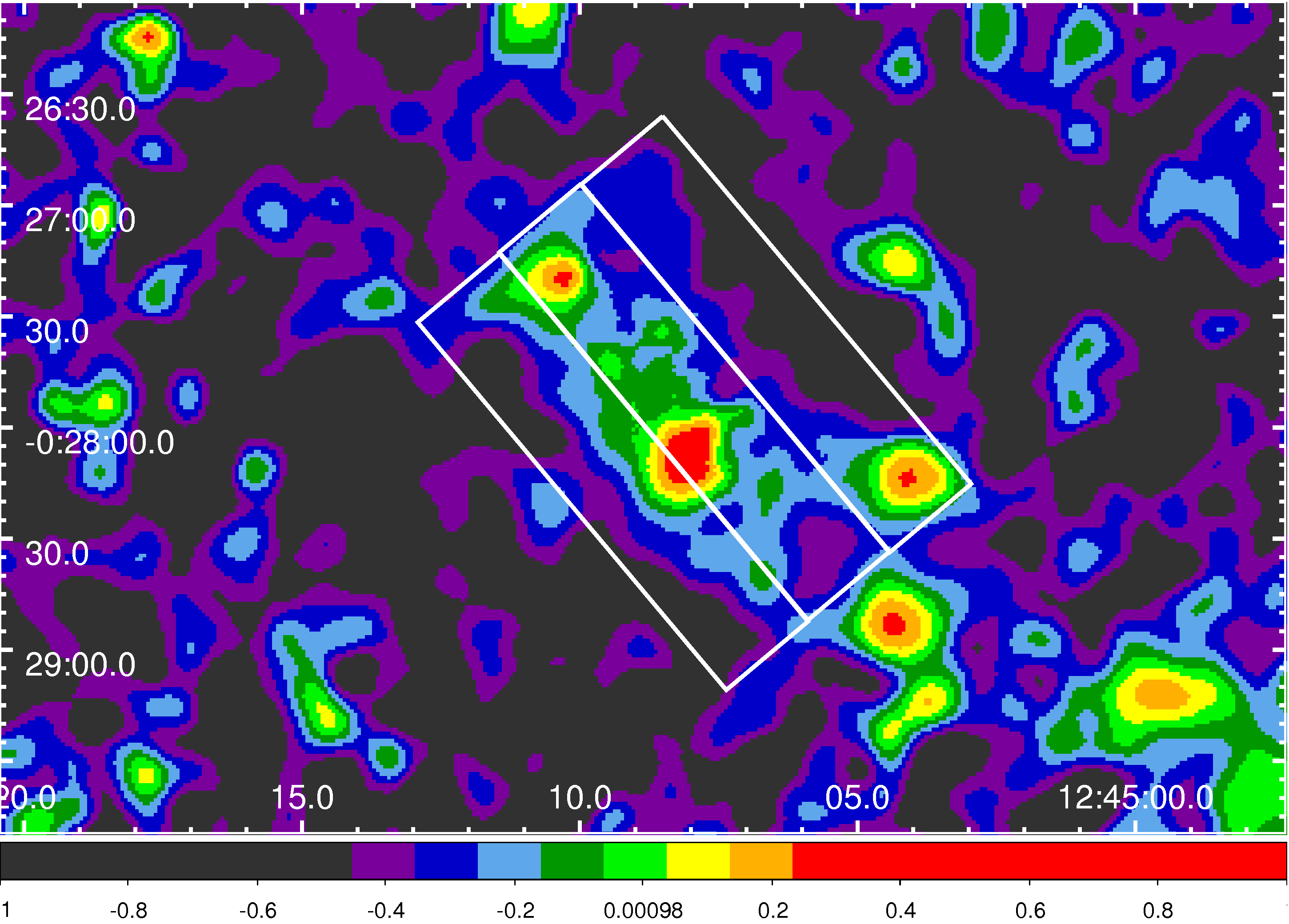}
        \caption {Hardness ratio map of NGC\,4666 with the spectral regions marked.}
        \label{xhr}
 \end{minipage}
 \begin{minipage}[c]{0.01\textwidth}
 \hfill
 \end{minipage}
 \begin{minipage}[c]{0.56\textwidth}
 
 \begin{threeparttable}
 \captionof{table}{Properties of the studied regions of NGC~4666}
\label{parametersmag_par}
\centering
\begin{tabular}{lcccccccc}
\hline\hline
\                       & n$_e$                                    & $\upeta$            &  L     & RM    &  B$_{\parallel}$ \\
\                         &[10$^{-3}$cm$^{-3}$] &                &  [kpc] & [rad m$^{-2}$]  &  [$\upmu$G]         \\
\hline
\vspace{5pt}
Disk    & 15~$\pm$~4                    &  0.5~$\pm$~0.1    & 6~$\pm$~2    &     150~$\pm$~50  & 2.1~$\pm$~0.9    \\
\vspace{5pt}
East    & 2.4~$\pm$~0.3                 &  0.6~$\pm$~0.2    & 20~$\pm$~5   &     90~$\pm$~50       & 2.3~$\pm$~1.4         \\
\vspace{5pt}
West    & 2.3~$\pm$~0.2                   &  0.6~$\pm$~0.2    & 20~$\pm$~5   &     125~$\pm$~50    & 3.4~$\pm$~1.7         \\
\hline 
\end{tabular}
\begin{flushleft}
\begin{tablenotes}
\item Notes: The columns give the region name, electron number density $n_e$, the volume filling factor $\eta$, line of sight L, the median RM value, and the magnetic field along the line of sight $B_{\parallel} = RM/(0.81 n_e L)$.
\end{tablenotes}
\end{flushleft}
\end{threeparttable}
 \end{minipage}
\end{figure*}

The parameters from the {\it mekal} components of the model fit were used to calculate electron densities, gas masses, thermal energies, and thermal energy densities of the hot gas. In our calculations we follow the widely accepted assumption that the cooler component describes the hot gas in the halo, while the hotter corresponds to the hot gas in the disk \citep[e.g.,][]{tuellmannI06}. Another important assumption is the emitting volume of the hot gas. In our calculations we assumed a disk thickness of 1\,kpc. For the halo we needed to take into account the inclination of the galaxy. The position of the spectral regions (Fig.~\ref{ximg2}) suggests that about half of both east and west regions still include emission from the disk. While in the latter region we expect the emission from both the disk and the halo (we see the top side of the disk), the east region might show a lower contribution from the soft emission (below ~1\,keV) of the halo below the disk. The hardness ratio map (Figure~\ref{xhr}) confirms our expectations, the eastern side of the disk being significantly harder than the western part. Consequently, we assumed the path length through the halo of NGC\,4666 to be 20\,kpc for the west region and 10\,kpc for the east region, which allows to account for different emitting volumes. All derived parameters of the hot gas both in the disk and in the halo are presented in Table~\ref{gasparams}.

Under the assumption that hot gas and magnetic fields are responsible for the observed Faraday RMs (Fig.~\ref{fig:N4666_C_RM}), we used the number densities of the hot gas derived from the X-ray spectra and calculated the strengths of the large-scale field parallel to the line of sight (Table~\ref{parametersmag_par}). The resulting field strengths of 2--3\,$\upmu$G are several times smaller than the strengths of the total field (see Fig.~\ref{fig:N4666_bfeld_genau}), as expected from the action of a large-scale dynamo.

\begin{table*}
\centering
\captionsetup{justification=centering}
\begin{threeparttable}
\caption{Derived parameters of the hot gas in the studied regions of NGC\,4666}
\label{gasparams}
\centering
\begin{tabular}{lcccccc}
\hline\hline
\ & Disk & Eastern disk & Western disk& Inner halo & Eastern halo&Western halo\\
\hline
\vspace{5pt}
n$_e$\,$\eta^{-0.5}$ [10$^{-3}$cm$^{-3}$] & 10.60$^{+1.00}_{-0.90}$ & 5.10$^{+1.48}_{-0.45}$ & 4.41$^{+1.03}_{-1.37}$& 
2.20$^{+0.32}_{-0.27}$ & 1.84$^{+0.27}_{-0.45}$ & 1.74$^{+0.19}_{-0.21}$ \\
\vspace{5pt}
M$_{gas}\,\eta^{0.5}$   [10$^6$M$_\odot$]  &  10.81$^{+0.97}_{-0.95}$   & 5.19$^{+1.51}_{-0.46}$  & 4.48$^{+1.05}_{-1.39}$&
44.80$^{+6.53}_{-5.60}$& 18.68$^{+2.77}_{-4.54}$& 35.46$^{+3.74}_{-4.33}$ \\
\vspace{5pt}
E$_{th}\,\eta^{0.5}$ [10$^{54}$\,erg]    & 18.27$^{+2.66}_{-3.01}$ & 6.39$^{+3.78}_{-1.79}$  & 7.19$^{+3.90}_{-4.27}$&
26.42$^{+3.92}_{-6.42}$  &9.10$^{+3.81}_{-3.83}$        & 20.32$^{+6.64}_{-5.16}$\\
\vspace{5pt}
$\epsilon_{th}\,\eta^{-0.5}$ [10$^{-12}$\,erg\,cm$^{-3}$]&15.10$^{+2.20}_{-2.49}$&5.28$^{+3.12}_{-1.48}$ &5.94$^{+3.22}_{-3.53}$&
35.46$^{+3.74}_{-4.33}$& 0.75$^{+0.31}_{-0.32}$ &  0.84$^{+0.27}_{-0.21}$\\
\hline
\end{tabular}
\begin{flushleft}
\begin{tablenotes}
\item Notes: $\eta$ is the volume filling factor. The lines give the region name, electron number density, gas mass, thermal energy, thermal energy density.
\end{tablenotes}
\end{flushleft}
\end{threeparttable}
\end{table*}

\normalsize

\subsection{Scale heights}
The intensity profile perpendicular to the major axis of edge-on galaxies can be best fitted by exponential or Gaussian functions. We follow the scale height determination undertaken for 13 CHANG-ES galaxies in \citet{krauseetal2018} to fit two-component exponentials to the radio intensity profiles. The scale height analysis was carried out on the combined C-band and L-band data. To compare the results in both bands the maps were smoothed to the same beam size of 13.3"~$\times$~13.3". We used the "BoxModels" tool within the NOD3 software package \citep{mulleretal2017}. The parameters of the analysis are presented in Table~\ref{tab:N4666nod3para}. 

\begin{figure*}
         \begin{minipage}[t]{0.49\textwidth}
\includegraphics[width=\hsize]{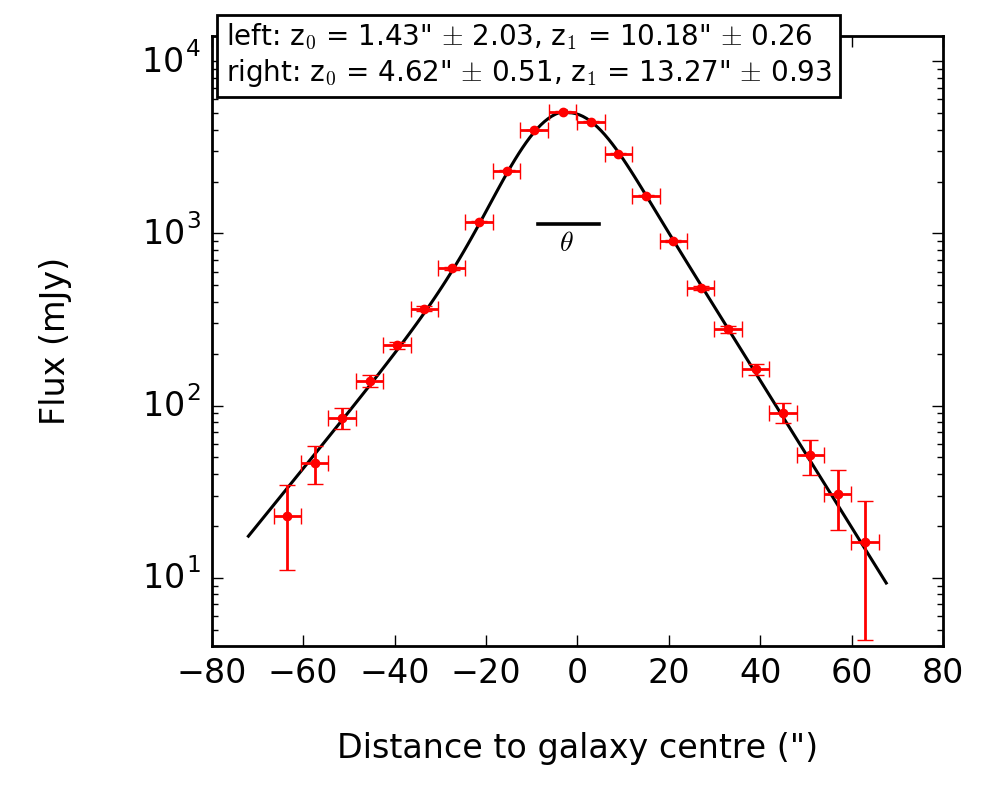}
        \caption{NGC~4666 scale height analysis C-band with different model fits for the northern (positive distances) and southern (negative distances) halo.}
        \label{fig:N4666_scaleheights_Cband}    \end{minipage}
\hspace{0.01\textwidth}
 \begin{minipage}[t]{0.49\textwidth}
\includegraphics[width=\hsize]{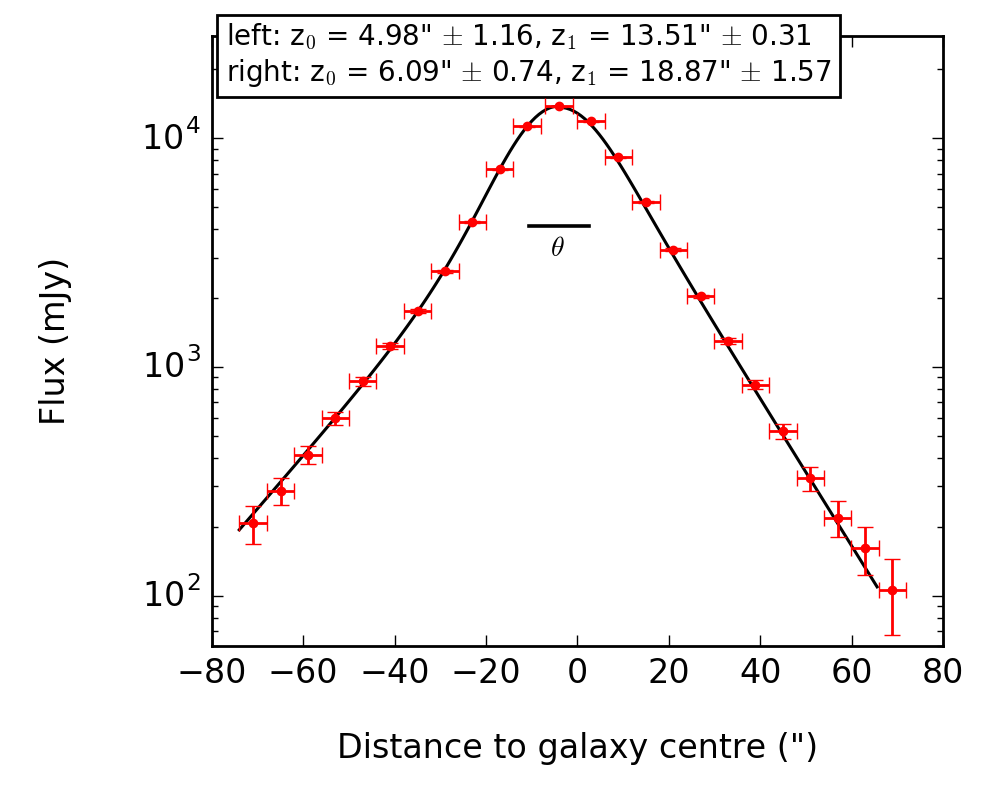}
                \caption{NGC~4666 scale height analysis L-band with different model fits for the northern (positive distances) and southern (negative distances) halo.}
        \label{fig:N4666_scaleheights_Lband}
                \end{minipage}
\end{figure*}

\begin{figure}
        \centering
                \includegraphics[width=\hsize]{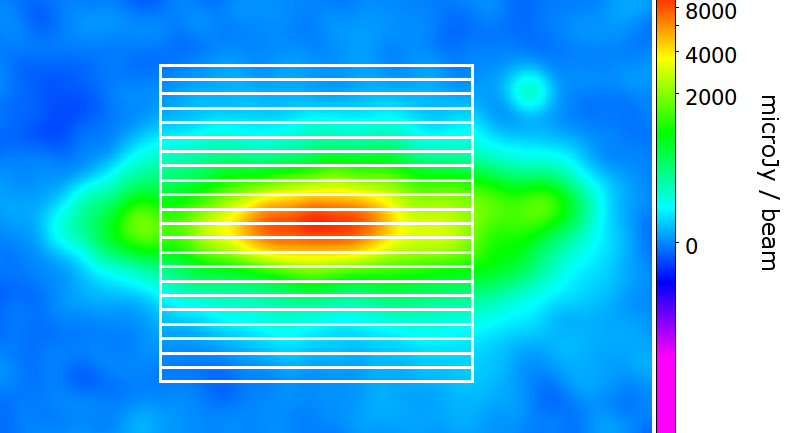}
                \caption{NGC~4666 boxes with a size of 130"~$\times$~6" for the scale height analysis in C-band with a beam of 13.3"~$\times$~13.3".}
        \label{fig:N4666_boxes_Cband}
\end{figure}

For NGC~4666 it is not easy to find good fit solutions with two-component exponentials along the whole disk if the galaxy is divided into different strips along the major axis. This is mainly due to its asymmetry between the two sides of the major axis. Therefore we used a box size of 130" $\times$ 6" to fit one intensity profile for the central galaxy, omitting the asymmetric outer part (see Figure~\ref{fig:N4666_boxes_Cband}). Furthermore, an asymmetry was found above and below the plane between the eastern and the western halo. Therefore fits were undertaken separately for above and below the plane. The resulting fits and scale heights are presented in Figures~\ref{fig:N4666_scaleheights_Cband} and \ref{fig:N4666_scaleheights_Lband} as well as in Tables~\ref{tab:N4666scaleheightsonestripecband} and \ref{tab:N4666scaleheightsonestripelband}. We note that with the inclination of 70$^\circ$ from \citet{walteretal2004} we were not able to fit the scale heights. Despite the fact that in optical images of this galaxy the disk is clearly seen and thus cannot be nearly 90$^\circ$, we conclude from our analysis that the inclination is rather $85^{\circ}$~$\pm$~2$^\circ$.

The resulting mean scale heights of the thin disk are 0.41~$\pm$~0.18\,kpc and 0.74~$\pm$~0.12\,kpc for C-band and L-band, respectively. The mean thick disk scale height in C-band is 1.57~$\pm$~0.21\,kpc and in L-band is 2.16~$\pm$~0.36\,kpc.

\begin{table}
\centering
\caption{NGC~4666 parameters}  
\label{tab:N4666nod3para}
\begin{tabular}{lcc}
\hline \hline
Parameter                                               &C-band                 & L-band\\
\hline
Beam (")                                                &       13.3                    & 13.3 \\
Effective beam (")                              & 16.9                  & 17.4 \\
Inclination     ($^\circ$)                      &  85                    &85                       \\
Position Angle ($^\circ$)               &  40                    &40              \\     
rms     (mJy/beam)                                      &  10                            &30             \\
Galaxy diameter (")                             &       240                     &260                    \\
Box width (")                                   & 130                           &130              \\
Box height (")                                  &  6                            &6                           \\
Number of boxes in X                     &  1                             &1               \\
Number of boxes in Y                            &  22                    &24               \\
\hline
\end{tabular}
\end{table}

\begin{table}
\centering
\caption{NGC~4666 scale heights of C-band}  
\label{tab:N4666scaleheightsonestripecband}
\begin{tabular}{llcc}
\hline \hline
                        &        &\multicolumn{2}{c}{Scale height C-band}        \\
                        &       &        ["]                                               & [kpc]                                       \\
\hline
\multirow{2}{*}{Left}    & thin component  &  1.43 $\pm$ 2.03  & 0.19 $\pm$ 0.31    \\
                                                & thick component &  10.18 $\pm$ 0.26 & 1.36  $\pm$ 0.03       \\
\hline
 \multirow{2}{*}{Right} & thin component  &  4.62 $\pm$ 0.51  & 0.62 $\pm$ 0.07    \\
                                                &thick component &  13.27 $\pm$ 0.93 & 1.77 $\pm$ 0.12  \\
\hline
\multirow{2}{*}{Mean} & thin component  & 3.03 $\pm$ 1.27 & 0.41 $\pm$ 0.18         \\
                                          &thick component & 11.73  $\pm$ 0.59  & 1.57 $\pm$ 0.21         \\
\hline
\end{tabular}
\end{table}

\begin{table}
\centering
\caption{NGC~4666 scale heights of L-band}  
\label{tab:N4666scaleheightsonestripelband}
\begin{tabular}{llcc}
\hline \hline
                                &        &\multicolumn{2}{c}{Scale height L-band} \\
                                &        &               ["]                             & [kpc]                 \\
\hline
\multirow{2}{*}{Left}    & thin component  &   4.98 $\pm$ 1.16     &  0.66 $\pm$ 0.15                              \\
                 &       thick component &  13.51 $\pm$ 0.31     & 1.80  $\pm$ 0.04        \\
\hline
 \multirow{2}{*}{Right}  &  thin component  &   6.09 $\pm$ 0.74     &  0.81 $\pm$ 0.10                              \\
                   &  thick component &   18.87 $\pm$ 1.57     &  2.52 $\pm$ 0.20          \\
\hline
\multirow{2}{*}{Mean} & thin component  & 5.54 $\pm$ 0.95 & 0.74 $\pm$ 0.12             \\
                                          &thick component  & 16.19 $\pm$ 0.94 & 2.16 $\pm$ 0.36                  \\
\hline
\end{tabular}
\end{table}

\begin{table*}
\begin{threeparttable}
\centering
\caption{Mean scale heights of NGC~4666}  
\label{tab:N4666para}
\begin{tabular}{ccccccc}
\hline \hline
\multicolumn{2}{c}{Mean scale height z$_C$}             & \multicolumn{2}{c}{Mean scale height z$_L$}                     & Galaxy diameter       d$_{r,\ C-band}$         &       MSD     &       Normalized scale height $\tilde{\text{z}}_C$ \\
 \ ["]                                                                          & [kpc]               & ["]                &        [kpc]                         & ["]                                                                                 & [10$^7$ M$_{\odot}$]          &               \\
\hline
11.73   $\pm$ 1.54              & 1.57 $\pm$ 0.21        &      16.19 $\pm$ 2.68   &  2.16 $\pm$ 0.36      & 240 $\pm$ 10              & 14.66                                                       &       4.89 $\pm$ 0.67                                                                                          \\
\hline
\end{tabular}
\small
\begin{tablenotes}
\item Notes. z$_\text{C}$ and z$_\text{L}$ are the mean values of Tables~\ref{tab:N4666scaleheightsonestripecband} and ~\ref{tab:N4666scaleheightsonestripelband}. MSD is the total mass surface density defined in \citet{krauseetal2018} as MSD~=~M$_T$/$\pi$~(d$_{25}$/2)$^2$ with the total mass M$_T$~=~1.3~$\times$ 10$^{11}$~M$_{\odot}$ and d$_{25}$~=~33.6\,kpc from \citet{irwinetal2012}. The normalized scale height is $\tilde{\text{z}}_\text{C}$~=~z$_C$/d$_r$~$\cdot$~100 as defined in \citet{krauseetal2018}.\\
\end{tablenotes}
\end{threeparttable}
\end{table*}

\begin{figure}
        \includegraphics[width=1.0\hsize]{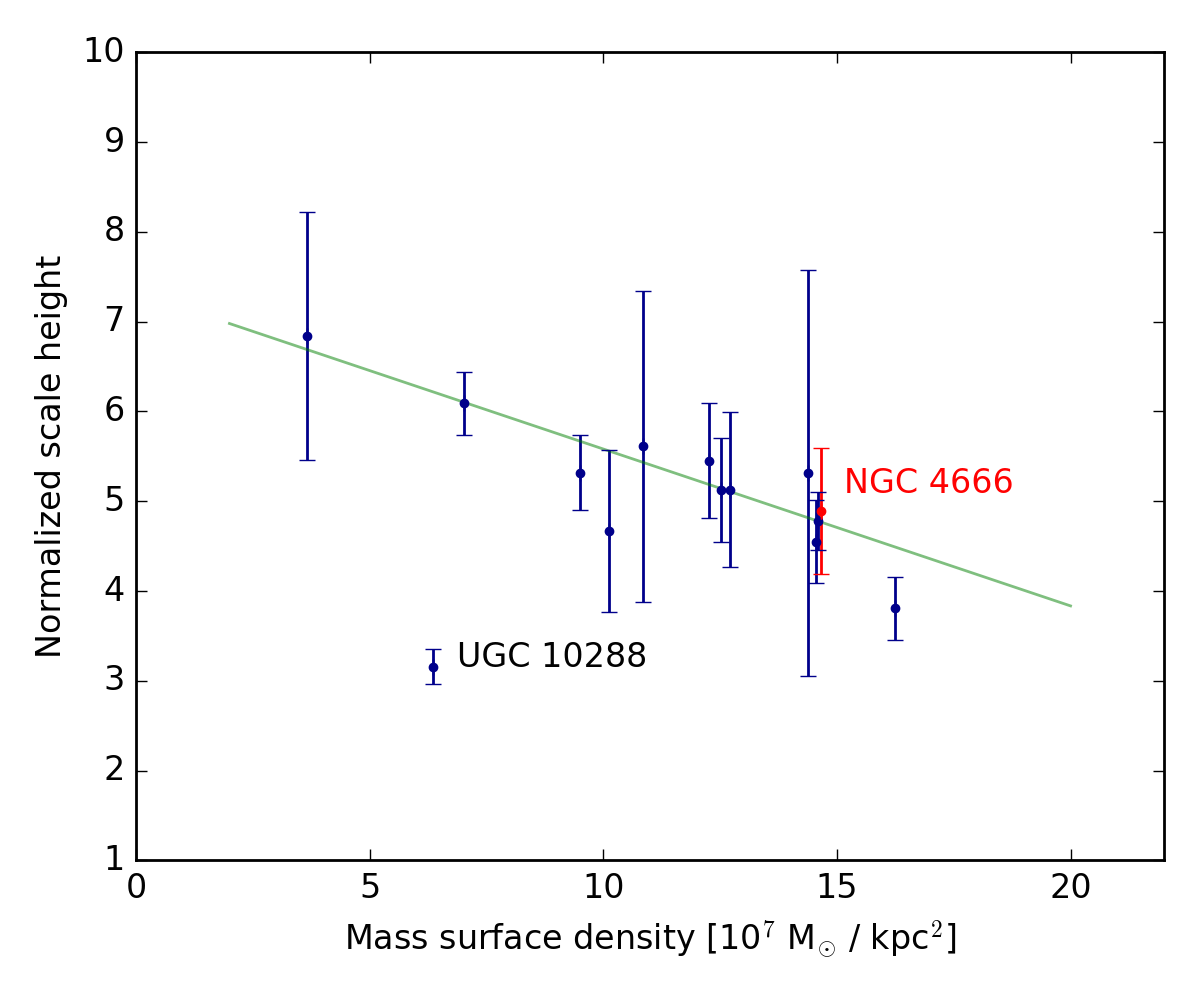}
        \caption{Plot of mass surface density versus normalized scale height for the CHANG-ES galaxies from \citet{krauseetal2018} with additional data point for NGC~4666 from Table~\ref{tab:N4666para}, which is marked in red. The data can be described by a linear fit (excluding UGC 10288).}
        \label{fig:N4666_normscaleheight_MSD}
\end{figure}

In Figure~\ref{fig:N4666_normscaleheight_MSD} the mass surface density (MSD) is plotted against the normalized scale height for 13 CHANG-ES galaxies from \citet{krauseetal2018}. The data point for NGC~4666 is added in red. The MSD is defined in \citet{krauseetal2018} as MSD~=~M$_T$/$\pi$~(d$_{25}$/2)$^2$ with the values for NGC~4666 of the total mass of M$_T$~=~1.3~$\times$ 10$^{11}$~M$_{\odot}$ and d$_{25}$~=~33.6\,kpc from \citet{irwinetal2012}. The normalized scale height is defined as $\tilde{\text{z}}_\text{C}$~=~z$_C$/d$_r$~$\cdot$~100 in \citet{krauseetal2018}. All derived values are presented in Table~\ref{tab:N4666para}. The galaxy fits nicely within the trend of an anticorrelation between those two parameters showing that with lower mass surface densities the normalized scale heights are increasing.

\subsection{Synchrotron polarization}
\subsubsection{Magnetic field orientation}
\begin{figure}
        \includegraphics[width=1.01\hsize]{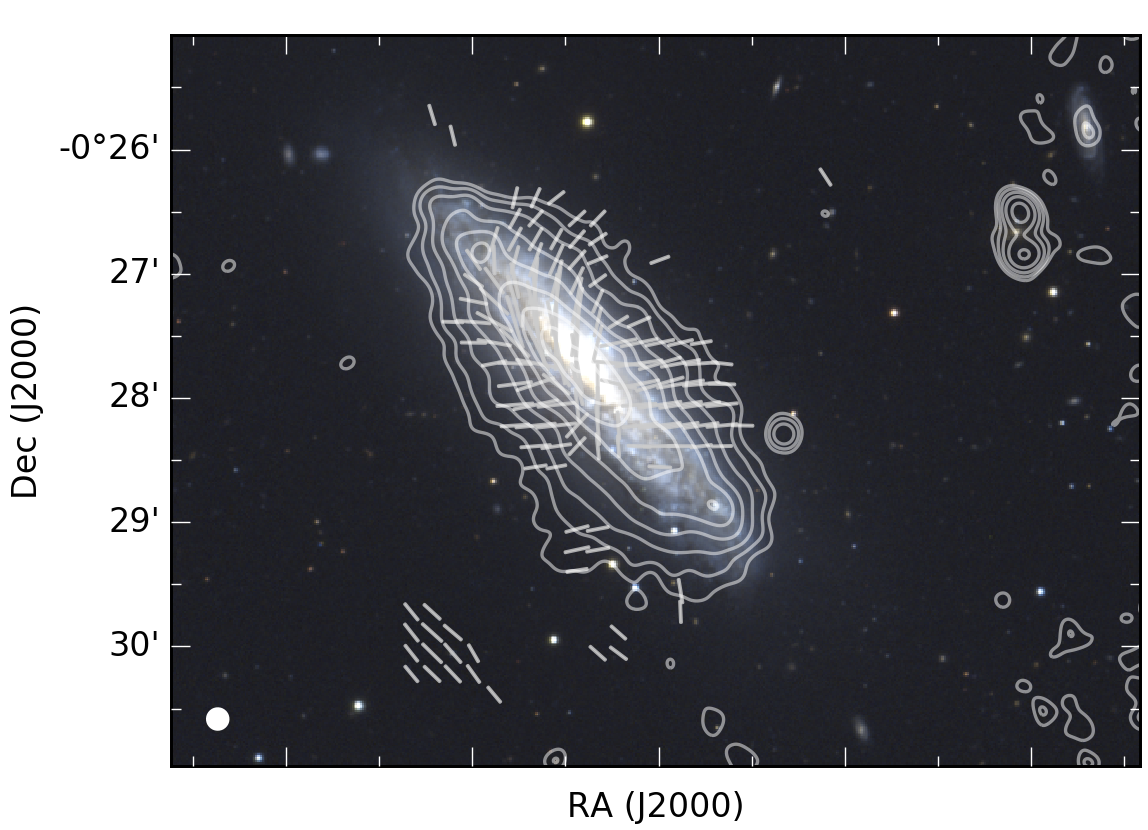}
        \caption{Color image of NGC~4666 produced from SDSS with Stokes I contours from C-band starting at a 3$\sigma$ level with a $\sigma$ of 11.0\,$\mu$Jy/beam and increase in powers of 2 (up to 128) with a robust zero weighting. The corresponding beam size is 10"~$\times$~10" (see bottom left). The apparent magnetic field orientations are shown in white.}
        \label{fig:N4666_Ccomb_synth_rob0_color}
\end{figure}

Figure~\ref{fig:N4666_Ccomb_synth_rob0_color} shows the magnetic field orientations with Stokes I contours from C-band overlayed on the optical SDSS image.  

\begin{figure*}
         \begin{minipage}[t]{0.49\textwidth}
        \includegraphics[width=\textwidth]{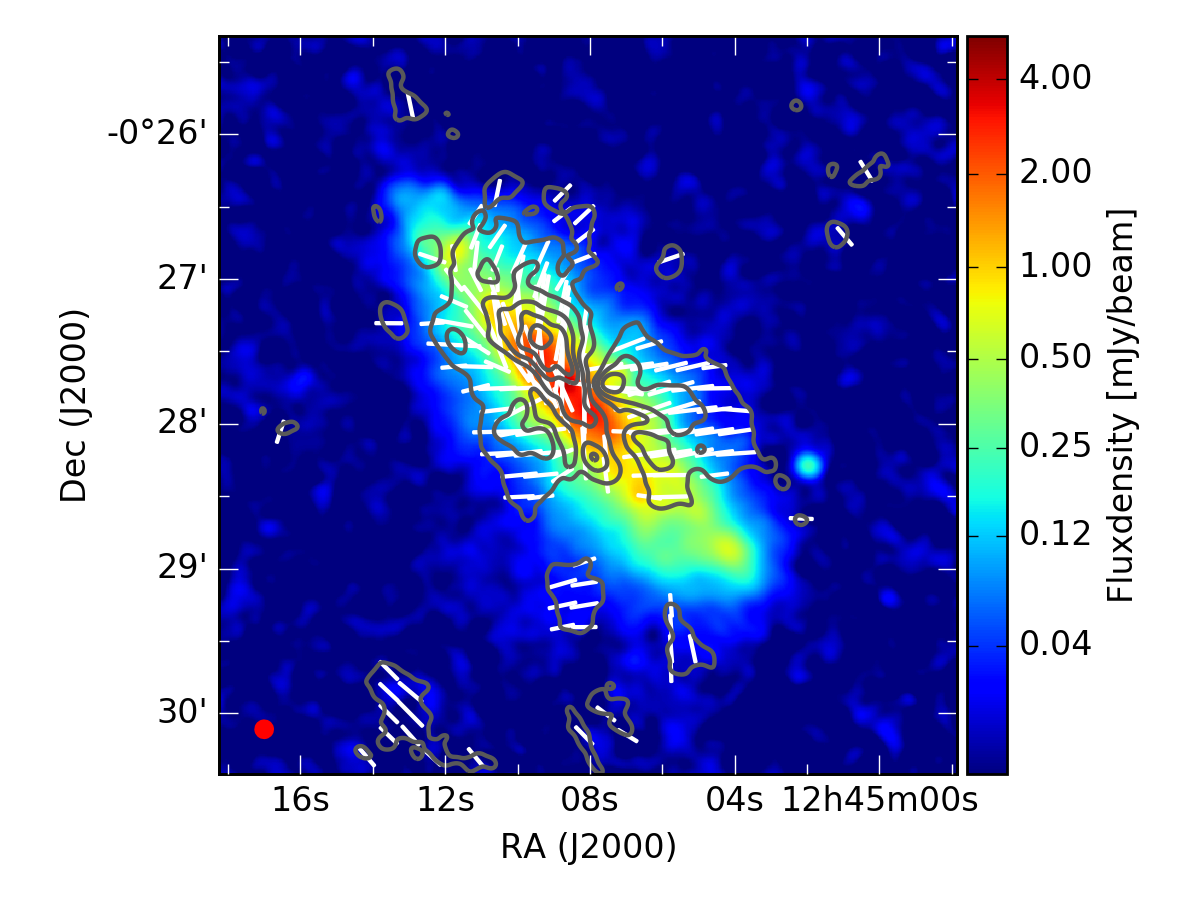}
        \caption{Total intensity as a color image of NGC~4666 from C-band with robust zero weighting, with a beam of 7"~$\times$~7" in red, obtained by Gaussian smoothing to fit the resolution of the polarization map, $\sigma$ is 8.1\,$\mu$Jy/beam. Gray polarization contours are at 3, 6, 9, 12, 15, 18 $\sigma$ levels with $\sigma$ of 7.0\,$\mu$Jy/beam with a robust two weighting. The apparent magnetic field orientations are shown in white.}
        \label{fig:N4666_Ccomb_synth_rob0}
        \end{minipage}
\hspace{0.01\textwidth}
 \begin{minipage}[t]{0.49\textwidth}
                \includegraphics[width=\textwidth]{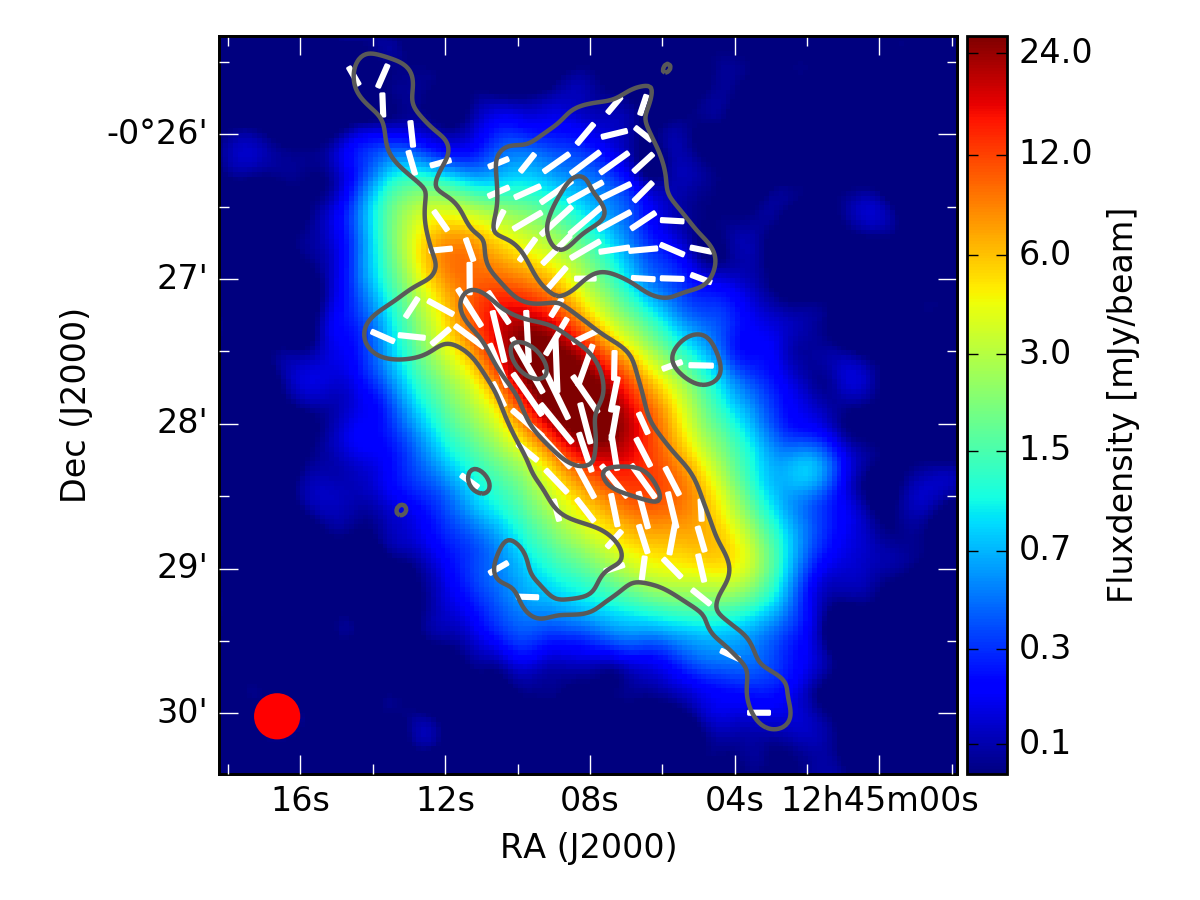}
        \caption{Total intensity as a color image of NGC~4666 from L-band with robust zero weighting, with a beam of 18"~$\times$~18" in red, obtained by Gaussian smoothing to fit the resolution of the polarization map with $\sigma$ of 40.1\,$\mu$Jy/beam. Gray polarization contours from RM synthesis are at 3, 6, 9 $\sigma$ levels with $\sigma$ of 20.0\,$\mu$Jy/beam with a robust two weighting. The magnetic field orientations are shown in white.}
        \label{fig:N4666_Lcomb_synth_rob2}
        \end{minipage}
        \end{figure*}
        \begin{figure*}
         \begin{minipage}[t]{0.49\textwidth}
        \includegraphics[width=\textwidth]{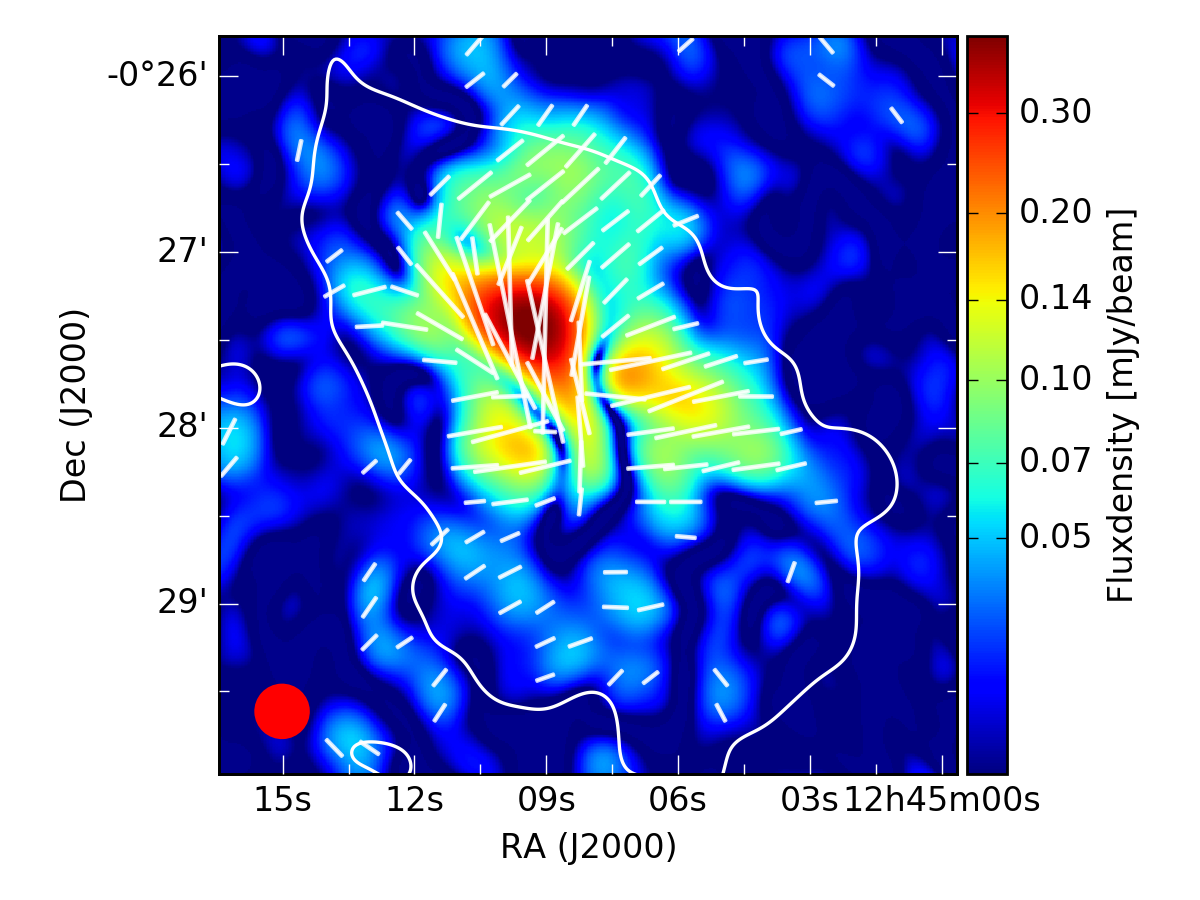}
        \caption{Polarized intensity as a color image of NGC~4666 from C-band with robust zero weighting and a uv-taper of 12k$\lambda$ and Gaussian smoothing to obtain the beam of 18"~$\times$~18" in red, $\sigma$ is 11\,$\mu$Jy/beam. The white total intensity contour correspond to 0.04\,mJy with robust zero weighting and smoothing. The apparent magnetic field orientations are shown in white.}
        \label{fig:N4666_Ccomb_rob0_15}
        \end{minipage}
\hspace{0.01\textwidth}
 \begin{minipage}[t]{0.49\textwidth}
                \includegraphics[width=\textwidth]{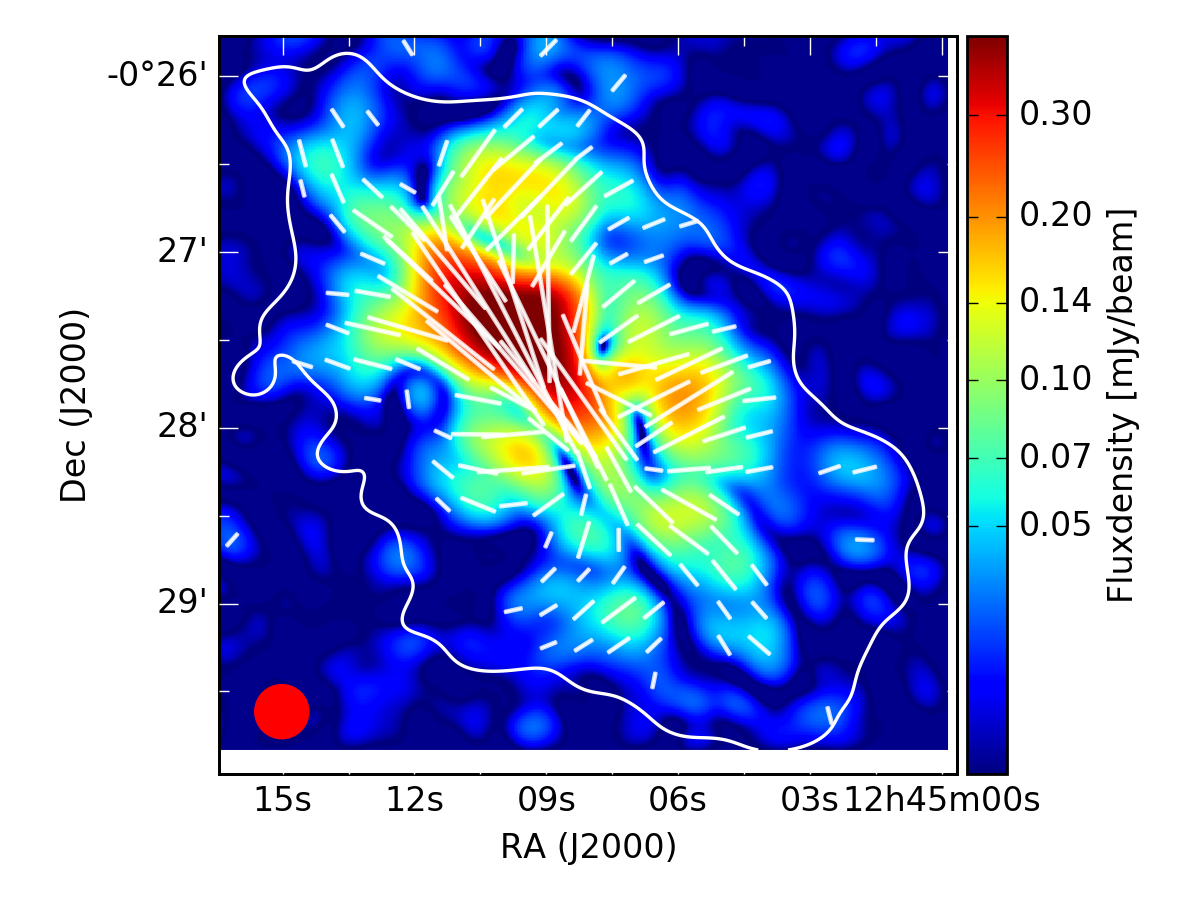}
        \caption{Polarized intensity as a color image of NGC~4666 from archival VLA C-band data with robust zero weighting, originally published in \citet{soidaetal2005}, with a beam of 18"~$\times$~18" in red, $\sigma$ is 9.3\,$\mu$Jy/beam. The white total intensity contour correspond to 0.04\,mJy with robust zero weighting and smoothing. The apparent magnetic field orientations are shown in white.}
        \label{fig:N4666_soida}
        \end{minipage}
        \end{figure*}

The polarization data of NGC~4666 for C-band are shown in Figure~\ref{fig:N4666_Ccomb_synth_rob0} (and Figure~\ref{fig:N4666_Ccomb_rob2_pol}). The polarized intensity contours and magnetic field orientations generated by imaging Stokes Q and U are shown on Stokes I (Stokes I contours on polarized intensity). For C-band, when imaging Stokes Q and U with robust two weighting compared to using RM synthesis with robust two weighting, we detect a similar distribution of extended polarized emission but a little more polarized flux. Therefore, Figure~\ref{fig:N4666_Ccomb_synth_rob0} shows the apparent magnetic field orientations from Q and U without RM synthesis. 
With the RM maps obtained from RM synthesis (as discussed in Section 3.6.3) the intrinsic magnetic field orientations may deviate from the apparent magnetic field vectors by less than $20\degr$ in most parts of NGC~4666.

In C-band, polarized intensity from nearly the entire galaxy is visible. The magnetic field orientations show an X-shaped structure and the intensity contours extend far into the halo as well as into the disk. There is no polarized intensity above 3$\sigma$ from the southern side of the galaxy, which is the receding side.


In Figure~\ref{fig:N4666_Lcomb_synth_rob2} (and Figure~\ref{fig:N4666_Lcomb_synth_rob2_pol}) the L-band polarization data are presented with intensity contours and magnetic field orientation on Stokes I using RM synthesis (the L-band Stokes I contours on polarized intensity). 

In L-band, the Faraday rotation effect is quite strong, meaning that some polarization is expected to be depolarized and not visible in the map. Without RM synthesis, polarized intensity in L-band is only observed in a small fraction of the disk. After applying RM synthesis, a factor of 1.4 more polarized intensity can be recovered, which is seen in Figure~\ref{fig:N4666_Lcomb_synth_rob2}. The polarized emission is distributed over the whole galaxy disk reaching into the halo. 

Specifically, in the halo to the northwest of the galaxy, a large polarized emission region is located with vertical field components. This region is also seen in the C-band data, but the emission does not reach as far out into the halo as in L-band. This indicates a large-scale ordered magnetic field extending into the halo. In summary, an X-shaped magnetic field structure is again visible in the polarization map of L-band using RM synthesis.

\subsubsection{Large-scale magnetic field in C-band -- comparison between L-band and the archival VLA map}

In Figure~\ref{fig:N4666_Ccomb_rob0_15} the large-scale polarization from the CHANG-ES C-band data is presented. The halo emission seen in this map is consistent with the emission in the L-band image (Fig.~\ref{fig:N4666_Lcomb_synth_rob2_pol}). Especially the polarized intensity to the southeast of the galaxy is recovered (RA~12h~45m~10s DEC~-0$^\circ$~29'~10"). The origin of this polarization structure could be the shell-like feature marked in the total intensity image with a higher resolution (Fig.~\ref{fig:N4666_Ccomb_bw}). Also, the extended polarized emission region in the northwestern halo of the galaxy is similar to the L-band image (Fig.~\ref{fig:N4666_Lcomb_synth_rob2}), whereas the polarized emission region in the southeastern halo in Figure~\ref{fig:N4666_Ccomb_rob0_15} is only recovered as a small feature in the L-band image.

To compare the CHANG-ES C-band data with the archival VLA C-band data, the CHANG-ES data were imaged with robust zero weighting. Due to the fact that the archival VLA data have nearly 380 minutes observing in C-band D-configuration compared to only 40 minutes of CHANG-ES C-band data, but 3 hours of C-configuration observations, a comparison is only possible by higher weighting the CHANG-ES data D-configuration using uv tapering. In general, the Stokes I background images, the polarized intensity, and the magnetic field vectors are comparable between the old VLA \citep[Fig.~\ref{fig:N4666_soida}, original published in][]{soidaetal2005} and the new CHANG-ES image (Fig.~\ref{fig:N4666_Ccomb_rob0_15}). Looking closer at the maps of NGC~4666 from C-band it is visible that in the CHANG-ES map there is less extended polarized flux, especially in the south (receding side) of the galaxy. In C-band this is probably not due to depolarization effects. Bandwidth depolarization can be ruled out because the polarization map produced with RM synthesis also shows only polarized emission in this region below 3$\upsigma$. If we compare both maps (Fig.~\ref{fig:N4666_soida} and Fig.~\ref{fig:N4666_Ccomb_rob0_15}) more carefully, we have to consider the different central frequencies of 4.86 GHz and 6 GHz for the archival VLA and the CHANG-ES observations, respectively. This leads to 20\% higher intensities (assuming an average spectral index of $-0.8$) in the archival VLA maps. For the comparison of polarized emission in the halo, especially in the south, we take the difference in central frequency into account by dividing the 4.86\,GHz image by 1.2 (assuming an average spectral index of $-0.8$). The difference between the polarized intensities in the south is less than 5$\upsigma$.

Significant differences between the two observations are the uv distribution and weighting as well as the integration time. The much larger bandwidth of the CHANG-ES observations does not compensate for the much shorter observation times compared to the archival data.

\begin{figure*}
         \begin{minipage}[t]{0.49\textwidth}
                \includegraphics[width=\textwidth]{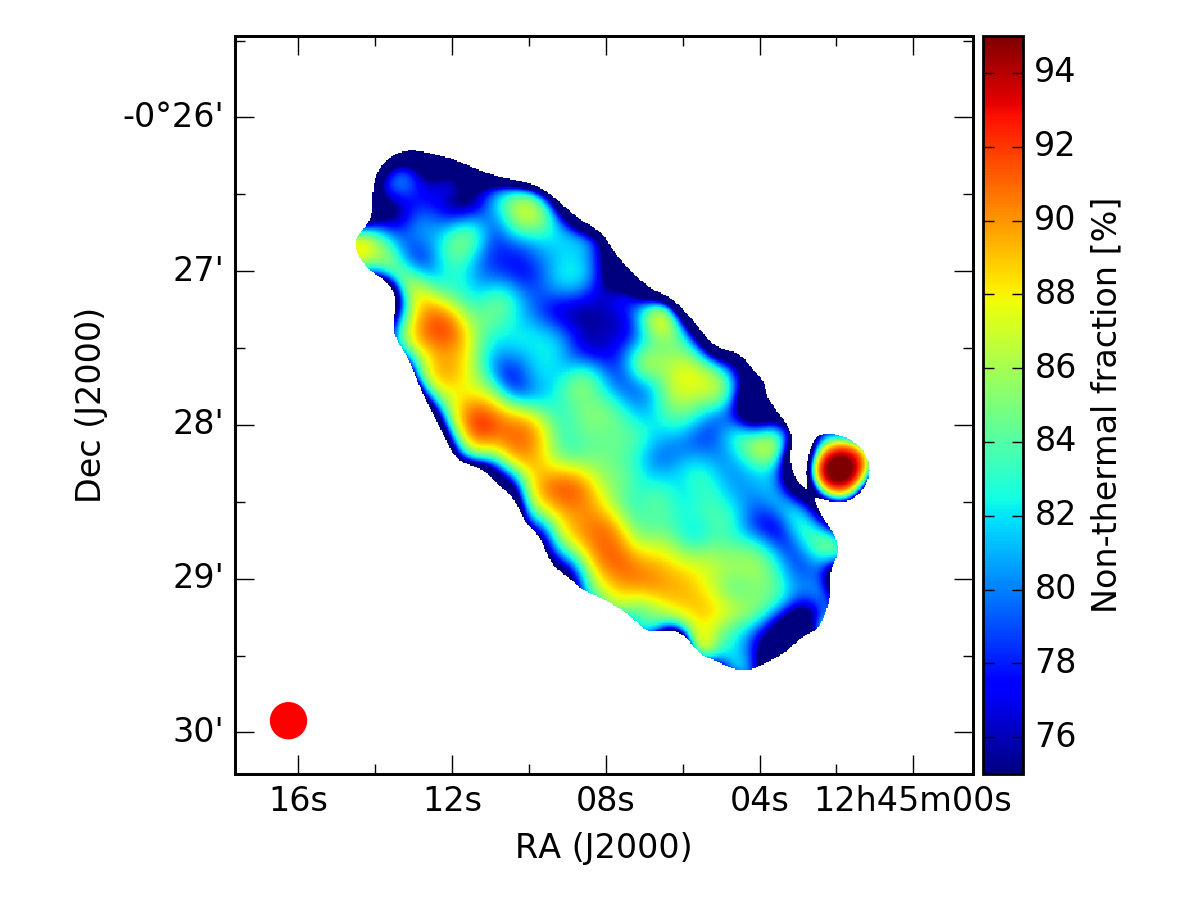}
        \caption{NGC~4666 nonthermal fraction map at C-band. The beam of 13.3"~$\times$~13.3" is shown in red (left bottom).}
        \label{fig:N4666_nonthermalfracC}
        \end{minipage}
\hspace{0.01\textwidth}
 \begin{minipage}[t]{0.49\textwidth}
                \includegraphics[width=\textwidth]{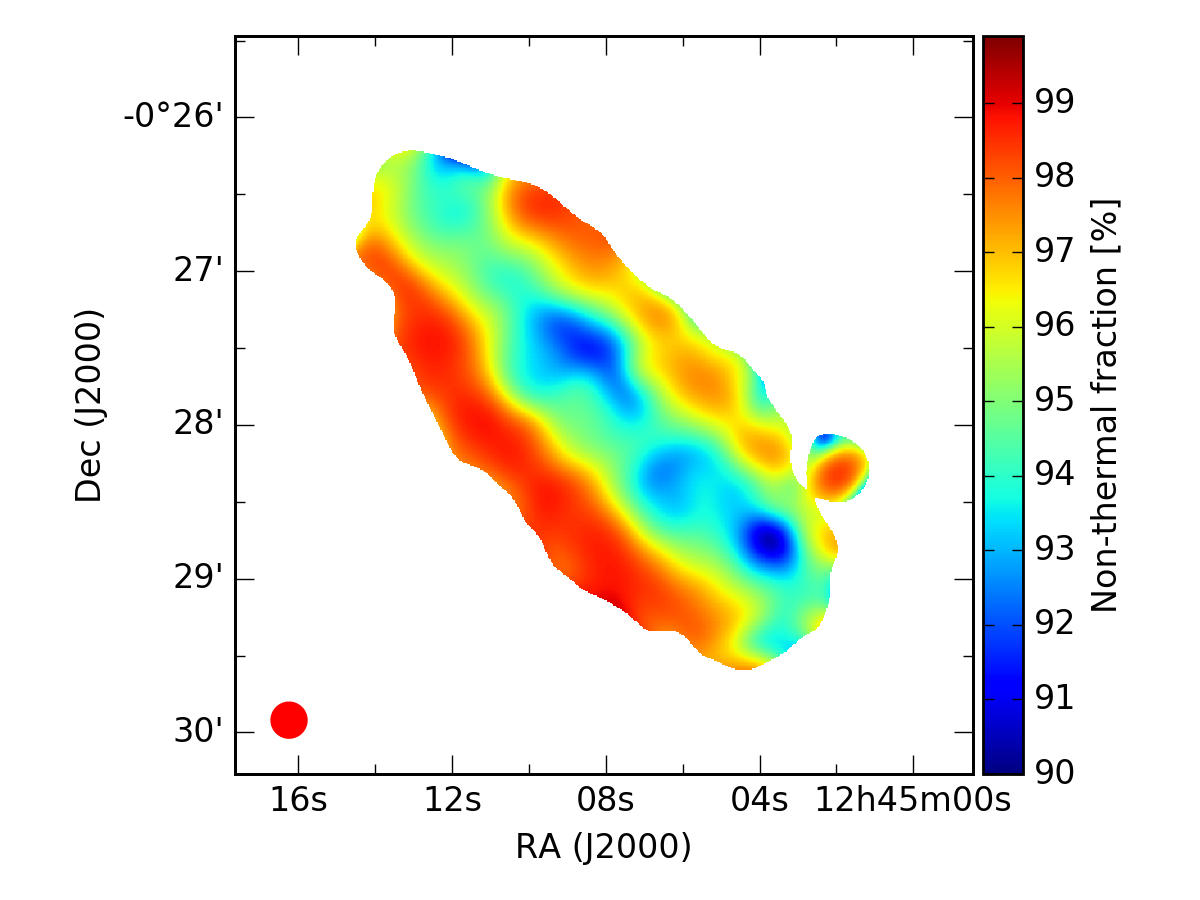}
        \caption{NGC~4666 nonthermal fraction map at L-band. The beam of 13.3"~$\times$~13.3" is shown in red (left bottom).}
        \label{fig:N4666_nonthermalfracL}
        \end{minipage}
        \end{figure*}

\subsubsection{RM maps}

\begin{figure}
        \includegraphics[width=0.49\textwidth]{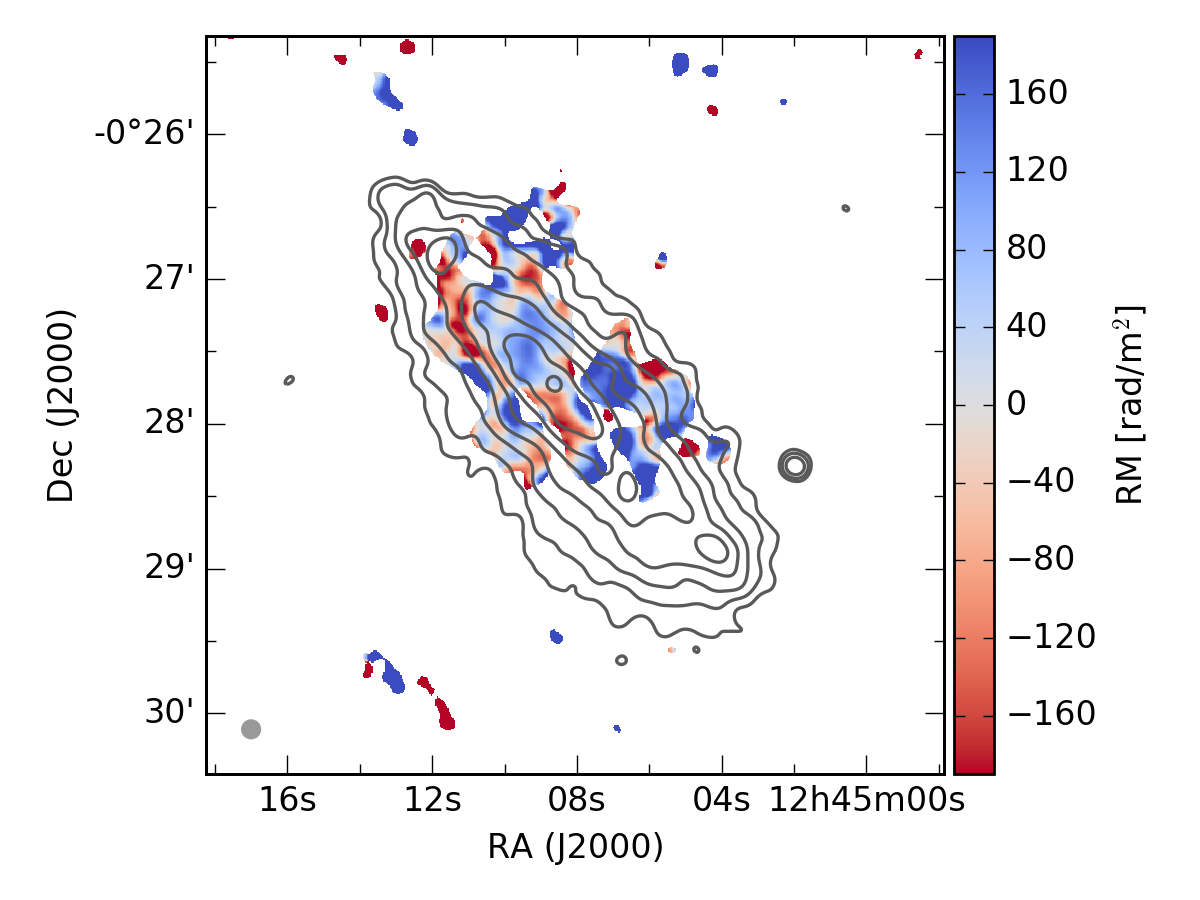}
        \caption{RM map of NGC~4666 from C-band on Stokes I contours from Figure~\ref{fig:N4666_Ccomb_synth_rob0} and a beam of 7"~$\times$~7" in gray. Contours start at a 3$\upsigma$ level with a $\upsigma$ of 9.3\,$\upmu$Jy/beam and increase in powers of 2 (up to 64) with an additional contour at 3.5\,mJy/beam to mark the central source. The RM map is cut below the 3$\upsigma$ level of 25$\upmu$Jy/beam of the polarized intensity map from RM synthesis.  The mean error is 22\,rad/m$^2$.}
        \label{fig:N4666_C_RM}
\end{figure}

A further outcome of the RM synthesis is the RM map, which represents the fitted peak position along the cube of each pixel. It represents the mean magnetic field component along the line of sight, where the value is positive for a field pointing towards the observer and negative for a field pointing away from the observer. The RM map of C-band is shown in Figure~\ref{fig:N4666_C_RM} and the one of L-band in Figure~\ref{fig:N4666-L-RM}. The RM values in C-band lie between $-200$ and $+200$\,rad/m$^2$. The L-band RM values are very different and between $-14$ and $+26$\,rad/m$^2$. In addition to the different RM values between the bands, the different beam sizes influence the scale of the local variations of these values. In the disk, the sign of RM values are comparable in both frequency bands.

\subsection{Thermal/nonthermal separation}
Following \citet{vargasetal2018}, we performed a thermal/nonthermal separation to get the nonthermal maps of C-band and L-band. 

We used the H$\upalpha$ emission derived from the H$\upalpha$ map, which was provided by \citet{dahlemetal1997}. The H$\upalpha$ flux density was corrected from 38\% [NII] contamination.

Then the [NII]-corrected H$\upalpha$ flux density ($L_{H\upalpha, \text{obs}}$) was absorption corrected via infrared WISE data at 22~$\upmu$m ($\upnu L_{\nu}(22\upmu$m)) with the calibration factor of 0.042 from \citet{vargasetal2018}:\\
\begin{equation}
L_{H\upalpha,\text{corr}}[\text{erg s}^{-1}] = L_{H\upalpha, \text{obs}}[\text{erg s}^{-1}] + 0.042 \cdot \upnu L_{\upnu}(22\upmu\text{m})[\text{erg s}^{-1}]
.\end{equation}
Following \citet{vargasetal2018}, the thermal emission was derived using the corrected H$\upalpha$ flux density (L$_{H\upalpha, \text{corr}}$) as:
\begin{align}
F_{\text{thermal}} & [\text{erg s}^{-1} \text{Hz}^{-1} \text{pix}^{-1}] = \\ \nonumber
                                                                        & 1.167 \times 10^{-14} \cdot \left(\frac{\text{T}_e}{10^4 K}\right)^{0.45} \cdot \left(\frac{\upnu}{GHz}\right)^{-0.1} \cdot L_{H\upalpha, \text{corr}} [\text{erg s}^{-1}].
\label{eq:murphy}
\end{align}

An electron temperature of T$_e$ = 10,000\,K is assumed, with $\upnu$ being the central frequency of 6\,GHz and 1.5\,GHz for C-band and L-band, respectively. The resulting thermal map was then subtracted from the radio map to derive the nonthermal map. To apply the thermal/nonthermal separation, all maps were smoothed to the same beam of 13.3"~$\times$~13.3". We refer to \citet{vargasetal2018} for a detailed analysis on the thermal/nonthermal separation in edge-on spiral galaxies.

\subsection{Nonthermal fractions}

The maps of nonthermal fraction are shown in Figure~\ref{fig:N4666_nonthermalfracC} for C-band and Figure~\ref{fig:N4666_nonthermalfracL} for L-band. The mean nonthermal fractions in the disk, the eastern halo, and the western halo are presented in Table~\ref{tab:nonthermalfrac}. In the disk these are 83.0\,\% in C-band and 93.8\,\% in L-band. The nonthermal fractions in the halo are higher below the disk in comparison to the nonthermal fraction above the disk.  In C-band, these are 88.0\,\% in the southeastern halo and 81.0\,\% in the northwestern halo;  in L-band these are 97.7\,\% in the southeastern halo and 96.7\,\% in the northwestern halo. The halo below the disk is closer to the observer. The interaction with NGC~4668 could effect the distributions of CRs as well as of magnetic fields. As seen before in the C-band radio image (Figure~\ref{fig:N4666_Ccomb_bw}) as well as in the scale height analysis, NGC~4666 seems to be different above and below the disk.

\begin{table}
\centering
\caption{Nonthermal fractions of NGC~4666}
\begin{tabular}{lccc}
\hline
Frequency &     Disk              & East &  West \\
\ [GHz]     &  [\%]   &   [\%] &   [\%] \\
\hline
1.5 (L-band) & 93.8  & 97.7  & 96.7\\
6 (C-band   & 83.0   & 88.0  & 81.0\\
\hline
\end{tabular}
\label{tab:nonthermalfrac}
\end{table}


\subsection{Nonthermal spectral index}

The nonthermal spectral index ($\upalpha_{\text{nt}}$) map was calculated from the nonthermal maps of the two bands at the observed frequencies of 1.5\,GHz and 6\,GHz of the CHANG-ES data, giving the spatial distribution of the spectral index in this galaxy.
The nonthermal SPI map was derived using the following equation:
\begin{equation}
\upalpha_{\text{nt}} = \frac{\text{log} I_{nth}(\upnu_1) - \text{log} I_{nth}(\upnu_2)}{\text{log} \upnu_1 - \text{log} \upnu_2} \, ,
\label{eq:spi}
\end{equation}  
where $\upnu_1$ and $\upnu_2$ are the central frequencies of C-band and L-band, respectively. From Equation~\ref{eq:spi}, the error $\Delta\upalpha_{\text{nt}}$ with respect to $\Delta I_1$ and $\Delta I_2$ can be determined using error propagation: 
\begin{align}
\Delta \upalpha_{\text{nt}} = \frac{1}{\text{log} \upnu_1 - \text{log} \upnu_2} \cdot \frac{1}{\text{ln} 10} \cdot \sqrt{\left(\frac{\Delta I_{nth}(\upnu_1)}{I_{nth}(\upnu_1)}\right)^2 + \left(\frac{\Delta I_{nth}(\upnu_2)}{I_{nth}(\upnu_2)}\right)^2} \, .
\label{eq:spierror}
\end{align}

\begin{figure}
        \centering
                \includegraphics[width=0.92\hsize]{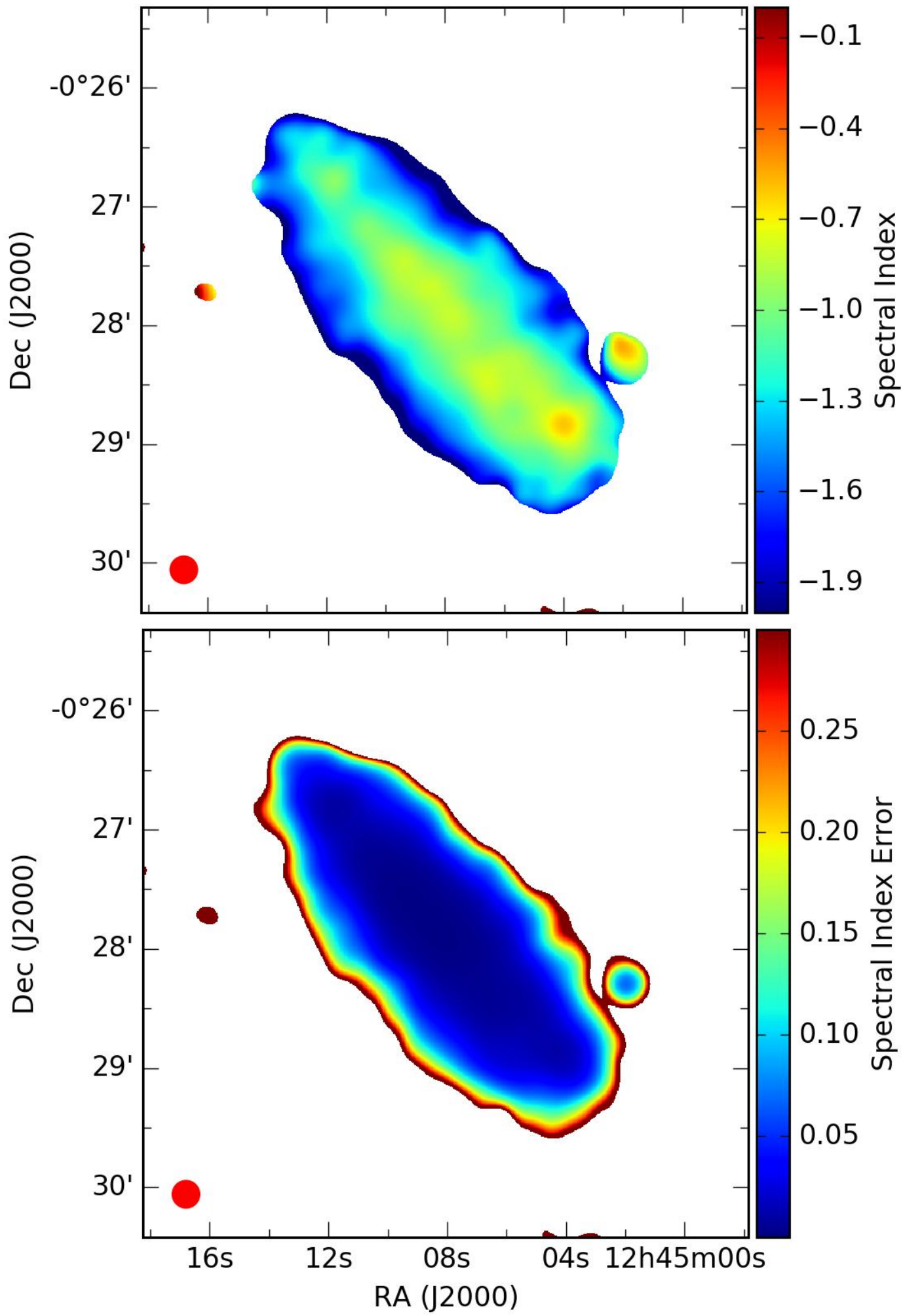}
        \caption{Top: NGC~4666 nonthermal spectral index ($\upalpha_{\text{nt}}$),\newline Bottom: Spectral index error map. The beam is 13.3"~$\times$~13.3".}
        \label{fig:N4666_SPIandError}
\end{figure}

Both resulting maps are displayed in Figure~\ref{fig:N4666_SPIandError}. The error map displays uncertainties of below 0.05 in the disk, which increase to 0.3 towards the edges. Based on the error map, the spectral index is cut off in order to only include trustable values. A mean nonthermal spectral index in the disk of $\upalpha_{\text{nt}}~=~-0.90~\pm~0.05$ is found, which is in good agreement with synchrotron radiation being the dominant radiation process. The spectral index steepens towards the edges and reaches values of $\upalpha_{\text{nt}}~=~-1.8~\pm~0.3$. The huge star formation region in the south of the galaxy is clearly visible with a flatter spectral index of $\upalpha_{\text{nt}}~=~-0.6~\pm~0.05$, which suggests synchrotron emission from young CRs.

\subsection{Magnetic field}
The magnetic field structure is obtained via RM synthesis to gain the intrinsic magnetic field vectors for the field component perpendicular to the line of sights as well as for the parallel line of sight component. The magnetic field strength is determined with the assumption of energy equipartition with the CR electrons.
 
\subsubsection{Magnetic field configuration of the disk field}
The analysis of disk magnetic fields of spiral galaxies makes it possible to reveal the mode and direction of large-scale fields. In face-on spiral galaxies the analysis is done via a sector integration using an RM map, where the RM values are plotted against the azimuth angle in degrees in the plane of the galaxy \citep{Krause1990}. Hereby, the mode of the disk field is quantified, that is, whether the field is axisymmetric (m=0), bisymmetric (m=1) or can be described by higher modes. However, this is not applicable to edge-on galaxies as the emission from the disk is observed in projection. 
Additionally, to distinguish between an even disk magnetic field parity (S, same direction above and below the mid plane) with a corresponding quadrupolar halo field and an odd disk magnetic field parity (A, opposite orientation above and below the mid plane) with a corresponding dipolar halo field, RM data at high resolution and sensitivity, and especially with a wider range in $\lambda^2$-space, have to be obtained. Here, we present a modified analysis method for RM maps of edge-on galaxies for the disk magnetic field and apply it to NGC 4666.

\begin{figure}
        \centering
                \includegraphics[width=0.5\textwidth]{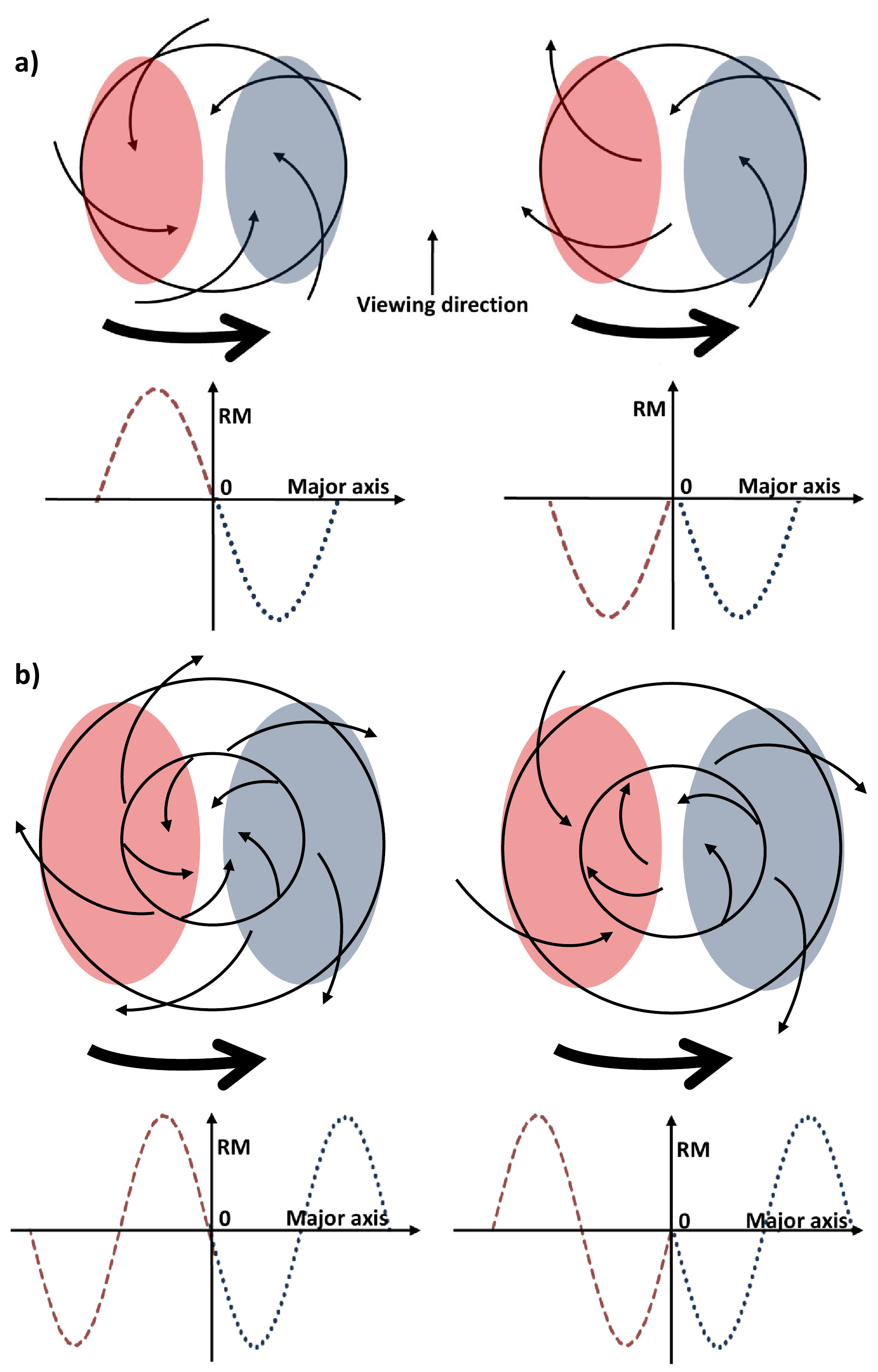}
        \caption{Expected RM variation for an edge-on view of a simplified axisymmetric disk magnetic field (left) and a simplified bisymmetric disk magnetic field (right) with trailing spiral arms, which are shaded red for the approaching side and blue for the receding side. The black arrows show the direction of the magnetic field vectors. The direction of the rotation of the disk is given by the bold black arrow below the four disks.
                \newline
    a) The upper part shows the face-on view of the two different disk fields. In the case of an axisymmetric disk field (left) all vectors are pointing inwards or outwards; here just one configuration is shown. In case of the bisymmetric magnetic field (right) the vectors from one side of the galaxy are pointing inwards and on the other side of the galaxy outwards (or the other way around). In the second row, the expected RM behavior is plotted along the major axis of each galaxy model for an observer watching the system edge-on. The red curve represents the RM values from the approaching side, the blue curve from the receding side. \newline  b) The upper part shows again the face-on view of the two different disk fields as above (left: axisymmetric disk field, right: bisymmetric magnetic field). In a simplified approach, a radial reversal is assumed at half of the disk radius. The last row shows the expected and simplified RM value variation in case of a radial field reversal at half the radius of the disk. }
        \label{fig:RM_expected}
\end{figure}

In this new approach a rectangle along the disk is used to obtain a profile of RM values along the major axis. Figure~\ref{fig:RM_expected} shows the expected RM behavior in projection for this analysis of an edge-on galaxy, where in the first two rows, the edge-on view of an axisymmetric disk field (left) leads to positive RM values on one side of the disk and to negative RM values on the other side. The edge-on view of a bisymmetric disk field (right) leads to negative RM values on both sides for the plotted field configuration. Both RM behaviors would be mirrored across the x-axis (major axis) if the magnetic field vectors were to point in the opposite direction. In the third and fourth rows of Fig.~\ref{fig:RM_expected}, the same is shown for a disk magnetic field with one radial reversal at half of the disk radius. 

Here we assume an even disk magnetic field parity with a quadrupolar halo field, which seems to be the most readily excited dynamo mode in dynamo models \citep[e.g.,][]{ruzmaikinetal1988}, especially considering a wind \citep[e.g.,][]{mossetal2010}. Additionally, the even disk parity is found in the Milky Way \citep{sunetal2008} and in the external galaxies NGC~253 \citep{heesenetal2009a}, NGC~891 \citep{krause2009}, and NGC~5775 \citep{soidaetal2011}.

\begin{figure*}
         \begin{minipage}[c]{0.44\textwidth}
                        \includegraphics[width=\hsize]{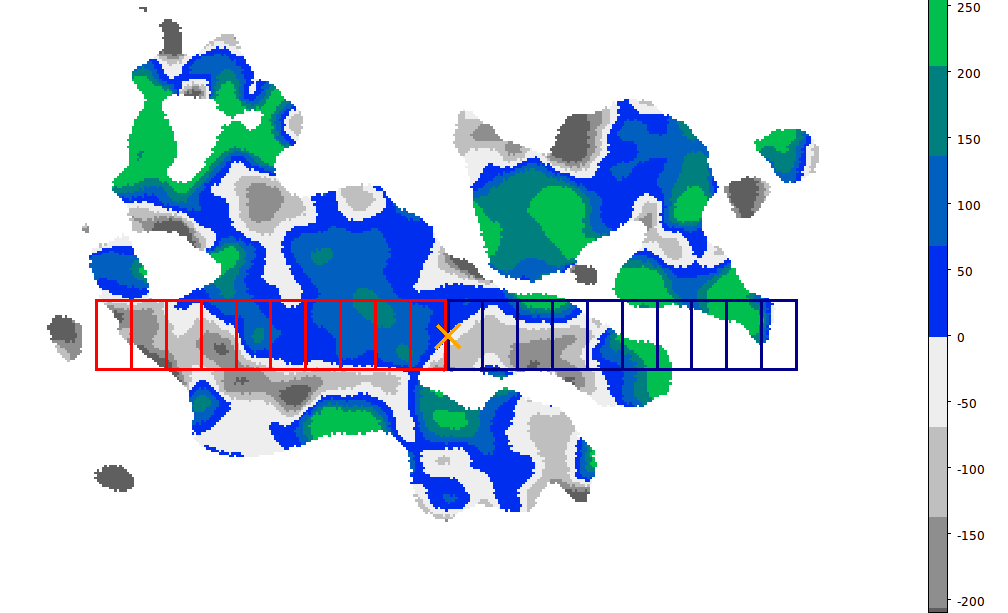}
        \end{minipage}
\hspace{0.01\textwidth}
        \begin{minipage}[c]{0.54\textwidth}
                \includegraphics[width=\hsize]{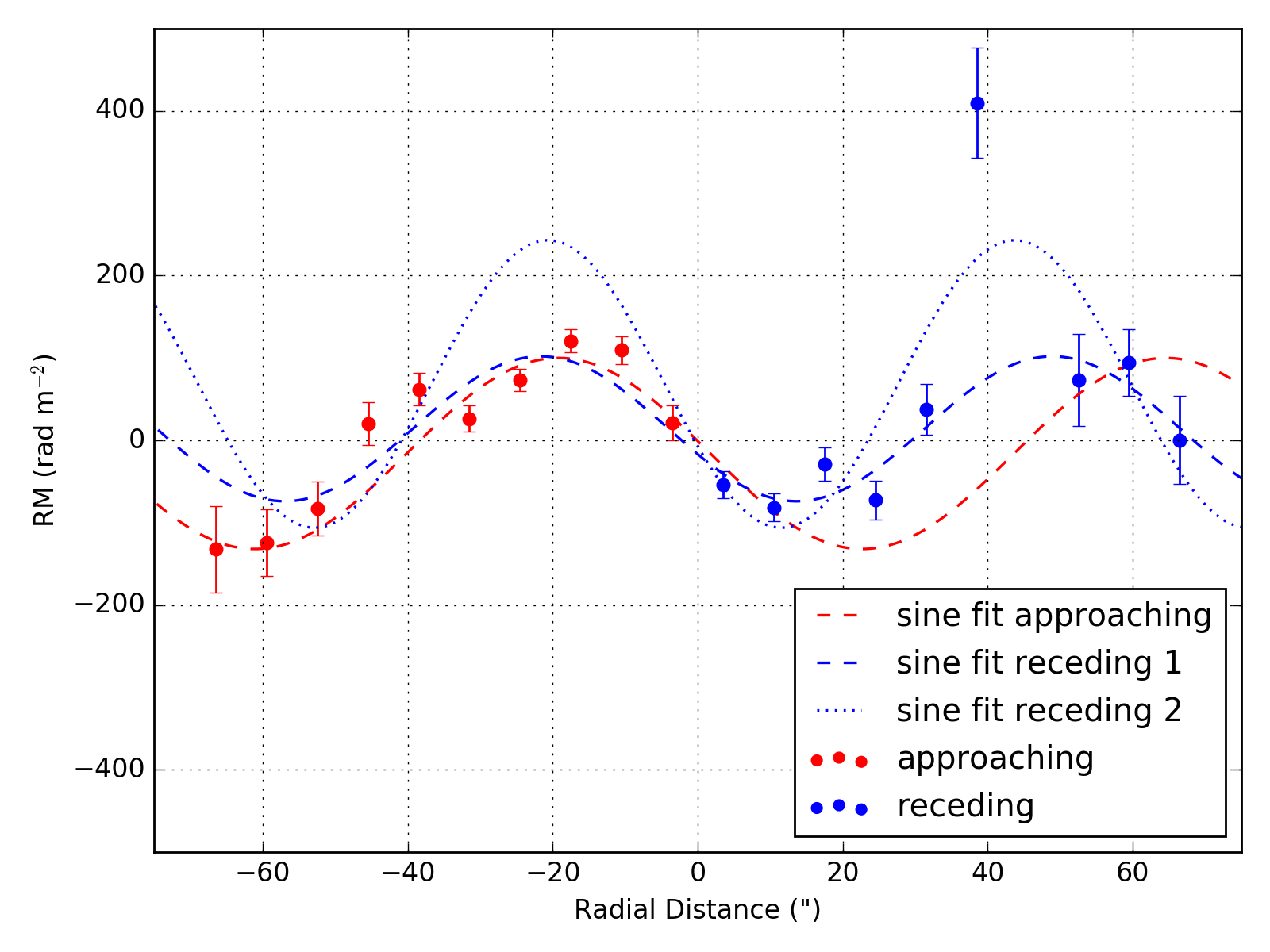}
        \end{minipage}
        \caption{Left: NGC 4666 C-band RM map, rotated by $50^\circ$ on sky, with a beam of 7.2" $\times$ 7.2". The rectangle boxes of 7"~$\times$~14" were used for averaging. The orange cross marks the center. Right: RM values from boxes along the major axis for both sides of the galaxy. The sinusoidal fit was made for the different sides separately indicated by the color. The sinusoidal fit for the receding side excludes the data point at 38.5" in Fit 1 and includes this data point in Fit 2.}
        \label{fig:N4666_Box_RM}
    \label{fig:N4666_sinefit}
\end{figure*}

The analysis described above was carried out on the C-band RM map. Figure~\ref{fig:N4666_Box_RM} shows the rectangle along the major axis with a box size of 7" $\times$ 14" (0.93\,kpc $\times$ 1.87\,kpc) on the RM map to the left. For this analysis RM synthesis was applied in a different way. The mean Q and U values within each box were determined from the Q and U maps of each spw. For each of these boxes there is one mean value from the Q map and one mean value from the U map. Then a cube was made and RM synthesis applied. The distance to the center against the resulting RM value of each box is plotted separately for the approaching side (left half of the galaxy, red) and the receding side (right half of the galaxy, blue). The RM error was calculated with RMSF/(2 PI/$\upsigma$) \citep{schnitzelerlee2017}, where RMSF is the resolution in Faraday space, PI is the polarized flux density, and $\upsigma$ the noise measured when integrating over the entire frequency band. This is a good error approximation for PI/$\upsigma$ > 5 \citep{schnitzelerlee2017}. With the RMSF in C-band of $\sim$ 1800\,rad\,m$^{-2}$ from Table~\ref{tab:RMpar} and a 20$\upsigma$ signal, a mean RM error of 50 rad\ m$^{-2}$ is derived.

\begin{figure*}
 \begin{minipage}[c]{0.45\textwidth}
        \includegraphics[width=1.15\hsize]{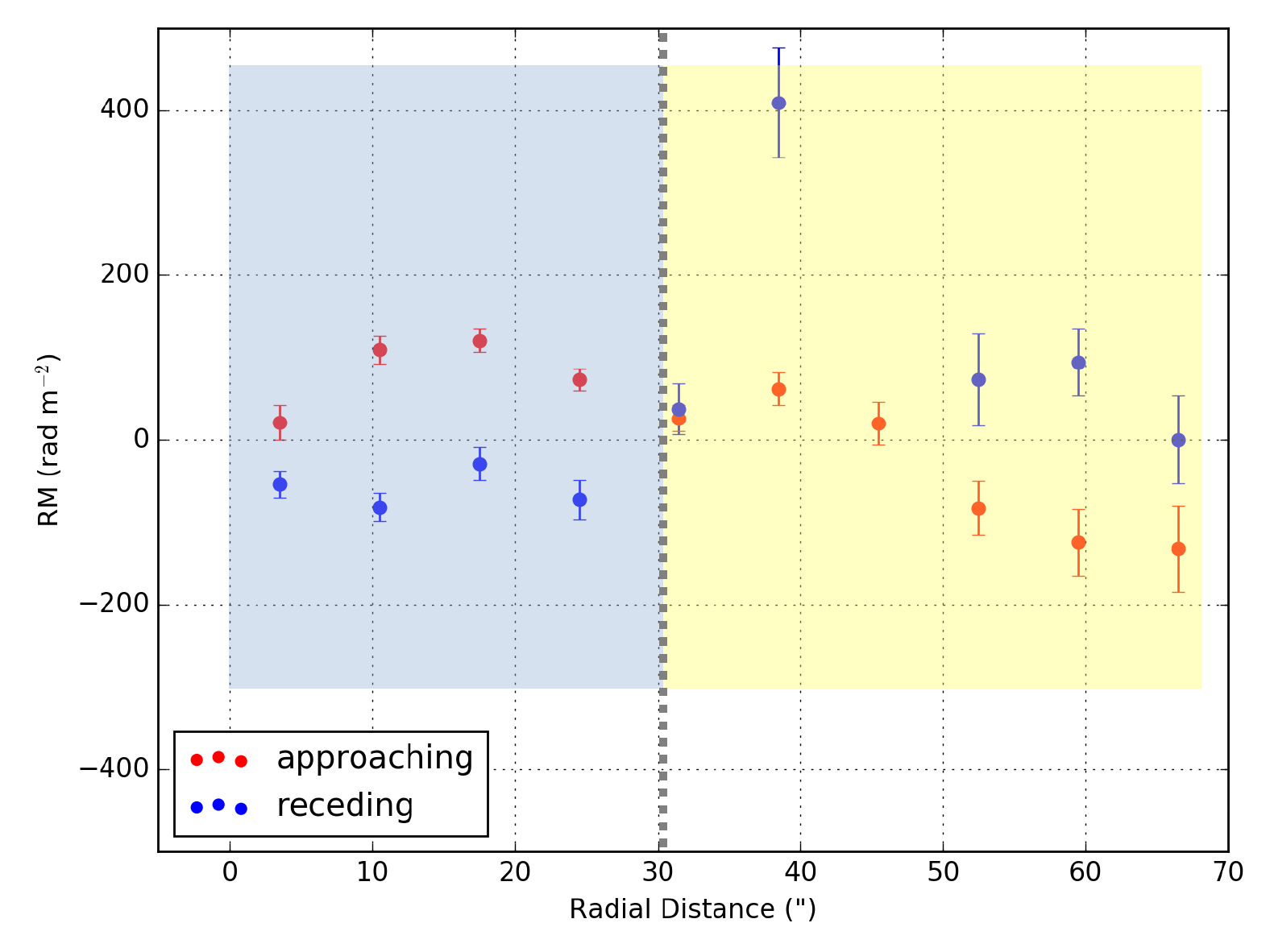}
 \end{minipage}
 \begin{minipage}[c]{0.15\textwidth}
 \hspace{0.07\textwidth}
 \end{minipage}
  \begin{minipage}[c]{0.45\textwidth}
     \includegraphics[width=0.5\textwidth]{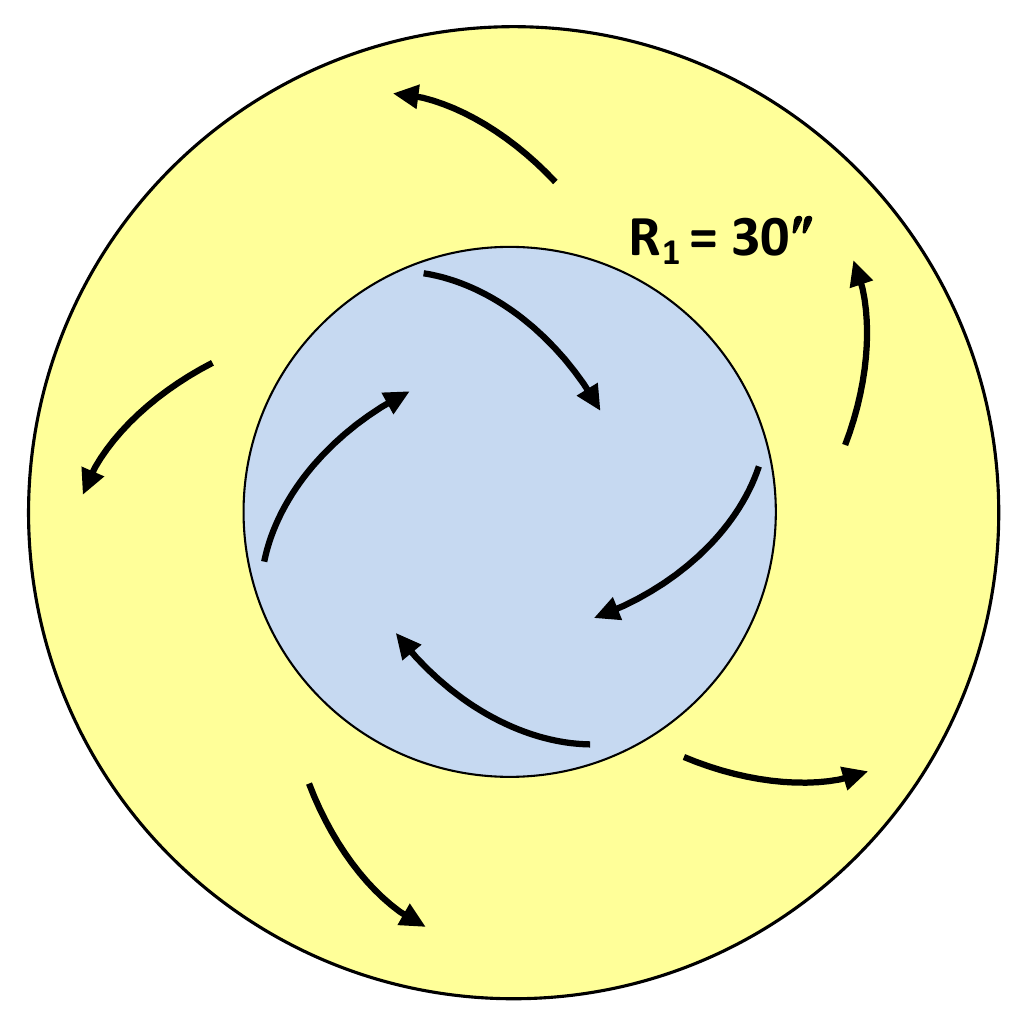}
  \end{minipage}  
        \caption{Left: NGC~4666 C-band RM values for boxes from the center outwards to the edge of the disk for both sides of the galaxy; as in the right plot of Figure~\ref{fig:N4666_Box_RM} but with the approaching side mirrored. The left blue-shaded part of the plot indicates the axisymmetric disk magnetic field with magnetic field vectors pointing inwards, the yellow-shaded part of the plot indicates the axisymmetric disk magnetic field with magnetic field vectors pointing outwards, and the grey dashed line marks the radius of a possible field reversal. Right: NGC~4666 simplified disk model.}
        \label{fig:N4666_diskmodell}
        \label{fig:N4666_RM_outwards}
\end{figure*}

The result is displayed on the right side of Figure~\ref{fig:N4666_Box_RM} and on the left side of Figure~\ref{fig:N4666_RM_outwards}. The RM pattern on both sides is approximately axisymmetric with respect to the x-axis, where a maximum of the approaching curve corresponds to a minimum of the receding curve and vice versa. The resulting curve of Figure~\ref{fig:N4666_RM_outwards} can be divided into two parts. The first part is up to $\pm$~30" (blue shaded) and the second part is from $\pm$~30" to the end (yellow shaded). In the first part we see that the approaching side shows mainly positive RM values and the receding side shows mainly negative RM values. This behavior is comparable to the expected RM behavior in the fourth row on the left-hand side of Figure~\ref{fig:RM_expected} for the axisymmetric case (m=0) with magnetic field vectors pointing inwards in the inner part and pointing outwards in the outer part. At the radius of about 30" (4\,kpc) of Figure~\ref{fig:N4666_RM_outwards}, the behavior of the RM values of the two halves of the galaxy of the blue shaded part of the plot change to the opposite. From a radial distance of 30" further out (in the yellow shaded part of the plot), the approaching side shows mainly negative values whereas the receding side shows mainly positive RM values. This indicates an axisymmetric field pointing outwards.

We further tried to fit a sinusoid ($f(x) = (a_0\ sin(a_1\ x+a_2)) + a_3$) to the data of both sides separately, which is shown with the dashed and dotted lines in Figure~\ref{fig:N4666_sinefit}. The receding side is fitted twice; Fit 1 excludes the data point at the radius of 38.5", while Fit 2 includes all data points. In Table~\ref{tab:sineparameter} we present the fit parameters. The period P of the sinus along r is given by P~=~2$\uppi / a_1$ which results in P~=~84"~$\pm$~7" for the approaching side and P~=~71"~$\pm$~7" for the receding side (fit 1). These two values are slightly different.

The parameter $a_2$ is the phase (in rad). The a$_2$ values of the two sides are the same within the errors and show that the sinusoidal functions have similar phases. A phase of $\uppi$ would correspond to an ASS field with vanishing pitch angle. The average a$_2$ value of $\simeq3.45$ indicates a pitch angle of about $20\degr,$ which is reasonable for spiral galaxies. Hence, the RM variations due to the disk magnetic field of both sides are consistent with each other and originate from the large-scale disk magnetic field.

\begin{table}
\centering
\caption{Fitting parameters for the sinusoidal fit $f(x) = (a_0\ sin(a_1\ x+a_2))+a_3$ for Figure~\ref{fig:N4666_sinefit}. Receding side 1 is the fit to the receding side without the data point at the radius of 38.5",  Receding side 2 is the fit to the receding side including all data points.}
\begin{tabular}{lccc}
\hline
         & Approaching side  &  Receding side 1 &  Receding side 2 \\
\hline
a$_0$ & 116 $\pm$ 12     &  88 $\pm$ 19       & 175 $\pm$ 56   \\
a$_1$ & 0.075 $\pm$ 0.006&  0.089 $\pm$ 0.006 & 0.097 $\pm$ 0.010\\
a$_2$ & 3.3 $\pm$ 0.2  &  3.5 $\pm$ 0.2     & 3.6 $\pm$ 0.4     \\
a$_3$ & -16 $\pm$ 10     &  14 $\pm$ 12       & 68 $\pm$ 36   \\
\hline
\end{tabular}
\label{tab:sineparameter}
\end{table}

A simplified model is shown in Figure~\ref{fig:N4666_diskmodell}, where one radial field reversal is assumed. The color code is equivalent to the modeled disk field of \citet{moss2012}. The inner magnetic field points inwards and then outwards. The orientation of the magnetic field vector can be determined using the simplified model of Figure~\ref{fig:RM_expected}. In the case of trailing spiral arms, which is suggested for NGC~4666, and positive RM values on the approaching side, the magnetic field vectors point inwards. Positive RM values on the receding side correspond to magnetic field vectors that point outwards \citep[see][]{krausebeck1998}. 

In conclusion, an axisymmetric magnetic field is discovered in NGC~4666. One field reversal along the radius of the orientation of the magnetic field is indicated by the C-band data at a radial distance from the center of the galaxy of about 4\,kpc. This radius coincides with the optical end of the bar of NGC~4666. The L-band data (see Fig.~\ref{fig:N4666-L-RM}) may suggest a second reversal at about 90" (12\,kpc) radius, but these data suffer from depolarization effects and may not be reliable. The presented disk model is very simple and should be improved with better data, maybe on other galaxies. The quality of the present data is not sufficient to analyze the halo parity, higher magnetic field modes, and pitch angles. A significant deviation in phase shift between the two sides of the galaxy would indicate that the assumption of a constant pitch angle between the two sides of the galaxy and/or between the inner and outer parts may not be fulfilled.

\subsubsection{Magnetic field strength via equipartition}
\label{subsec:magnetic_field_strength_via_equipartition}
The maps of the magnetic field strength were determined with the revised equipartition formula by \citet{beckkrause2005}.

\begin{figure}
         \includegraphics[width=\hsize]{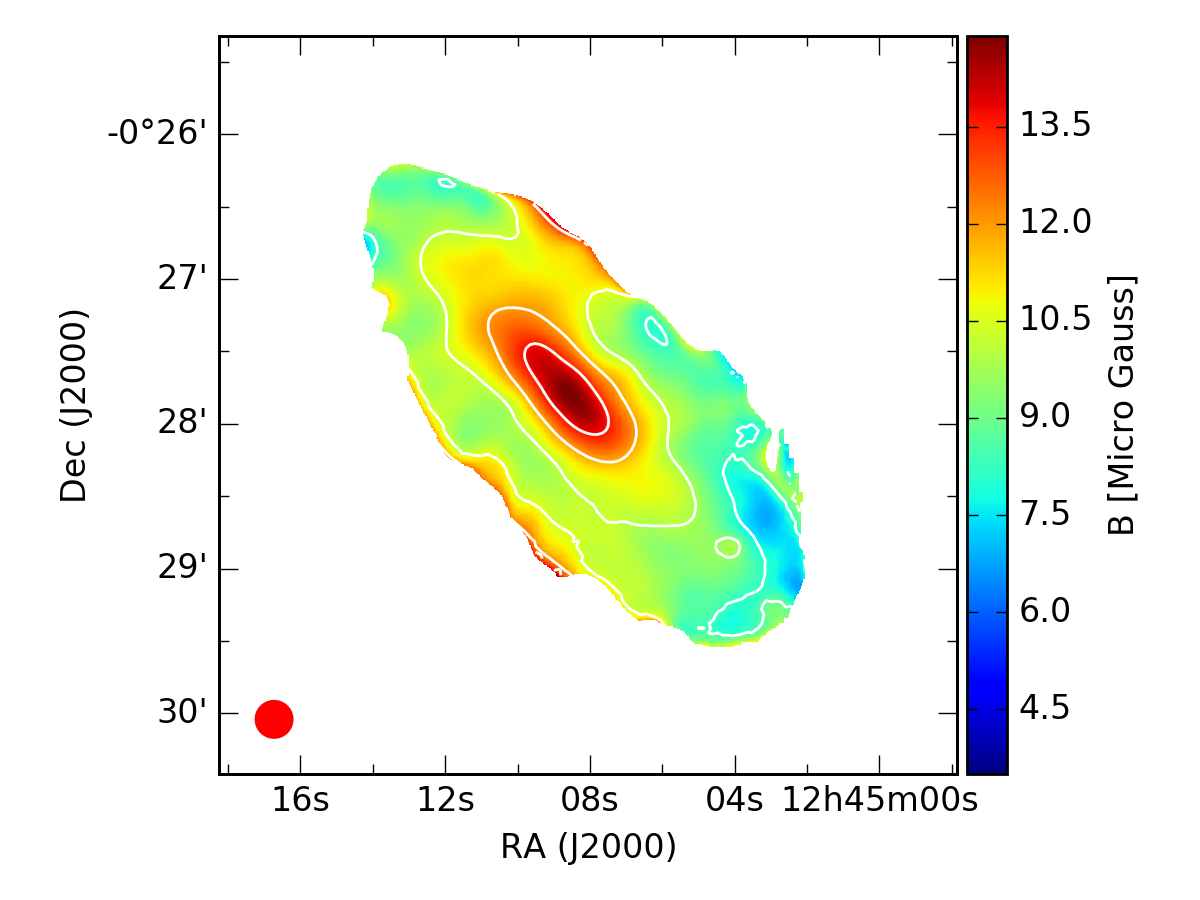}
                \caption[NGC~4666 magnetic field strength pixel based]{NGC~4666 magnetic field strength calculated pixel-based via the equipartition formula using the resolved nonthermal spectral index map and the intensity map with a beam of 13.3"~$\times$~13.3" (bottom left).}
        \label{fig:N4666_bfeld_genau}
\end{figure}

The calculations were done pixel-by-pixel by taking the nonthermal intensity map and the resolved nonthermal spectral index map into account. The pixel values of the magnetic field strength B$_{eq}$ were calculated according to the revised equipartition formula by \citet{beckkrause2005}. We assumed a spheroid path length model varying between l~=~19~kpc in the center and 1= 6~kpc at the edges and K$_0$~=~100, which is consistent with CR data from the local Milky Way. 

The resulting map is presented in Figure~\ref{fig:N4666_bfeld_genau}. The total magnetic field strength in the center reaches 15\,$\mu$G. The field strength decreases towards the halo to 8.5\,$\mu$G. The mean disk magnetic field strength is 12.3\,$\mu$G, which is in agreement with the values found for 13 CHANG-ES galaxies by \citet{krauseetal2018}. The mean halo magnetic field strength is 10.1\,$\mu$G in the eastern halo and 9.6\,$\mu$G in the western halo. The smallest magnetic field strength of 7\,$\mu$G is seen in the south of the galaxy. It appears that the magnetic field strength is smoothly distributed over the whole galaxy.

\citet{dahlemetal1997} found a total field strength of 14.4\,$\mu$G in the disk and 7.1\,$\mu$G in the halo, using C-band data and a fixed spectral index of $\upalpha \cong -$0.79 in the disk and $\upalpha \cong -$0.9 in the halo. The mean value of the disk and the halo total field strength derived here are somewhat higher in comparison to those in \citet{dahlemetal1997}. The reason is the difference of the application of the equipartition formula. The use of a fixed spectral index leads to a faster decline of the magnetic field strength towards the halo.

\subsection{One-dimensional cosmic ray transport model}
\label{subsec:spinteractive}

The transport processes of CRs into the halo (perpendicular to the galaxy disk) are derived by using a 1D CR transport model with {\small SPINNAKER} \citep[SPectral INdex Numerical Analysis of K(c)osmic-ray Electron Radio-emission;][]{heesenetal2016, heesenetal2018} with a fixed inner boundary condition of $N(E,0) = N_0 E^{-\gamma_0}$ and a corresponding nonthermal spectral index $\upalpha_\mathrm{nt} = (1-\gamma_0)/2$. 
Either pure advection is modeled,  with an advection speed $V$ assumed to be
constant, or pure diffusion with a diffusion coefficient of $D = D_0$ E$^{\mu}_{GeV}$, assumed to be along the magnetic field lines. Here, the cosmic ray electron (CRe) energy is $E_{\text{GeV}}$ in units of GeV and $\mu$ is between $0.3$ and $0.6$, as measured in the Milky Way \citep{strongetal2007}. The combined synchrotron and inverse compton radiation energy loss rate for a single CRe is taken into account with one loss term.

This model is used on the total intensity profile of the pure synchrotron maps, which represent the synchrotron intensity emitted by the CRe as well as on the corresponding spectral index map of NGC~4666.

An interactive {\small PYTHON} wrapper around {\small SPINNAKER} \linebreak ({\small SPINTERACTIVE}; Miskolczi 2018, private communication) was used to fit the data.\footnote{SPINNAKER and SPINTERACTIVE are available at: www.github.com/vheesen/Spinnaker} We varied the input parameters manually, while checking the quality of the fit on a graphical user interface, and thus arrived at the best-fitting solution. The main input parameter is the transport process, choosing between diffusion and advection. Depending on this choice, either the diffusion coefficient or the advection speed can be changed. Further parameters to be fitted are the CRe injection spectral index, $\gamma_0$, the magnetic field strengths in total and of the thin disk component ($B_0$, $B_1$), and magnetic field scale heights $hb1$ and $hb2$, for the thin and thick disk, respectively. {\small SPINNAKER} models the magnetic field distribution as the superposition of two exponential functions. Generally speaking, synchrotron intensity profiles can be best described by a Gaussian function for diffusion and by an exponential function for advection.

The intensity distribution of NGC~4666 is produced via {\small BoxModels} \citep{mulleretal2017}
using one stripe covering the width of the major axis of the galaxy (box size of 200"~$\times$~6"). The intensity value of two boxes at the same distance to the midplane ($z$) of the two halo sides are averaged to get the plotted intensity distribution for NGC~4666. 

\begin{figure*}
\begin{minipage}[c]{0.63\textwidth}
 \includegraphics[width=1.0\textwidth]{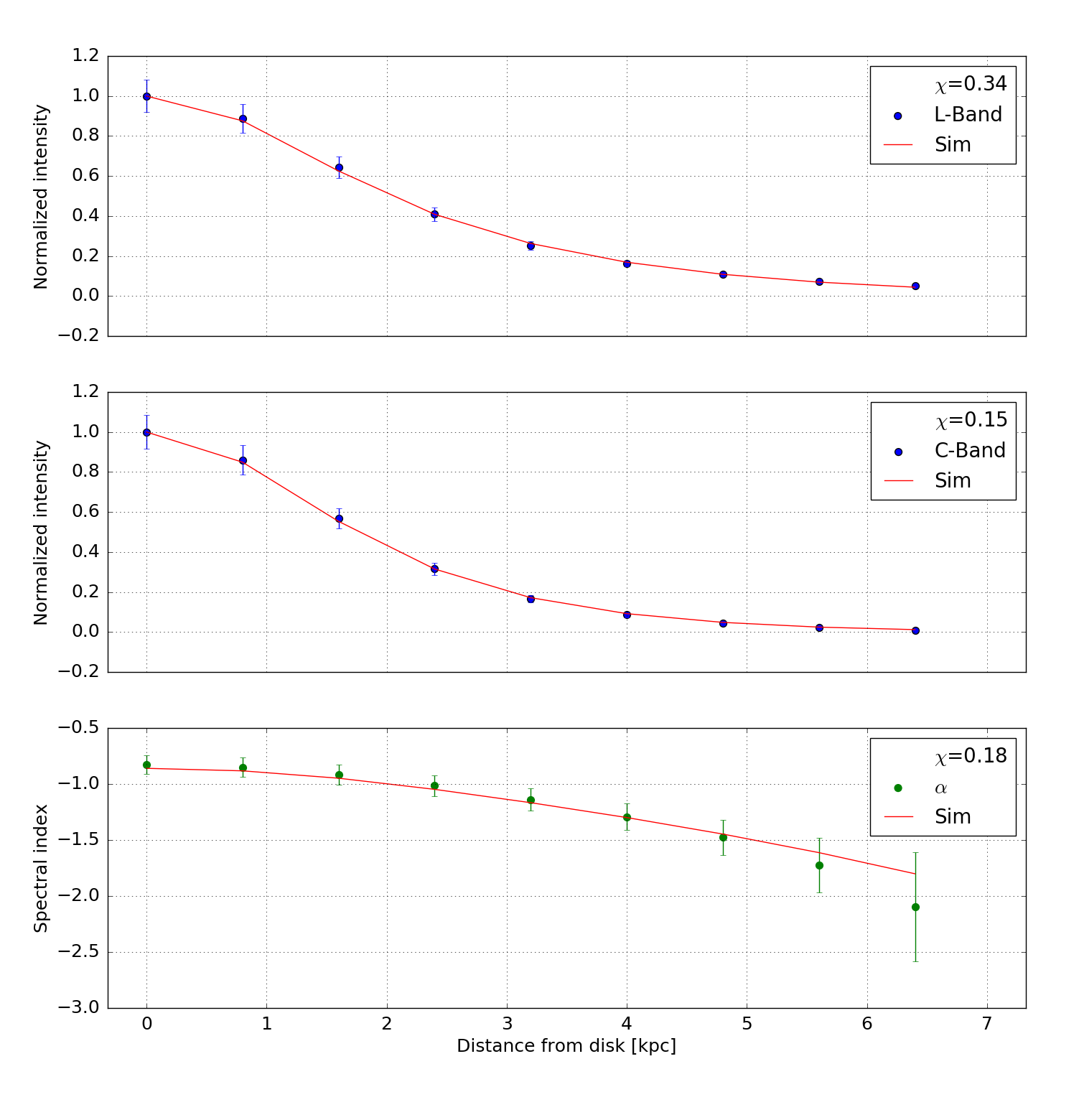}
 \captionof{figure}{CR advection transport model for NGC~4666 with {\scriptsize SPINTERACTIVE}. The data points are the observed values from 1.5\,GHz (L-band), 6\,GHz (C-band), and the spectral index; the red line represents the model simulation. The errors correspond to a weighted standard deviation. The basic parameters are listed in Table~\ref{tab:spinnakerpar_adv}.}
 \label{fig:N4666-spinteractive}
\end{minipage}
\begin{minipage}[c]{0.07\textwidth}
\hfill
\end{minipage}
\begin{minipage}[c]{0.3\textwidth}
\begin{threeparttable}
\captionof{table}[{\small SPINNAKER} parameters.]{{\small SPINNAKER} parameter for advection.}
\vspace{-2.2cm}
\begin{tabular}[c]{lc}
\hline \hline
Parameter       &  Value \\ \hline
$\gamma_0$    & 2.6  \\
$B_0$ [$\upmu$G]   & 13.1 \\
$B_1$  [$\upmu$G]  &3.2  \\
$hb1$ ["] &  2.5 \\
$hb2$ ["]   & 5.0\\
\hline
Advection: & \, \\
$V$ [km \, s$^{-1}$]& 310 \\
Galaxy mode\tnote{1} &1  \\
Adiabatic losses& No\\
Velocity field\tnote{2}&  0 \\
$\upchi^2_{\text{adv}}$ & 0.22  \\
\hline
\end{tabular}
\begin{tablenotes}
\footnotesize
\item \textbf{Notes:}
\item[1] galaxy mode 1: The magnetic field is defined from Eq.~\ref{magfeld_spinnaker}.
\item[2] velocity field 0: The advection speed is constant, which gives the best fit to the data.
\end{tablenotes}
\normalsize
\label{tab:spinnakerpar_adv}
\end{threeparttable}
\end{minipage}
\end{figure*}

We found a good solution for advection ($\chi_{adv}^2$=0.22), the result of which is shown in Figure~\ref{fig:N4666-spinteractive}. The top panel shows the C-band data, the middle the L-band data and the bottom panel shows the spectral index calculated from the corresponding data points of the C-band and L-band data. All the important adjustable parameters are listed  in Table~\ref{tab:spinnakerpar_adv}. Figure~\ref{fig:N4666-spinteractive} also contains the information about the final $\chi^2$ values for the L-band data, C-band data, and spectral index. \\

The advection speed of 310\,km~s$^{-1}$ can be compared with the results from the sample of galaxies modeled with {\small SPINNAKER} by \citet{heesenetal2018}, which included NGC~4666. They found higher advection speeds between 500 and 700\,km~s$^{-1}$, but this can be explained by their higher equipartition magnetic field strength of $B_0=18.2~\mu\rm G$ as a model input; that strength was derived using a simple thin disk model with a short path length of 1~kpc, whereas we now employ a more sophisticated model (Section~\ref{subsec:magnetic_field_strength_via_equipartition}). If we scale their advection speeds to the magnetic field strength used here of $B_0=13.1~\mu\rm G$, assuming the CRe loss time-scale is $\propto B_0^{-3/2}$ \citep[Equation~(5) in][]{heesenetal2016}, their advection speeds are now between 260 and 550\,km~s$^{-1}$, including the uncertainties, in fair agreement with our results. Our new data, however, allow us to measure the advection speed much more precisely.

The magnetic field strength of 13.1\,$\upmu$G for the thin component of the magnetic field is in very good agreement with the mean disk value from equipartition of 12.3\,$\upmu$G. The nonthermal spectral index found with {\small SPINTERACTIVE} follows from $\gamma_0$ = 2.6. This results in $\upalpha_\mathrm{nt}$~=~$-$0.8. This value is in good agreement with the mean value of the spectral index map (Fig.~\ref{fig:N4666_SPIandError}).

A diffusion model was tested for comparison; results of which are shown in Appendix~\ref{app:additional_figures}. Figure~\ref{fig:N4666-spinteractive-dif} shows that the best-fitting diffusion model performs worse than the advection model. The total $\chi^2_{dif}$~=~5.6 is much higher than for advection, which mainly originates from the spectral index profile fit. Advection leads to a linearly steepening spectral index profile, whereas diffusion leads to a parabolic one. Our new results confirm that advection is the main transport process of CRs in NGC~4666. This distinction is based purely on the goodness of the fit, whereas for the old VLA data used by \citet{heesenetal2018} this was not possible.
In summary, advection appears to be the main transport process of CRs in NGC~4666.

\subsection{Magnetic field strength comparison}

\begin{figure}
        \centering
                \includegraphics[width=0.45\textwidth]{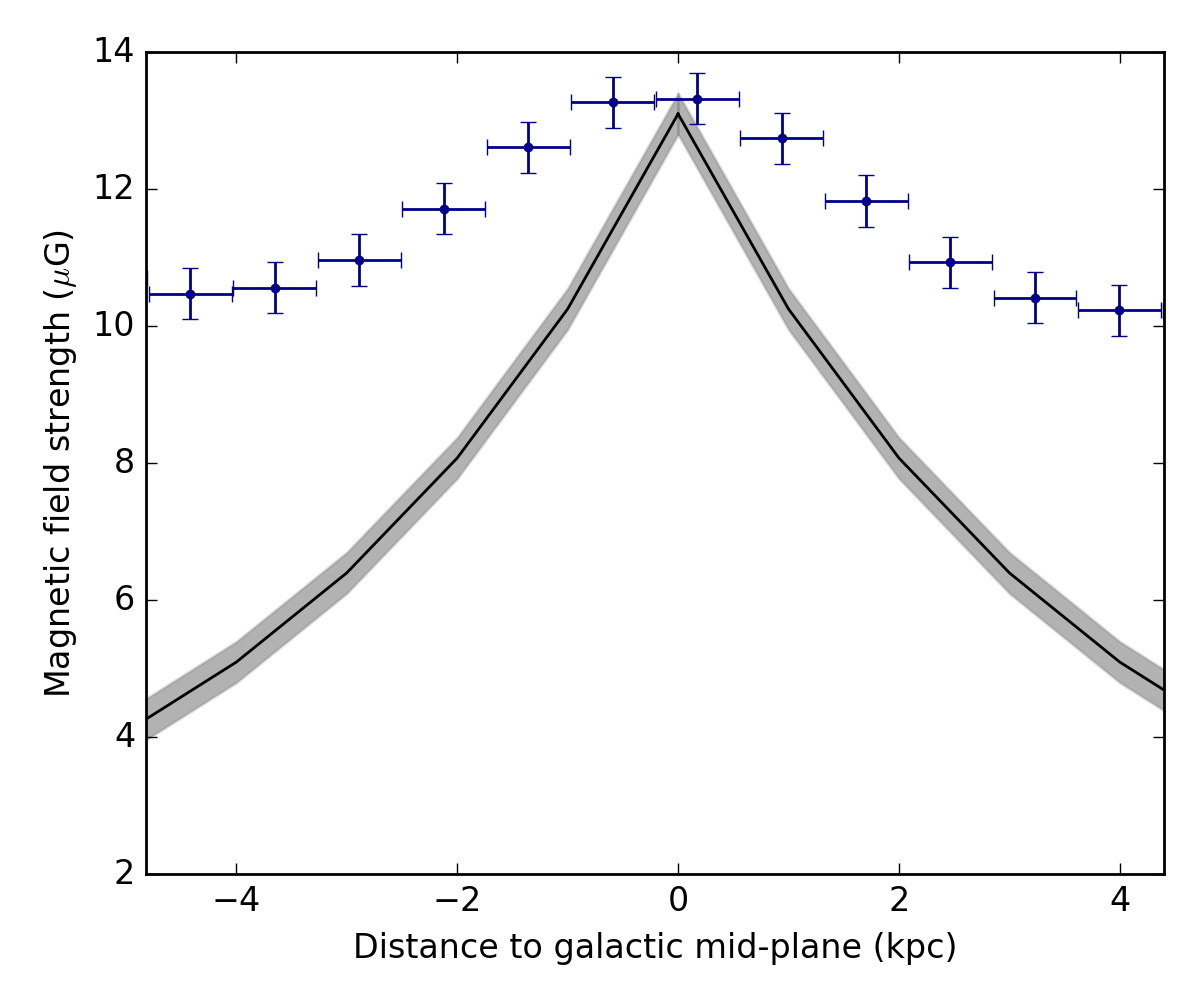}
        \caption{Intensity profile of equipartition magnetic field strength with dark blue data points from Fig.~\ref{fig:N4666_bfeld_genau}, magnetic field strength distribution from {\scriptsize SPINNAKER} with the black line, and gray shaded errors.}
        \label{fig:N4666-spintbfeld}
\end{figure}

In Figure~\ref{fig:N4666-spintbfeld}, the magnetic field strength distribution of NGC~4666 determined by using the equipartition assumption are shown and compared to the magnetic field profile from {\small SPINNAKER}. The magnetic field strength profile in the vertical direction determined with the equipartition formula using the resolved spectral index map from Figure~\ref{fig:N4666_bfeld_genau} is presented by dark blue data points with a box size of 6"~$\times$~180". The black line with shaded gray errors represents the magnetic field model from {\small SPINNAKER}. The magnetic field is modeled as a two-component exponential function:
\begin{align}
B(z) = B_1 \ \text{exp}(-|z|/hb1) + (B_0 - B_1) \, \text{exp}(-|z|/hb2) \, ,
\label{magfeld_spinnaker}
\end{align}
with the magnetic field scale heights for the thin disk $hb1$ and the thick disk $hb2$, the magnetic field in the central midplane $B_0$ and the magnetic field in the halo $B_1$.

Both distributions show similar values of the magnetic field in the midplane of the galaxy, but large discrepancies in the halo. In the halo, the {\small SPINNAKER} magnetic field strength is about 2.5 times less than the equipartition value in the halo at 5\,kpc from the center.
To achieve the observed synchrotron flux density, the number density of CRe (which is proportional to the CR energy density) needs to be larger by a factor of $2.5^{1-\upalpha_\mathrm{nt}} \approx 5$ compared to the equipartition value. Together, the ratio of energy densities of cosmic rays and magnetic fields is $2.5^{3-\upalpha_\mathrm{nt}} \approx 35$, which is a huge deviation from equipartition.

This deviation is a consequence of the cosmic ray advection equation used in {\small SPINNAKER}. For a constant advection velocity, the cosmic ray energy density is constant, neglecting radiation and adiabatic losses of the electrons. Hence, the decrease of the nonthermal radio continuum intensity is fully ascribed to a decrease of the magnetic field strength. The {\small SPINNAKER} magnetic field strength is therefore a lower bound for the actual field strength in the halo.

If we were allowing for an accelerating wind, the magnetic field strength would be closer to the equipartition value. However, our advection model with a constant speed already gives a relatively good fit to the data, and  therefore seems to be sufficient. In the near future, we hope to probe radio haloes to even larger distances from the star-forming disk using low-frequency radio continuum observations that reveal the oldest CRe. For instance, the 144\,MHz LOFAR study of \citet{miskolczietal2018} of the edge-on galaxy NGC~3556 detected the radio halo to heights of 15\,kpc. These authors could show that an accelerated wind model fits better than one with a constant wind speed.

\section{Summary and Conclusions}

The superwind and starburst galaxy NGC~4666 exhibits a boxy radio halo in radio total intensity. Its extent is up to 9\,kpc from the midplane into the halo in L-band, which is probably produced by a supernova-driven wind across almost the entire disk. The supplemental X-ray data show large-scale hot gas surrounding the galaxy, which is also box-shaped and reaches the same height as the L-band radio data. As in other CHANG-ES galaxies, such as for example NGC~4631 \citep{CHANGESII2012}, we find the magnetic field and X-ray morphologies to be similar, which indicates an important influence of the magnetic field on the hot gas dynamics.

For the first time a central point source of NGC~4666 was detected at radio wavelengths and further observed in our X-ray data with clear AGN fingerprints, which supports previous findings from X-ray observations that NGC~4666 harbors an AGN. A corresponding bubble-like structure was found below the midplane in the radio data, which was previously suggested by \citet{dahlemetal1997}, who found indications for a bipolar outflow with an associated galactic superwind within the central 6.5\,kpc. In our X-ray data the extension of the bubble-structure is seen, which reaches far into the halo.

NGC~4666 appears different above and below the disk. First, this is visible in the total intensity image of the combined C-band observations (Figure~\ref{fig:N4666_Ccomb_bw}). NGC~4666 shows extraordinary filamentary structures above the disk reaching into the halo like fingers. These radio filaments are comparable to H$\upalpha$ filamentary structures. In contrast to this, shell-like structures are visible below the disk. Secondly, the X-ray analysis revealed higher thermal energy densities in the western part (above the disk). Thirdly, the nonthermal fraction maps in C-band and L-band show an asymmetric fraction of the two halo sides, where the western halo (above the disk) is characterized by higher thermal fractions. The most likely explanation is the interaction of the galaxy with NGC~4668, which is also discussed to be the reason for the starburst happening in NGC~4666 \citep{walteretal2004}. The interacting galaxy NGC~4668 is located to the southeast of NGC~4666.

A scale-height analysis was applied to radio data smoothed to a beam of 13.3"~$\times$~13.3" in both bands. An asymmetry is found in the halo above and below the disk as well as between the two sides of the major axis. The mean scale height of the thick disk in C-band is 1.57~$\pm$~0.21\,kpc and the mean scale height of the thick disk in L-band is 2.16~$\pm$~0.36\,kpc.

The results of RM synthesis show that NGC~4666 is characterized by an X-shaped large-scale magnetic field structure with an observed magnetic field reaching far into the halo. In C-band, the polarized intensity and magnetic field orientation were determined by imaging Stokes Q and U with a robust two weighting. In comparison to RM synthesis with robust two weighting, the polarization map from Stokes Q and Stokes U and robust two weighting shows 1.2 times more polarized flux density. In the uv-tapered and smoothed images from archival VLA and from CHANG-ES data the large-scale X-shaped magnetic field in C-band is visible. In L-band, RM synthesis is useful and necessary to detect polarized intensity from the entire galaxy with a factor of 1.4 of enhanced polarized flux density in comparison to imaging Stokes Q and U (both with robust two weighting).

Large radio halos and large-scale magnetic fields are observed in many CHANG-ES galaxies \citep{wiegertetal2015} and in other edge-on galaxies \citep{krause2009,beck2016}. These findings suggest a general mechanism in spiral galaxies that produces and maintains the large-scale fields. A universal explanation is the mean-field dynamo theory, where a regular field is induced from seed fields. These were probably amplified beforehand by the small-scale dynamo, which amplifies turbulent magnetic fields.

Generally, the CHANG-ES galaxies tend to show less or no polarization on the receding side of the galaxy. Furthermore, the peak of polarized flux density is shifted towards the approaching side of the galaxy. In NGC~4666, polarized emission in L-band was recovered with RM synthesis also in the south side of the galaxy, which is the receding side. In C-band, RM synthesis was not able to recover the polarized emission in the south. By weighting and smoothing the C-band data, low-level polarized intensity was recovered also in the south. Differences in the C-band observations of NGC~4666 could originate from the short observing time and uv distribution leading to parts of the large scale faint polarized emission being missed. This could be the reason why less polarized intensity is detected on the receding side in C-band. Nevertheless it does not explain why the peak of polarized intensity is shifted towards the approaching side, which seems to be an intrinsic feature of the emission. 

The CR transport was modeled with the 1D {\small SPINNAKER} model and found to be advection dominated with an advection speed of 310 km s$^{-1}$. This velocity is slightly higher than the escape velocity of this galaxy estimated by \citet{heesenetal2018} to be v$_{\text{esc}}$~=~$\sqrt{2}$~v$_{\text{rot}}$~$\approx$~280\,km~s$^{-1}$, neglecting the contribution of a dark-matter halo. We therefore conclude that NGC 4666 has a galactic wind, as was already indicated by the box-shaped appearance of the radio halo. We assumed a constant wind speed for simplicity, but relaxing this to an accelerating wind, the escape velocity should be easily surpassed in the halo even if there is an extended dark-matter halo.

The magnetic field strength map was then calculated using the equipartition formula from \citet{beckkrause2005} with the C-band synchrotron map, the resolved spectral index map, and the spheroid path length model. The mean strength of the disk magnetic field is 12.3\,$\mu$G, which is in good agreement with the field strengths of other spiral galaxies \citep{beck2016}. Nevertheless, there are a few drawbacks of field strength determinations via equipartition. The result of the equation leads to an overestimate if the magnetic field varies along the line of sight or within the telescope beam, whereas an underestimate of the field strength happens in regions where energy losses of the CR electrons are important, for example in starburst regions or halos \citep{beck2016}.

The magnetic field strength profile determined from equipartition was compared to the {\small SPINNAKER} model of the magnetic field and found to be similar in the midplane (which is a consequence of setting the $B_0$ parameter to be similar to the equipartition field strength) and very different towards the halo with a factor of 2.5 in the magnetic field strength. In the halo, the {\small SPINNAKER} magnetic field strength is a lower limit for the actual field strength.

Analyses of the RM maps of NGC~4666 derived from RM synthesis reveals that the structure of the disk field is likely axisymmetric with indications of one magnetic field reversal at a radius of about 30" (4\,kpc). The data suggest that the field orientation points inwards and then outwards with increasing radial distance from the center. The large-scale field reversals between the central region and the disk that have been observed in M31 \citep{giessubelbeck2014} and IC342 \citep{beck2015}, as well as the reversal along the azimuth in NGC~4414 \citep{soidaetal2002}, are not comparable with the findings in this paper. This is the first time an indication of a radial field reversal within the disk of an external galaxy has been found. 

In our Milky Way, there is one large-scale magnetic reversal close to the solar radius at about 7\,kpc distance from the Galactic center, while the existence and location(s) of more reversals is still under debate \citep{haverkorn2015}.

The occurrence of radial field reversals is of fundamental importance for the theory of magnetic fields. In mean-field dynamo theory, there are different radial eigenmodes of the mean-field induction equation. Higher-order modes with many radial reversals may have a higher amplitude to begin with, but eventually the lowest order mode without reversals will win out because it grows fastest \citep{poezd1993}. Radial field reversals may also move outward in the nonlinear dynamo stage, which is governed, at least partly, by the slow process of radial diffusion \citep{chamandy2013}. Radial reversals may still persist in present-day galaxies hosting a dynamo in their evolutionary phase.

\citet{moss2012} were able to predict radial field reversals in the disk of a spiral galaxy using a numerical mean-field dynamo model with continuous injection of turbulent fields to represent the effect of supernova explosions in discrete star forming regions with ongoing small-scale dynamo action. In their paper, a face-on view of a spiral galaxy is shown with multiple field reversals, which is used as an example for the construction of the simplified disk model of NGC~4666 (Figure~\ref{fig:N4666_diskmodell}). Our finding of an indication of a reversal in NGC~4666 is in agreement with dynamo theory and supports the $\upalpha - \upomega$ dynamo theory acting in spiral galaxies. Furthermore, it reveals that the Milky Way is probably not an exception in showing a radial field reversal.

\begin{acknowledgements}
We thank the anonymous referee for help in improving this work. We thank B. Adebahr for providing the RM synthesis scripts and A. Basu for detailed discussions and helpful suggestions.  
This research is kindly supported and funded by the Hans-B\"ockler Foundation. This research was also supported and funded by the DFG Research Unit 1254 "Magnetisation of Interstellar and Intergalactic Media: The Prospects of Low-Frequency Radio Observations". The National Radio Astronomy Observatory is a facility of the National Science Foundation operated under cooperative agreement by Associated Universities, Inc. We made use of the NASA's Astrophysics Data System Bibliographic Services, and the NASA/IPAC Extragalactic Database (NED) which is operated by the Jet Propulsion Laboratory, California Institute of Technology, under contract with the National Aeronautics and Space Administration. This research has made use of the VizieR catalogue access tool, CDS, Strasbourg, France. The original description of the VizieR service was published in A\&AS 143, 23.

Funding for the Sloan Digital Sky Survey IV has been provided by the Alfred P. Sloan Foundation, the U.S. Department of Energy Office of Science, and the Participating Institutions. SDSS-IV acknowledges support and resources from the Center for High-Performance Computing at the University of Utah. The SDSS web site is www.sdss.org. SDSS-IV is managed by the Astrophysical Research Consortium for the 
Participating Institutions of the SDSS Collaboration including the 
Brazilian Participation Group, the Carnegie Institution for Science, 
Carnegie Mellon University, the Chilean Participation Group, the French Participation Group, Harvard-Smithsonian Center for Astrophysics, 
Instituto de Astrof\'isica de Canarias, The Johns Hopkins University, 
Kavli Institute for the Physics and Mathematics of the Universe (IPMU) / 
University of Tokyo, Lawrence Berkeley National Laboratory, 
Leibniz Institut f\"ur  Astrophysik Potsdam (AIP),  
Max-Planck-Institut f\"ur  Astronomie (MPIA Heidelberg), 
Max-Planck-Institut f\"ur  Astrophysik (MPA Garching), 
Max-Planck-Institut f\"ur  Extraterrestrische Physik (MPE), 
National Astronomical Observatories of China, New Mexico State University, 
New York University, University of Notre Dame, 
Observat\'ario Nacional / MCTI, The Ohio State University, 
Pennsylvania State University, Shanghai Astronomical Observatory, 
United Kingdom Participation Group,
Universidad Nacional Aut\'onoma de M\'exico, University of Arizona, 
University of Colorado Boulder, University of Oxford, University of Portsmouth, 
University of Utah, University of Virginia, University of Washington, University of Wisconsin, Vanderbilt University, and Yale University.

\end{acknowledgements}

\bibliography{Bibliography}
\bibliographystyle{aa}

\begin{appendix} 
\section{Additional figures}
\label{app:additional_figures}

\subsection{Radio data}
\begin{figure}[h]
        \centering
        \includegraphics[width=0.48\textwidth]{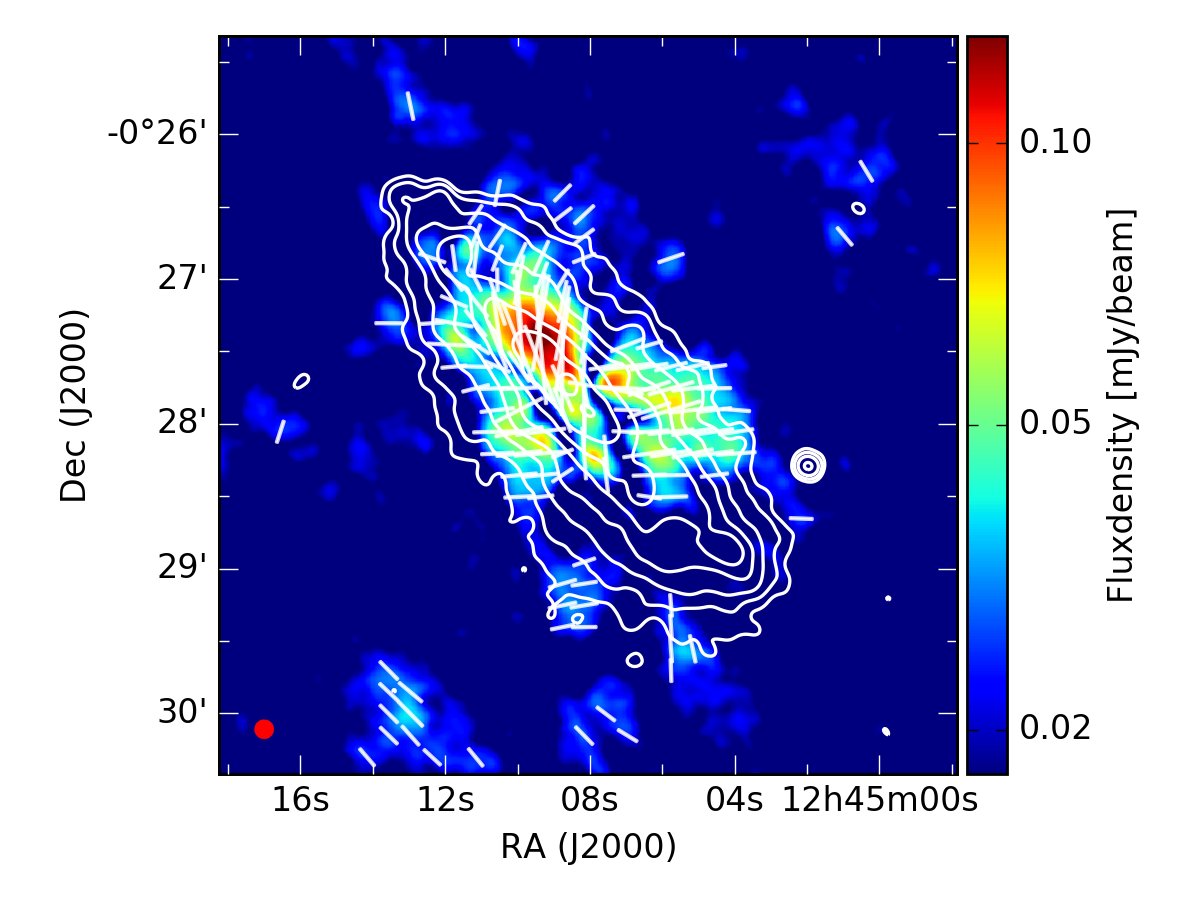}
\caption{Polarized intensity image of NGC~4666 from C-band with robust two weighting, a beam of 7"~$\times$~7" in red (bottom left), and a  $\sigma$ of 7.0\,$\mu$Jy/beam. White Stokes $I$ contours starting at a 3$\sigma$ level with a $\sigma$ of 8.1\,$\mu$Jy/beam and increase in powers of 2 (up to 128) with an additional contour at 6.2\,mJy to mark the central source with robust zero weighting. The magnetic field orientations are shown in white.}
        \label{fig:N4666_Ccomb_rob2_pol}
\end{figure}

\begin{figure}[h]
        \centering
        \includegraphics[width=0.48\textwidth]{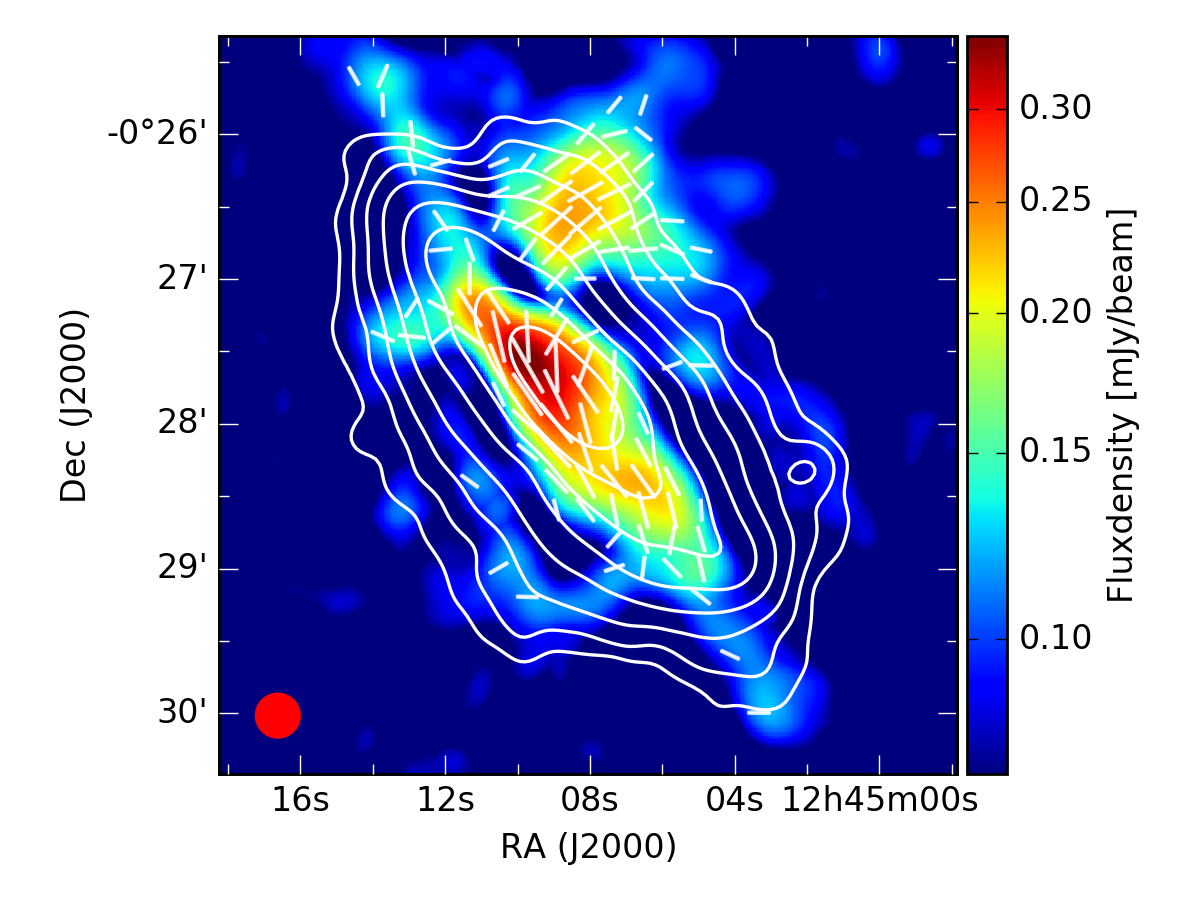}
        \caption{Polarized intensity image of NGC~4666 from L-band using RM synthesis with a robust two weighting, a beam of 18"~$\times$~18" in red (bottom left), and a $\sigma$ of 20.0\,$\mu$Jy/beam. White Stokes I contours starting at a 3$\sigma$ level with a $\sigma$ of 40.1\,$\mu$Jy/beam and increase in powers of 2 (up to 128) with robust two weighting. The magnetic field orientations are shown in white.}
        \label{fig:N4666_Lcomb_synth_rob2_pol}
\end{figure}

\begin{figure}
        \centering
        \includegraphics[width=0.52\textwidth]{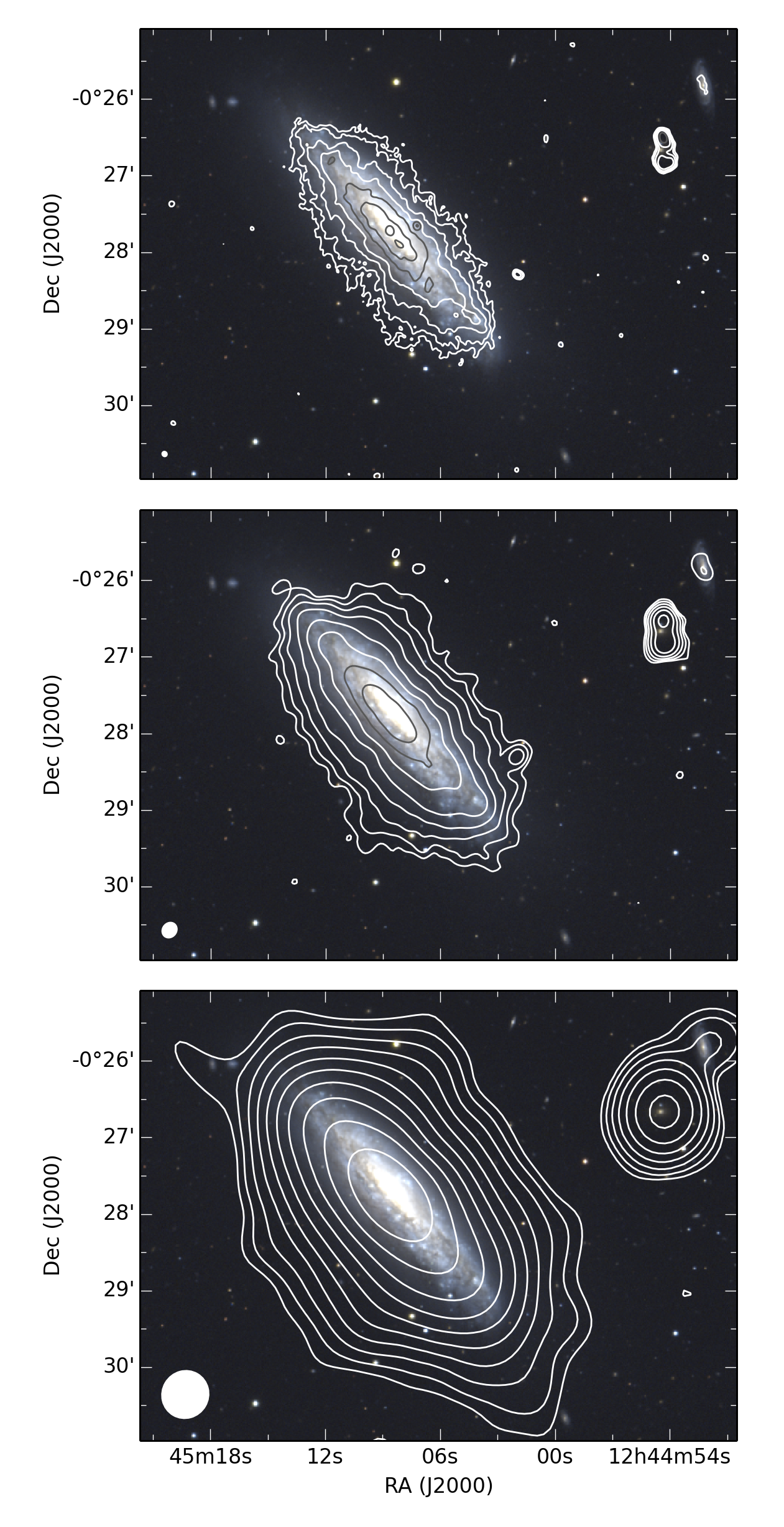}
        \caption{NGC~4666 Stokes I image of B-configuration L-band (BL), C-configuration L-band (CL), and D-configuration L-band (DL), contours start at a 3$\sigma$ level with a $\sigma$ of 18.0\,$\mu$Jy/beam (BL), 30\,$\mu$Jy/beam (CL) and 34\,$\mu$Jy/beam (DL) and increase in powers of 2 (up to 64, 128 and 512). The beam sizes of 3.9" $\times$ 4.6", 10.7" $\times$ 12.3", and 36.0" $\times$ 37.4" are shown in the bottom-left corner of each image. Cleaning was done with robust zero weighting.}
        \label{fig:N4666-BCDL_all}
\end{figure}

\begin{figure*}
         \begin{minipage}[t]{0.49\textwidth}
                        \includegraphics[width=\textwidth]{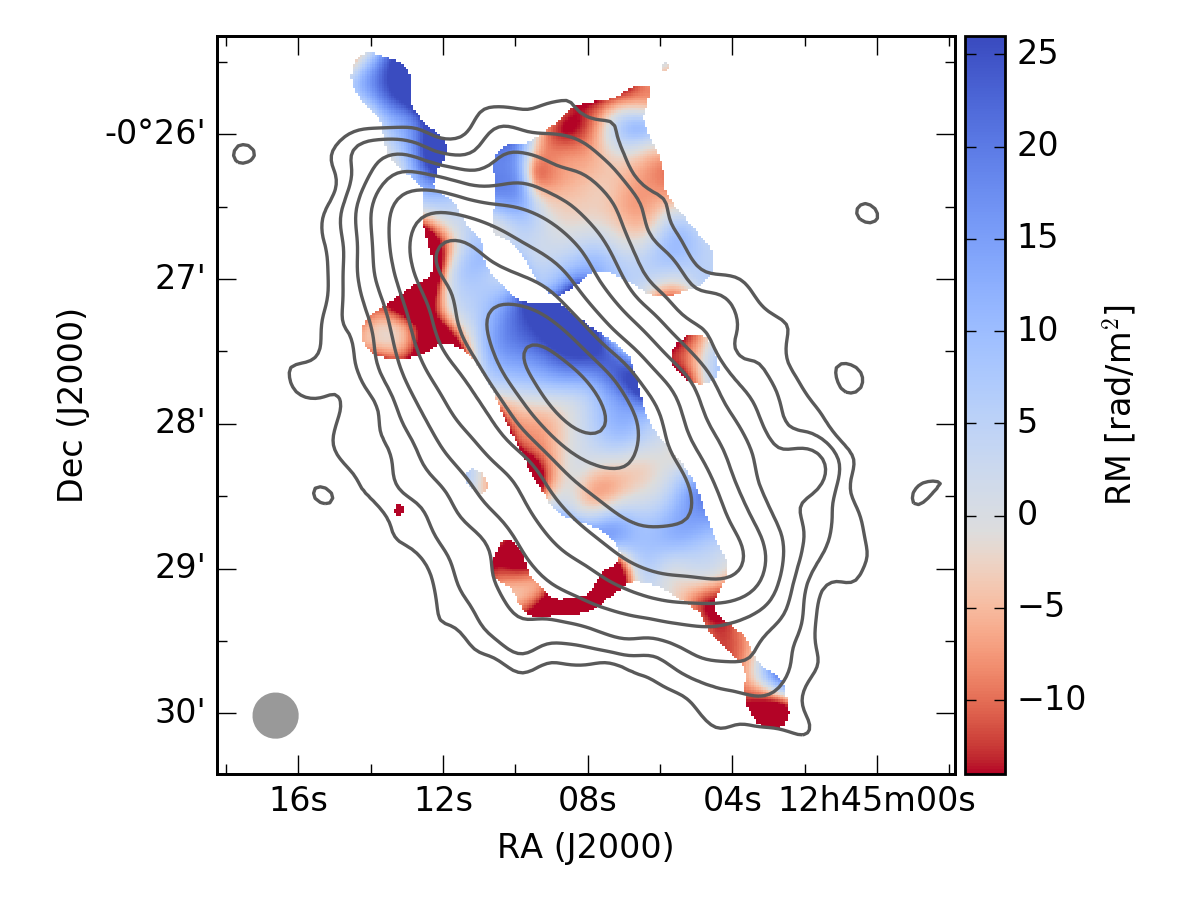}
        \caption{RM map of NGC~4666 from L-band on Stokes I contours from Figure~\ref{fig:N4666_Lcomb_synth_rob2} and a beam of 18"~$\times$~18" in gray. Contours start at a 3$\upsigma$ level with a $\upsigma$ of 40.1\,$\upmu$Jy/beam and increase in powers of 2 (up to 256). The RM map is cut below the 3$\upsigma$ level of 11.1$\upmu$Jy/beam of the polarized intensity map. The mean error is 0.5\,rad/m$^2$.}
        \label{fig:N4666-L-RM}
        \end{minipage}
 \begin{minipage}[t]{0.49\textwidth}
        \hspace{0.01\textwidth}
        \end{minipage}
        \end{figure*}

In Figs.~\ref{fig:N4666_Ccomb_rob2_pol} and \ref{fig:N4666_Lcomb_synth_rob2_pol}, polarized intensity is shown with Stokes I contours for both frequencies. In Fig.~\ref{fig:N4666-BCDL_all} images of the different configurations of L-band are shown. In Fig.~\ref{fig:N4666-L-RM} the RM map of L-band are presented.

\subsection{SPINNAKER}
\begin{figure*}
\begin{minipage}[c]{0.63\textwidth}
 \includegraphics[width=1.0\textwidth]{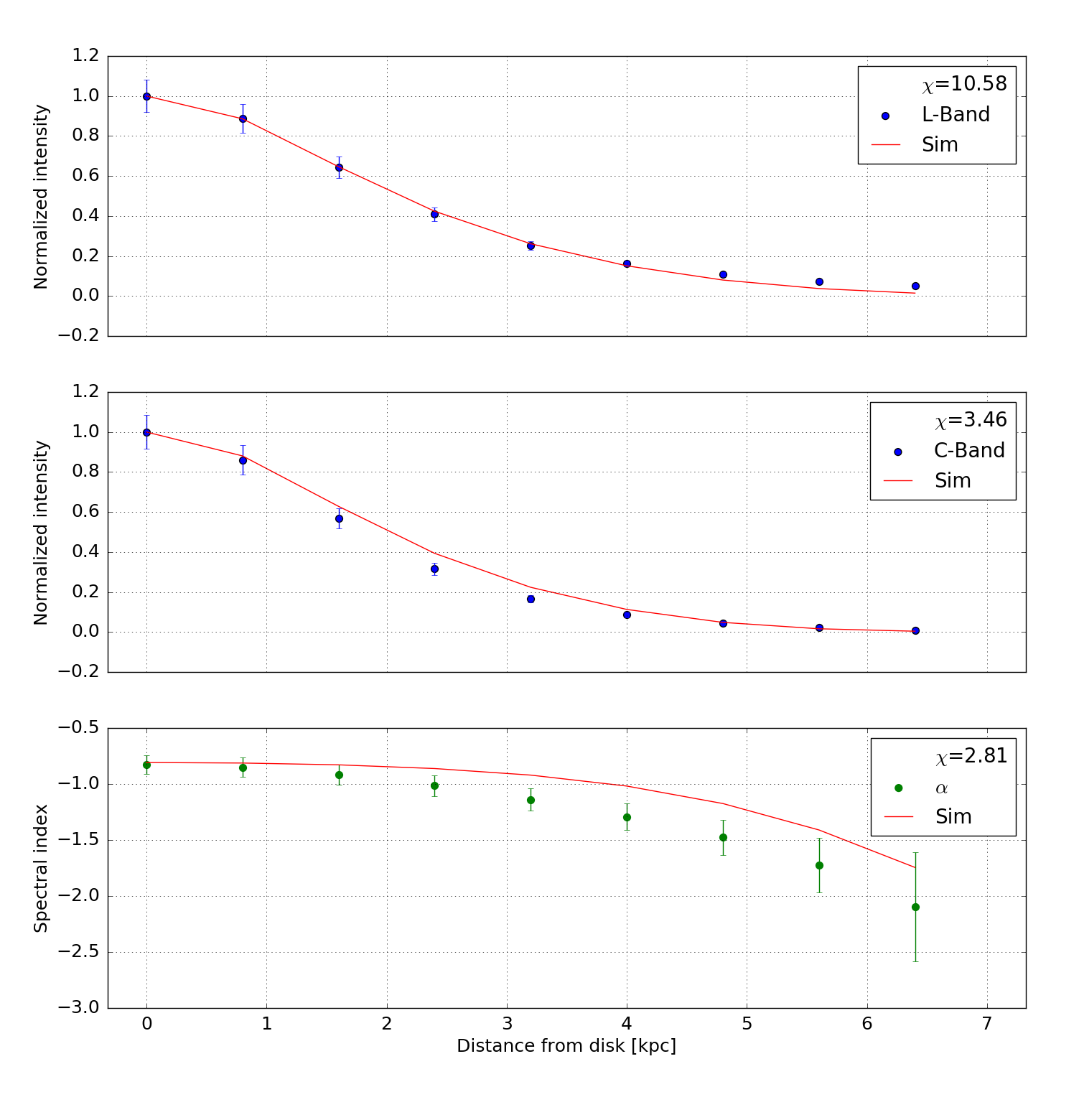}
 \captionof{figure}{CR transport model for NGC~4666 with {\scriptsize SPINTERACTIVE} with diffusion for comparison. The data points are the observed values from 1.5 GHz (L-band), 6GHz (C-band) and the spectral index, the red line represents the model simulation. The errors are represented by a weighted standard deviation. The adjustable parameters are listed in Table~\ref{tab:spinnakerpar_dif}.}
 \label{fig:N4666-spinteractive-dif}
\end{minipage}
\begin{minipage}[c]{0.07\textwidth}
\hfill
\end{minipage}
\begin{minipage}[c]{0.3\textwidth}
\captionof{table}[{\small SPINNAKER} parameters.]{{\small SPINNAKER} parameter for diffusion.}  
\begin{tabular}[c]{lc}
\hline \hline
Parameter       &  Value \\ \hline
$\gamma_0$    & 2.6  \\
$B_0$ [$\upmu$G]   & 13.1 \\
$B_1$  [$\upmu$G]  &3.2  \\
$hb1$ ["] &  2.5 \\
$hb2$ ["]   & 5.0\\
\hline
Diffusion: & \, \\
D$_0$ [10$^{28}$ cm$^2$~s$^{-1}$] & 7.0 \\
$\upmu$ & 0.5 \\
$\upchi^2_{\text{dif}}$ & 5.6 \\
\hline
\end{tabular}
\normalsize
\label{tab:spinnakerpar_dif}
\end{minipage}
\end{figure*}

In Figure~\ref{fig:N4666-spinteractive-dif} we show the {\small SPINNAKER} result with diffusion, which was done for comparison and not used.

\end{appendix}
\end{document}